\documentclass[journal=jpccck,manuscript=article]{achemso}
\usepackage[english]{babel}
\usepackage[latin1]{inputenc}
\usepackage[T1]{fontenc}
\usepackage{graphicx}
\usepackage{hyperref}
\usepackage[normalem]{ulem}
\usepackage{latexsym}
\usepackage{url}
\usepackage{amsmath} 
\usepackage{amsthm}
\usepackage{multirow}
\usepackage{pstricks}
\usepackage{pgfplots}
\usepackage{amsfonts}
 \usepackage{times}
 \usepackage{units} 
\usepackage{doi}
\newcommand{\superscript}[1]{\ensuremath{^{\textrm{#1}}}}
\newcommand{\subscript}[1]{\ensuremath{_{\textrm{#1}}}}

\author{Conn O'Rourke}
\email{ucapcor@live.ucl.ac.uk}
\affiliation[LCN]
{London Centre for Nanotechnology, 17-19 Gordon St, London, WC1H 0AH}
\alsoaffiliation[UCL]
{Department of Physics \& Astronomy, University College London, Gower St,
London, WC1E 6BT}
\alsoaffiliation[UCL Satellite, MANA]
{UCL Satellite, International Centre for Materials Nanoarchitectonics (MANA), National Institute for Materials Science (NIMS), 1-1 Namiki, Tsukuba, Ibaraki 305-0044, Japan}
\author{David R. Bowler}
\affiliation[LCN]
{London Centre for Nanotechnology, 17-19 Gordon St, London, WC1H 0AH}
\alsoaffiliation[UCL]
{Department of Physics \& Astronomy, University College London, Gower St,
London, WC1E 6BT}
\alsoaffiliation[UCL Satellite, MANA]
{UCL Satellite, International Centre for Materials Nanoarchitectonics (MANA), National Institute for Materials Science (NIMS), 1-1 Namiki, Tsukuba, Ibaraki 305-0044, Japan}
\title[]{DSSC Anchoring Groups: A Surface Dependent Decision}

\begin{document}
\begin{abstract}
Electrodes in dye sensitised solar cells (DSSCs) are typically
nanocrystalline anatase TiO\subscript{2} with a majority (101) surface
exposed. Generally the sensitising dye employs a carboxylic anchoring
moiety through which it adheres to the TiO\subscript{2} surface.
Recent interest in exploiting the properties
of differing TiO\subscript{2} electrode morphologies,
such as rutile nanorods exposing the (110) surface and 
anatase electrodes with high percentages of the (001) surface exposed, 
begs the question of whether this anchoring
strategy is best, irrespective of the majority surface exposed. 
Here we address this question by presenting density 
functional theory calculations contrasting 
the binding properties of two promising anchoring groups, phosphonic 
acid and boronic acid, to 
that of carboxylic acid. Anchor-electrode
 interactions are studied for the prototypical anatase (101) surface, 
along 
with the anatase (001) and rutile (110) surfaces.  
Finally the effect of using these alternative anchoring groups
to bind a typical coumarin dye (NKX-2311) to these TiO\subscript{2}
substrates is examined. Significant differences in the binding
properties are found depending on both the anchor and surface,
illustrating that the choice of anchor is necessarily
dependent upon the surface exposed in the electrode.
In particular the boronic acid is found to show the potential
to be an excellent anchor choice for electrodes exposing the
anatase (001) surface.

\end{abstract}

\section{Introduction}

Titanium dioxide (TiO\subscript{2}) has a wide range of practical
applications; for example in 
photocatalysis \cite{chptr3_photoc,chptr3_photoc2}, as a white 
pigment \cite{chptr3_whitepig}, 
photo-degradation of molecules at its surface make it useful as 
an anti-bacterial agent\cite{chptr3_bacter} and in waste water 
treatment\cite{chptr3_waterp}, 
and of course it is used in dye sensitised solar cells\cite{OReagan_Grat}.
Due to the number of technological uses, considerable effort has 
been expended in  
characterising the properties of TiO\subscript{2}(for a review see
\cite{chptr3_diebold}). 
TiO\subscript{2} surface structure and its interaction with 
molecules is of fundamental importance to many potential applications,
 not least DSSCs. This fact has acted as the 
driving force behind the characterisation of titania surfaces, and 
the study of how these surfaces and molecules interact.  
In this paper we examine how the binding properties of molecules 
to TiO\subscript{2} surfaces depends on both the surface adhered to,
 as well as the anchoring group the molecule uses to bind to 
the surface. 

Dye sensitised solar cells' efficiency relies heavily on the 
interplay between sensitising dye and the TiO\subscript{2} surface
to which it binds. A strong interaction will ensure the dye remains
bound to the surface, and good electronic overlap between the 
surface and dye is essential for efficient charge transfer \cite{chpt3_durrant_charge}.
This stresses the importance played by the anchoring moiety in a 
sensitising dye. Ruthenium based record efficiency dyes 
(N719, N3, black dye \cite{chpt3_black_dye,chpt3_N719,chpt3_N3}) all contain one or more carboxylic
acid binding groups, and the vast majority of sensitising dyes 
have followed this anchoring strategy. 

Titanium dioxide exists in several polymorphs, two of which
are used in DSSCs, anatase and rutile. 
TiO\subscript{2} in nanocrystalline form is 
most thermodynamically 
stable in its anatase phase \cite{chpt3_ana}, with the (101) face 
dominating more than 94\% of the crystal surface \cite{chptr3_selloni}.
Coupling this with the fact that most sensitising dyes use a
carboxylic binding moiety, highlights the importance of the 
(101)-carboxylic acid interaction. Several experimental and
theoretical studies have been devoted to examining the 
interaction of carboxylic anchors with the TiO\subscript{2} (101) 
surface\cite{chptr3_formic_101,chptr3_formic_101_exp}.

However, carboxylic acid groups are not the only choice for 
anchoring dyes to TiO\subscript{2} surfaces, and bias towards 
them may be at the expense of other potentially useful candidates 
being neglected. Examples include phosphonic 
acid \cite{chptr3_phos_dye,chptr3_phos_expt}, boronic acid\cite{chptr3_bor_dye}, 
and cyano-benzoic acid\cite{chptr3_cyano_dye}, all 
of which have been used as binding groups in DSSCs. Notably 
dyes utilising phosphonic acids have shown a stronger 
binding interaction with TiO\subscript{2} than 
carboxylic acids\cite{chptr3_phos_expt,chptr3_phos_110}, suggesting an advantage over 
those utilising carboxylic acids. Increased binding strength
could lead to higher dye uptake and enhanced longevity over
carboxylic acid bound dyes, as these tend to slowly 
desorb from the TiO\subscript{2} surface.  

In a similar vein, while the prevalence of the anatase (101) surface
make it extremely important, the 
interaction of dyes with other TiO\subscript{2} surfaces and 
polymorphs should not be neglected. For example,
the use of rutile TiO\subscript{2} nanorods exposing the (110)
surface has
been explored as a potential avenue for increased electron transport
rates through the electrode, resulting from reduced grain boundaries
\cite{chptr3_NR_trans,chptr3_NR_110maj1,chptr3_NR_1}. Also interest in the less stable (001) anatase surface
is increasing due to recent work showing that electrodes exposing
the (001) face significantly improves device performance
as a result of enhanced light scattering and increased surface 
activity \cite{chptr3_001_DSSC1,chptr3_001_DSSC2,CHPTR3_001_DSSC3}. 

Interactions between any particular anchor group and 
TiO\subscript{2} will necessarily differ depending on the surface. 
The aim of this paper, therefore, is to assess the relative 
merits of three potential anchor groups when binding to these less 
utilised, but increasingly important, surfaces. Firstly we introduce
the two most important polymorphs of TiO\subscript{2},
with an examination of the three 
mentioned surfaces; anatase (101), anatase (001) and rutile (110). 
Taking each of these in turn we then investigate the 
adsorption of two potential anchoring
groups, phosphonic acid and boronic acid, at these surfaces and
contrast to that of carboxylic acid. Finally we examine the properties
of a full dye, the NKX-2311 coumarin dye, bound to each of these
surfaces through all three anchoring groups.

\section{Computational Details}


All the calculations in this paper have been performed using the 
plane wave DFT code VASP (version 5.3.3) \cite{chpt3_VASP}. 
Electron exchange and correlation was treated within the 
generalised gradient 
approximation of Perdew and Wang\cite{chpt3_PW91-2} and the
pseudopotential method was utilised in the form of ultrasoft
Vanderbilt pseudopotentials\cite{chpt3_vanderbilt} to treat core
electrons. For Titanium atoms the 4s\superscript{2}3d\superscript{2} electrons
are treated as valence electrons, for boron, oxygen, carbon and nitrogen the 
1s electrons are treated as being in the core. Phosphorous atoms are treated 
with the 1s\superscript{2}2s\superscript{2}2p\superscript{6} electrons 
considered to be in the core. Semi-local functionals such as the PW91
functional employed here are known to incorrectly describe defect states
in TiO\subscript{2}, such as oxygen vacancies. Hybrid functionals and
GGA corrected for on-site Coulomb interactions (GGA+U) have been 
shown to improve the desciption of these defects, however as we are 
examining the interaction of adsorbates with clean TiO\subscript{2} 
surfaces we restrict our approch to that of the pure GGA functional.


Sampling of the reciprocal space is performed using a 
Monkhorst-Pack k-point grid.  For the bulk calculations we have employed 
a k-point mesh of ($6\times6\times3$) and ($3\times3\times6$)
for anatase and rutile respectively.
Total fixed-volume energies at a cut-off of 450 eV are found to be 
converged to within 17 meV 
of those obtained at a cut-off of 650 eV for both rutile and anatase.
During the calculation of the lattice parameters the higher 
cut-off of 650 eV has been used to ensure accuracy. 
Structural relaxation is performed using the RMM-DIIS\cite{chpt3_RMM}
method until the forces on free ions were less than 0.005 eV/\AA\ for
bulk calculations. Calculated bulk lattice parameters a \& c show 
good agreement with experiment\cite{chpt3_anatase_exp}
 and previous computational work\cite{chptr3_selloni},
 and are 3.817\AA\ \& 9.737\AA\ for 
anatase and 4.602\AA\ \& 2.949\AA\ for rutile.

For the surface and molecular adsorption calculations 
the lesser 450 eV cut-off has been used, along with a less
 stringent maximum
force criterion of 0.03 eV\AA\superscript{-1}. Similarly the k-point
 mesh density has been reduced to 1 perpendicular to the surface.
In order to replicate the bulk, the bottom layer of 
the anatase slabs have been restrained to remain in the bulk
position. For rutile we have also performed calculations with the bottom
layer free to relax.

In all surface and adsorption
calculations, in order to prevent spurious interactions between adjacent
images, the cell size is such that at least 9\AA\ of 
vacuum separates periodic slab images (and adjacent molecules in
adsorption calculations).  

\section{TiO\subscript{2} Surfaces}



In order to obtain an accurate picture of the  molecular adsorption 
at TiO\subscript{2} surfaces, it is important to converge the
surface properties with respect to the number of layers in our 
slab model. 
In this section we characterise the anatase (101), 
anatase (001) and rutile (110) surfaces and converge the
surface properties to an acceptable level so as to accurately 
examine the more computationally demanding molecular adsorptions 
yet to be performed with minimum computational effort.


\subsection{Anatase (101)}

\begin{figure}[h]
 \begin{center}
  \begin{tabular}{c c}
    \begin{minipage}[h]{0.45\textwidth}
      \includegraphics[trim = 0mm 20mm 0mm 10mm, clip, width=1.\textwidth]{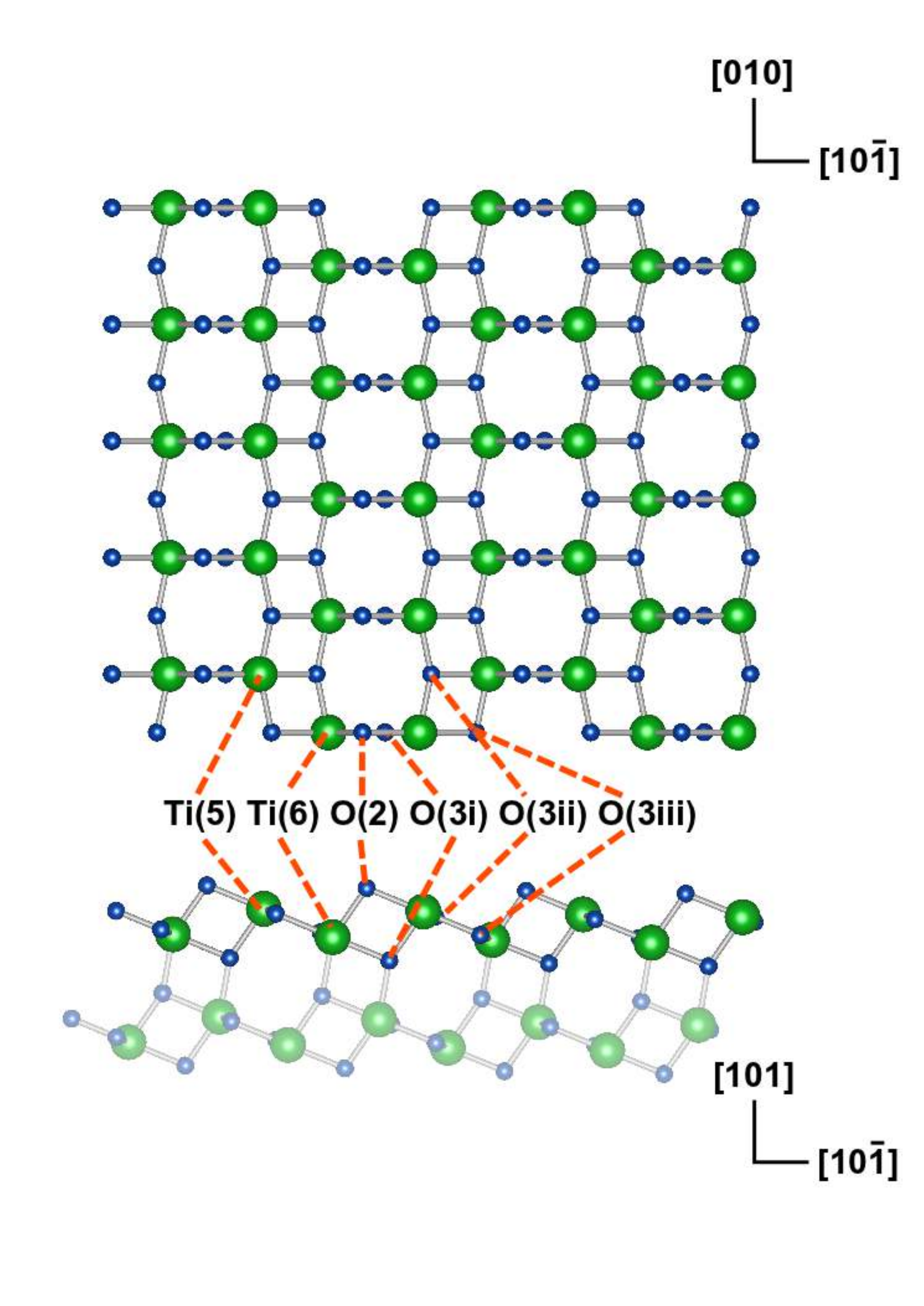}
    \end{minipage}
   &
    \begin{minipage}[h]{0.45\textwidth}
      \includegraphics[trim = 0mm 20mm 0mm 10mm, clip, width=1.\textwidth]{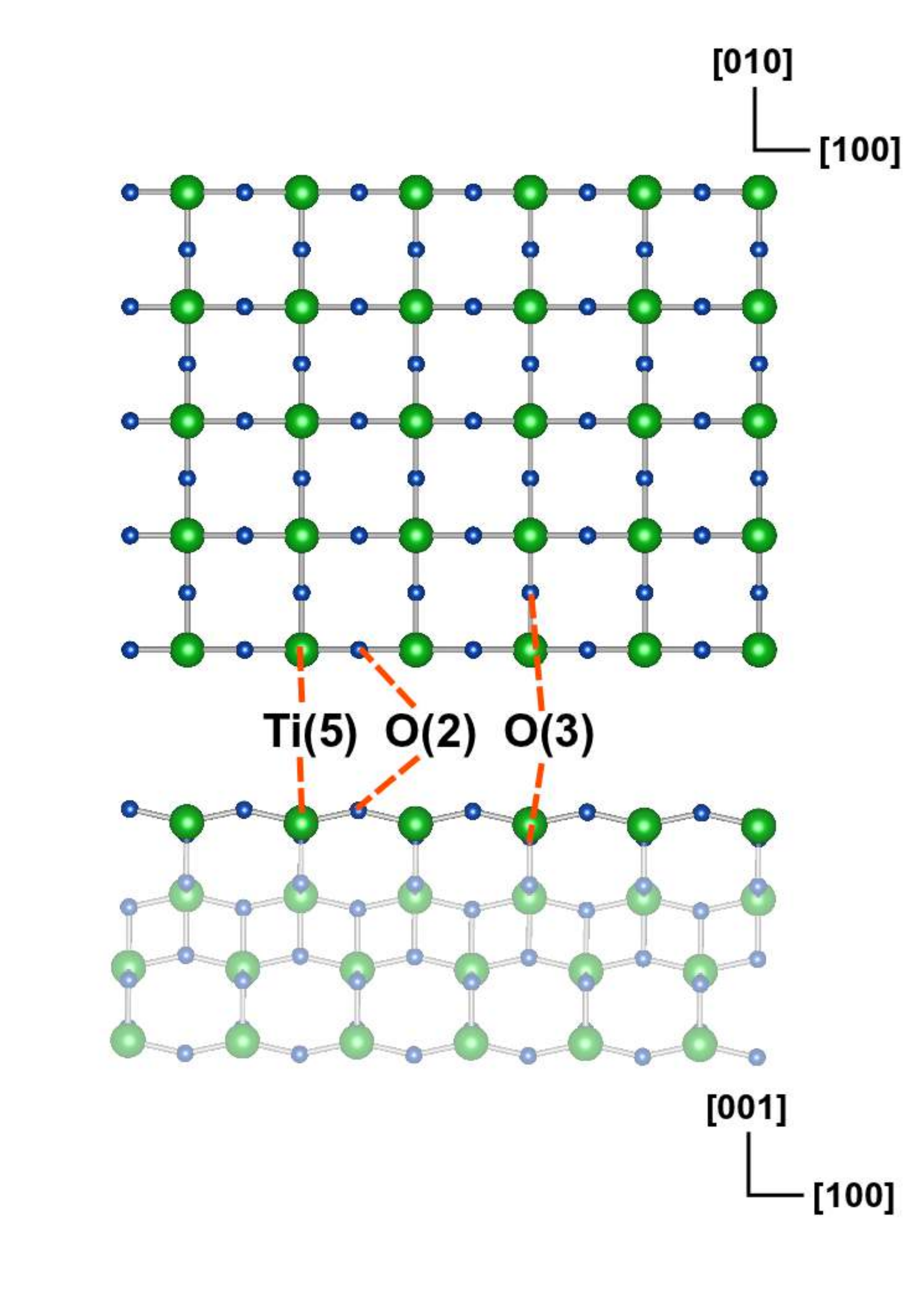}
    \end{minipage}
   \\
 \end{tabular}
 \caption{Anatase surfaces: (101) left and (001) right. Titanium atoms are green, oxygen atoms are blue. Both top views (top) and side views (bottom) are included, with coordination of surface atoms labelled.}
 \label{chpt3_ana_surf}
 \end{center}
\end{figure}

On formation of the stoichiometric (101) surface both fivefold-coordinated titanium (Ti(5)) and
twofold-coordinated oxygen (O(2)) atoms become exposed. Also present in the topmost layer
are O(3) and the less exposed Ti(6) atoms, figure \ref{chpt3_ana_surf}. 

The displacements resulting from surface relaxation in the [101] and [10$\bar{1}$] 
directions of the topmost atoms for three and six 
layer slabs can be seen in figure \ref{chpt3_101}. Most notable are the large outward
displacement of the O(3ii) and inward displacement of the Ti(5) atoms in the [101] direction.
A more jagged surface along the [010] direction, with the O(3ii) atom protruding beyond the Ti(5)
atom, results.

Important also is the reduction in length and change of direction of the O(2)-Ti(6) bond,
from 1.95 (\AA) to 1.87 (\AA), as a result of the substantial outward [101] relaxation of the
 Ti(6) atom and large displacement in the [10$\bar{1}$] direction of the O(2) atom. These surface
 characteristics reproduce well those found in other studies using the PBE functional \cite{chptr3_selloni}.

Surface formation energies have been calculated by the subtraction of the bulk energy per layer
 times the number of layers from the total energy obtained, and dividing by the total exposed area, 2A \eqref{chpt3_unrel}. Relaxed
  surface energies are then computed by subtraction of the energy change on relaxation per
   unit area from this unrelaxed surface energy \eqref{chpt3_rel}. Both relaxed and unrelaxed surface
   energies for slabs of varying thickness are presented in table \ref{chpt3_101}(i).

\begin{subequations}\label{chpt3_Surf}
\begin{align}
E_{unrel} & =\frac{1}{2 A} \lbrack E_{un}^{Tot} - n E_{bulk}\rbrack\label{chpt3_unrel}\\
E_{rel}   & =\lbrack E_{unrel} - \frac{\Delta E}{A}\rbrack\label{chpt3_rel}
\end{align}
\end{subequations}

For the six layer slab good agreement is found between the calculated unrelaxed (relaxed) surface
energies of 1.267 J/m\superscript{2} (0.537 J/m\superscript{2}) and that of previous work for a six layer slab using the
PBE functional, 1.28 J/m\superscript{2} (0.49 J/m\superscript{2}) \cite{chptr3_selloni}.

Unrelaxed (relaxed) surface formation energies for the three layer slab are
found to be converged to within 0.01 J/m\superscript{2} (0.015 J/m\superscript{2}) with respect to
that of the six layer slab. Although the convergence of relaxation displacements for the three layer
representation are not as tight, it can be seen that the three layer model does qualitatively
reproduce to a good degree the characteristic displacements found in the six layer slab. As a
trade-off between computational ease and accuracy we have opted for the three layer TiO\subscript{2} (101) slab for the calculations 
involving adsorption of molecules.

\begin{table}[h]
\begin{tabular}{c c}

  \begin{minipage}[t]{0.3\textwidth}
  \begin{center}
   \begin{tabular}{c c c }
   \hline
   \hline
   & \multicolumn{2}{c}{Surface Energy}\\
   N\subscript{lay} & E\subscript{unrel}& E\subscript{rel} \\
   & (J/m\superscript{2})&(J/m\superscript{2}) \\
   \hline
   3&1.259 &0.522 \\
   4&1.260 &0.514 \\
   5&1.260 &0.534 \\
   6&1.267 &0.537 \\
   \hline
   \hline
   \end{tabular}
  \end{center}
  \end{minipage}
&
  \begin{minipage}[t]{0.7\textwidth}
  \begin{center}
   \begin{tabular}{c c c}
   \hline
   \hline
   &\multicolumn{2}{c}{Atomic Disp. (\AA)}\\
   Label& [10$\bar{1}$]&[101]\\
   \hline
   Ti(5)   &-0.114 (-0.131)& -0.101  (-0.140)\\
   Ti(6)   & 0.027 ( 0.012)&  0.207  ( 0.154)\\
   O(2)    & 0.140 ( 0.117)&  0.017  ( 0.031)\\
   O(3i)   & 0.031 ( 0.030)& -0.005  (-0.051)\\
   O(3ii)  & 0.051 ( 0.030)&  0.262  ( 0.219)\\
   O(3iii) &-0.002 (-0.020)&  0.081  ( 0.034)\\
   \hline
   \hline
  \end{tabular}
 \end{center}
 \end{minipage}
\\
\textbf{(i)}&\textbf{(ii)}\\
 \end{tabular}
\caption{\textbf{Anatase (101) Surface:} \textbf{(i)} Relaxed and unrelaxed surface energies for different layer anatase slabs
\textbf{(ii)} Displacements from bulk position on relaxation of Anatase (101) surface for 3 layer slab (values for 6 layer slab are shown in brackets)}
\label{chpt3_101}
\end{table}

\subsection{Anatase (001)}

The unreconstructed (001) surface is reported to be much less stable than
the (101) surface\cite{chptr3_selloni,chptr3_vitadini_form_001}. 
However based on a Wulff construction it 
has been shown that for anatase nanocrystals, although comprising 
a much smaller area than the majority (101) surface, the (001) 
crystal face will still be exposed\cite{chptr3_selloni} in agreement with experiment.

Cleaving the anatase lattice perpendicular to the (001) surface
exposes both two-fold and three-fold coordinated oxygen atoms, as 
seen in figure \ref{chpt3_ana_surf}. 
In contrast to the (101) surface however, the O(2) 
atoms number \nicefrac{1}{2} of those oxygen atoms exposed in 
the surface (for the (101) surface O(2) atoms make up 
\nicefrac{1}{3}). Similarly the (001) surface exposes only Ti(5)
atoms, as opposed to the (101) surface which expose equal numbers of
Ti(5) and Ti(6), as seen in  \ref{chpt3_ana_surf}. 
The high proportion of undercoordinated atoms in the (001) surface 
goes some way to explain its reported high reactivity\cite{chptr3_001_reactive,chptr3_001_reactive2}.

It is this high reactivity that makes the (001) surface of interest
for many applications such as photocatalysis and 
photo-degradation of organic molecules \cite{chptr3_001_reactive2,chptr3_001_photoc,chptr3_001_photodegrad}. Recent work has shown 
that it is possible to increase the exposed percentage of 
the (001) surface by, for example, using hydrofluoric acid as a 
capping agent, thus improving 
its photocatalytic 
properties\cite{chptr3_001_photoc}. 
Similarly increasing the percentage of the 
exposed (001) face has been shown to improve DSSC 
performance\cite{chptr3_Anatase001_3}.
 Several reasons for this improvement have been
suggested, such as the increased reactivity leading to higher 
dye adsorption, improved light scattering 
properties,
and improved crystallinity leading to lower recombination 
rates \cite{chptr3_Anatase001_3,chptr3_001_dsscrecomb,chptr3_dssc_001_ads}.

Calculated surface energies for the (001) surface are presented in 
table \ref{chpt3_001_surf_en}, and can be seen to be significantly
higher than those of the (101) surface, as expected. Our calculated
value of 1.145 J/m\superscript{2} (1.06 J/m\superscript{2}) for the
 unrelaxed (relaxed) surface energy of the 4 layer slab is in good 
agreement with the result of 1.12 J/m\superscript{2} 
(0.98 J/m\superscript{2}) reported elsewhere \cite{chptr3_selloni}.
 It can be 
seen that the relaxed surface energies are converged to within 
0.004 J/m\superscript{2} for a four layer slab compared to those
calculated for a seven layer slab, and we employ the four layer 
slab for the calculations involving molecular adsorption.

\begin{table}[h]
\begin{tabular}{c c}
  \begin{minipage}[t]{0.3\textwidth}
  \begin{center}
   \begin{tabular}{c c c }
   \hline
   \hline
   & \multicolumn{2}{c}{Surface Energy}\\
   N\subscript{lay} & E\subscript{unrel}& E\subscript{rel} \\
   & (J/m\superscript{2})&(J/m\superscript{2}) \\
   \hline
   2& 1.172& 1.053\\
   3& 1.220& 1.054\\
   4& 1.145& 1.060\\
   5& 1.146& 1.060\\
   6& 1.144& 1.055\\
   7& 1.144& 1.056\\
   \hline
   \hline
   \end{tabular}
  \end{center}
  \end{minipage}
&
  \begin{minipage}[t]{0.7\textwidth}
  \begin{center}
   \begin{tabular}{c c c}
   \hline
   \hline
   &\multicolumn{2}{c}{Atomic Disp. (\AA)}\\
   Label& [001]&\\
   \hline
   Ti(5)   & -0.047 (+0.048)&\\
    O(2)   & +0.046 (+0.034)&\\
    O(3)   & +0.020 (-0.010)&\\
   \hline
   \hline
  \end{tabular}
 \end{center}
 \end{minipage}
\\
\textbf{(i)}&\textbf{(ii)}\\
 \end{tabular}
\caption{\textbf{Anatase (001) Surface:} \textbf{(i)} Relaxed and unrelaxed surface energies for different layer anatase slabs
\textbf{(ii)} Displacements from bulk position on relaxation of Anatase (001) surface for 3 layer slab (values for 7 layer slab are shown in brackets)}
\label{chpt3_001_surf_en}
\end{table}

Atomic displacements for the surface atoms in the [001] direction are
also given in table \ref{chpt3_001_surf_en}, and agree well with 
previous work\cite{chptr3_selloni}. 
Along the [100] direction we find that in order to obtain
the broken symmetry surface solution reported 
elsewhere\cite{chptr3_selloni},
with the Ti(5)-O(2)-Ti(5) bonds extending and shortening respectively to
2.20\AA\ and 1.76\AA, we require a larger simulation cell  
than the ($1\times 1\times 1$) here. 
Indeed this symetry breaking was also found in larger
supercells used for the adsorption of acidic binding 
groups, with the Ti(5)-O(2) bond lengths changing to 2.20 and
1.78\AA. 


\subsection{Rutile (110)}

Typical DSSC electrodes are composed of nanocrystalline 
TiO\subscript{2} particles, in anatase form. 
Grain boundaries between the crystals can lead to high rates 
of recombination and low electron diffusion coefficients.
One potential approach to minimise this effect has been 
to construct single crystalline rutile nanorods, reducing the
grain boundaries and improving electron 
transport\cite{chptr3_NR_trans2} and
also leading to increased surface areas, thereby improving dye
take-up\cite{chptr3_NR_eff}. Grown
along the [001] direction these nanorods expose a majority (110) 
surface for dye adsorption\cite{chptr3_NR_110maj1,chptr3_NR_110maj}, highlighting the importance of the
interaction between the (110) surface and potential dye anchors.

\begin{figure}[h]
 \begin{center}
    \begin{minipage}[h]{0.45\textwidth}
      \includegraphics[trim = 0mm 0mm 0mm 0mm, clip, width=1.\textwidth]{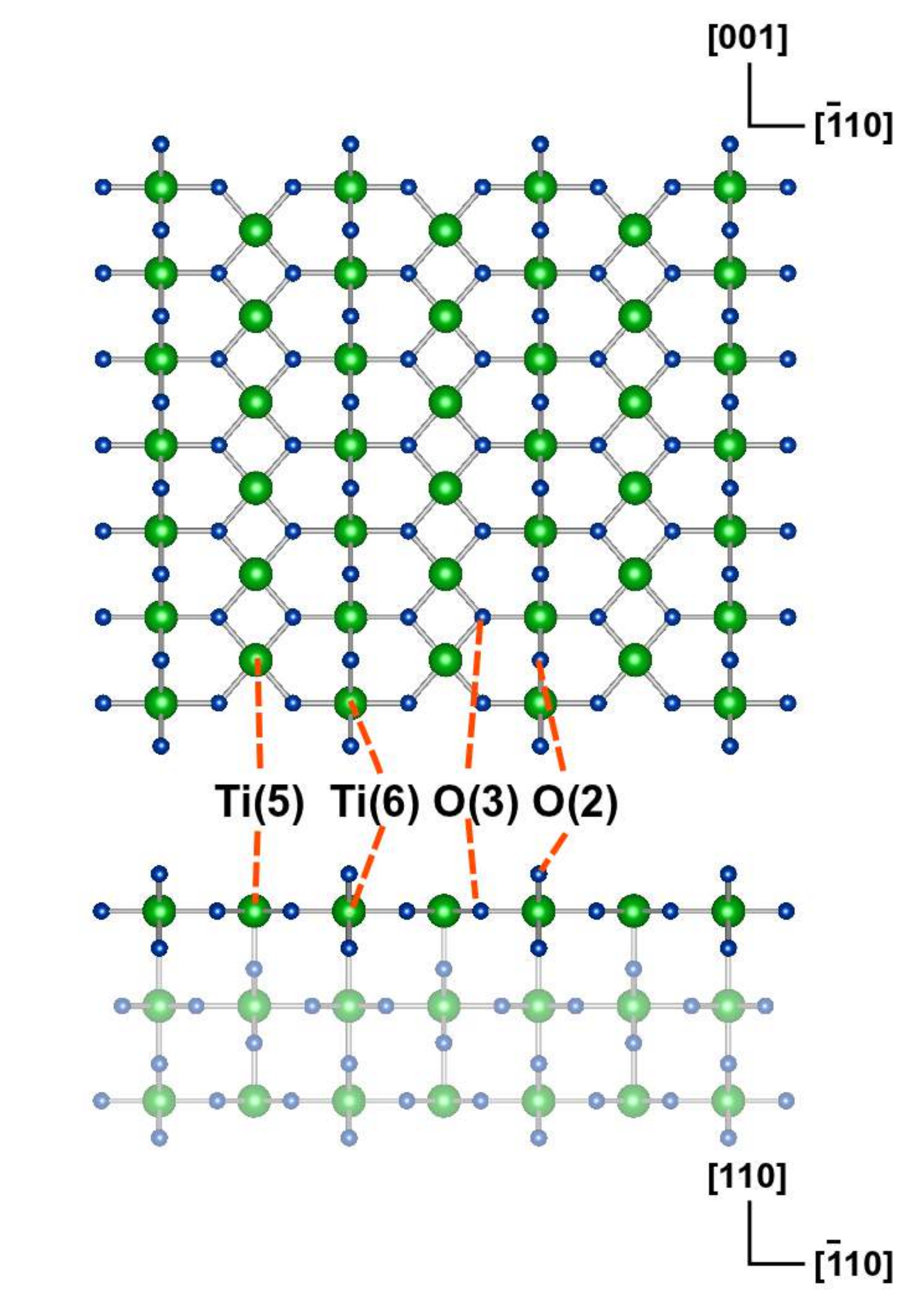}
    \end{minipage}
 \caption{Rutile (110) surfaces: Titanium atoms are green, oxygen atoms are blue. Both top view (top) and side view (bottom) are included, with coordination of surface atoms labelled.}
 \label{chpt3_110}
 \end{center}
\end{figure}

Forming a stoichiometric surface by truncating the bulk rutile 
structure along the [110] direction exposes both fully
coordinated O(3) and undercoordinated O(2) atoms, as can be seen
in figure \ref{chpt3_110}. Similarly both
fully coordinated and under-coordinated Ti(5) and Ti(6) atoms
are exposed. 

Examining the rutile slab we can see that it has an 
alternating bi-periodic layer structure; in the bottom
of figure \ref{chpt3_110} it can be seen that
the inclusion of each additional layer causes the position of
the exposed O(2) along the [1$\bar{1}$0] direction
to alternate, such that they coincide by layer with period two.
This alternating structure leads to 
the surface properties of rutile (110) being slower to converge
with respect to slab thickness
when compared to other TiO\subscript{2} surfaces. As such we have examined 
the effects of two methods; full relaxation of the entire slab (FR) and 
fixing the bottom surface to the bulk positions (BF). 

Calculated surface energies for both systems can be seen in 
table \ref{chpt3_110_surf_en}. As discussed elsewhere
there are a wide range of theoretical values reported for rutile
(110) surface energies\cite{rut_110_SE_struct}, hinting at the
difficulty in converging the properties of this surface. We see that the
surface energies for slabs with the bottom layer fixed 
converge to a value in the region of $\sim$
0.36-0.39 eV, with a significant number of layers
required in order to reach this convergence. The fully relaxed
slab converges more slowly, seemingly reaching a much higher
converged energy of $\sim$ 0.49-0.50 eV after 12 layers. We would
suggest that increasing the number of layers further should cause the 
surface energy to approach that of the fully relaxed slab, but comes
with significant computational expense. Another point to note is that
both relaxed surface energies for the FR and BF slabs converge
bi-periodically, following the same pattern as the surface structure.
Examining the atomic displacements for the BF and FR slabs in 
table \ref{chpt3_110_surf_en}(ii), we see
that both methods show the same displacement trends in the $[110]$ 
direction as experiment, and the 4 layer slab in both cases is 
well converged with respect to the 11 layer slab, illustrating that
the effect of the bi-periodic structure is minimised as the thickness
increases.

 

\begin{table}[h]
\begin{tabular}{c c}
  \begin{minipage}{0.3\textwidth}
  \begin{center}
   \begin{tabular}{c c c c }
   \hline
   \hline
   & \multicolumn{3}{c}{Surface Energy}\\
   N\subscript{lay} & E\subscript{unrel}& E\subscript{rel}(BF) &E\subscript{rel}(FR) \\
   & (J/m\superscript{2})&(J/m\superscript{2})&(J/m\superscript{2}) \\
   \hline
   3& 1.448& 0.681& 0.771\\
   4& 1.421& 0.333& 0.444\\
   5& 1.429& 0.495& 0.603\\
   6& 1.427& 0.345& 0.473\\
   7& 1.429& 0.426& 0.560\\
   8& 1.427& 0.360& 0.492\\
   9& 1.427& 0.402& 0.532\\
  10& 1.424& 0.363& 0.499\\
  11& 1.423& 0.392& 0.525\\
  12& 1.423& 0.373& 0.503\\ 
   \hline
   \hline
   \end{tabular}
  \end{center}
  \end{minipage}
&
\small
  \begin{minipage}{0.7\textwidth}
  \begin{center}
   \begin{tabular}{c c c c }
   \hline
   \hline
   &\multicolumn{3}{c}{Atomic Disp. (\AA)}\\
   Label& [110](BF)&[110] FR& Expt\cite{chpt3_110_pos}\\
   \hline
   Ti(6)   & +0.265(+0.247)& +0.292(+0.216)& +0.25 $\pm$0.03\\
   Ti(5)   & -0.154(-0.186)& -0.186(-0.193)& -0.19 $\pm$0.03\\
    O(2)   & +0.053(+0.037)& +0.079(+0.005)& +0.10 $\pm$0.05\\
    O(3)   & +0.188(+0.149)& +0.160(+0.142)& +0.17 $\pm$0.08\\
   \hline
   \hline
  \end{tabular}
 \end{center}
 \end{minipage}
\\
\textbf{(i)}&\textbf{(ii)}\\
 \end{tabular}
\caption{\textbf{Rutile (110) Surface:} \textbf{(i)} Relaxed and unrelaxed surface energies for different layer rutile slabs (BF is with the bottom fo the slab fixed, FR is total relaxation).
\textbf{(ii)} Displacements from bulk position on relaxation of Rutile (110) surface for 4 layer slab (values for 11 layer slab are shown in brackets)In both \textbf{(i)} \& \textbf{(ii)} BF stands for the bottom of the slab fixed, while FR is total relaxation.}
\label{chpt3_110_surf_en}
\end{table}

\normalsize

As mentioned, reported values for the relaxed and 
unrelaxed surface energies show a wide variation depending on the approach
and number of layers used\cite{rut_110_SE_struct}.
Previous work with the LDA functional for a fully relaxed slab
of 6 layer thickness reports a values of 1.78
J/m\superscript{2} and 0.84 J/m\superscript{2}
for the relaxed and unrelaxed surface energies, while
using GGA-PBE the values reported are 1.38 and 0.35 J/m\superscript{2}
respectively\cite{chptr3_selloni}. 
These values are in reasonable
agreement with those for our fully relaxed 6 layer slab of 
1.427 J/m\superscript{2} and 0.473 J/m\superscript{2}. 
Again highlighting the dependence of the (110) surface energies
on the approach used, in reference
\cite{rut_110_SE_struct} the reported values differ by around 0.1eV
between the GGA-PW91 functional result and that using the GGA-PBE
functional, with the PW91 surface energies converging to $\sim$ 1.48
and $\sim$ 0.57 J/m\superscript{2} respectively.


Given the apparent discrepancies between much of the reported results 
for surface energies of the (110) surface, and focusing on 
the objective behind this paper, to examine the interaction
of anchoring groups with the surface, we also study the convergence of 
two further properties. Firstly we have also looked at the 
convergence with slab thickness of calculated adsorption energies 
for two anchoring groups on the (110) surface. The two anchoring groups
chosen are the boronic and formic acids groups, which can be seen in 
figure \ref{CHPTR3_ACIDS}, and will be discussed further in later sections.
Calculated adsorption energies for both (in a bidentate 
binding configuration) can be seen in table \ref{ads_converge}.

\begin{table}[h]
  \begin{minipage}[t]{0.5\textwidth}
  \begin{center}
   \begin{tabular}{c c c c c}
   \hline
   \hline
   & \multicolumn{4}{c}{Adsorption Energy}\\
   & \multicolumn{2}{c}{FR}&\multicolumn{2}{c}{BF}\\
   N\subscript{lay} & Boronic & Formic & Boronic & Formic\\
   & (eV)&(eV)&(eV)&(eV)\\
   \hline
   2& -0.104& -0.725& -0.752& -1.297\\
   3& -3.193& -2.668& -2.523& -2.144\\
   4& -1.018& -1.326& -1.129& -1.363\\
   5& -2.075& -1.934& -1.707& -1.673\\
   6& -1.344& -1.567& -1.276& -1.432\\ 
   7& -2.117& -2.044& -1.469& -1.528\\
   \hline
   \hline
   \end{tabular}
  \end{center}
  \end{minipage}
\caption{Calculated adsorption energies for the boronic and formic acid
anchoring moieties on rutile (110) surfaces of differing thickness. FR
again stands for fully relaxed slabs, while BF stands for slabs with
the bottom layer fixed.}
\label{ads_converge}
\end{table}

We can see that the convergence of the adsorption energies
is also slow with respect to slab thickness, with the calculated
values exhibiting large variations as the number of layers increases
and also converging bi-periodically.
It can be seen that the relative absorption energies for the anchors 
on the fully relaxed slab vary with the slab thickness, with 
the boronic acid being most stable on slabs with odd layers while 
the formic acid is most stable on even. 
This is also the case for the BF slab up until around 5 layers, 
where there the adsorption energies are very similar. For six and
seven layer slabs the general relative binding strength is consistent, 
with the formic acid binding more strongly. 
Clearly converging the properties of the rutile (110) surface has its 
difficulties, although from this result we can say that fixing the bottom
layer aids convergence of the adsorption energies. 
The computational demand of simulating a slab of seven or 
more layers is less than ideal, and it is 
worth noting that a four layer slab, although not as well converged,
does get the relative ordering of the anchor binding strengths correct.

\begin{figure}[h]
 \begin{center}
  \begin{tabular}{c c}
    \begin{minipage}[h]{0.5\textwidth}
      \includegraphics[angle = -90,trim = 60mm 60mm 35mm 50mm, clip, width=1.\textwidth]{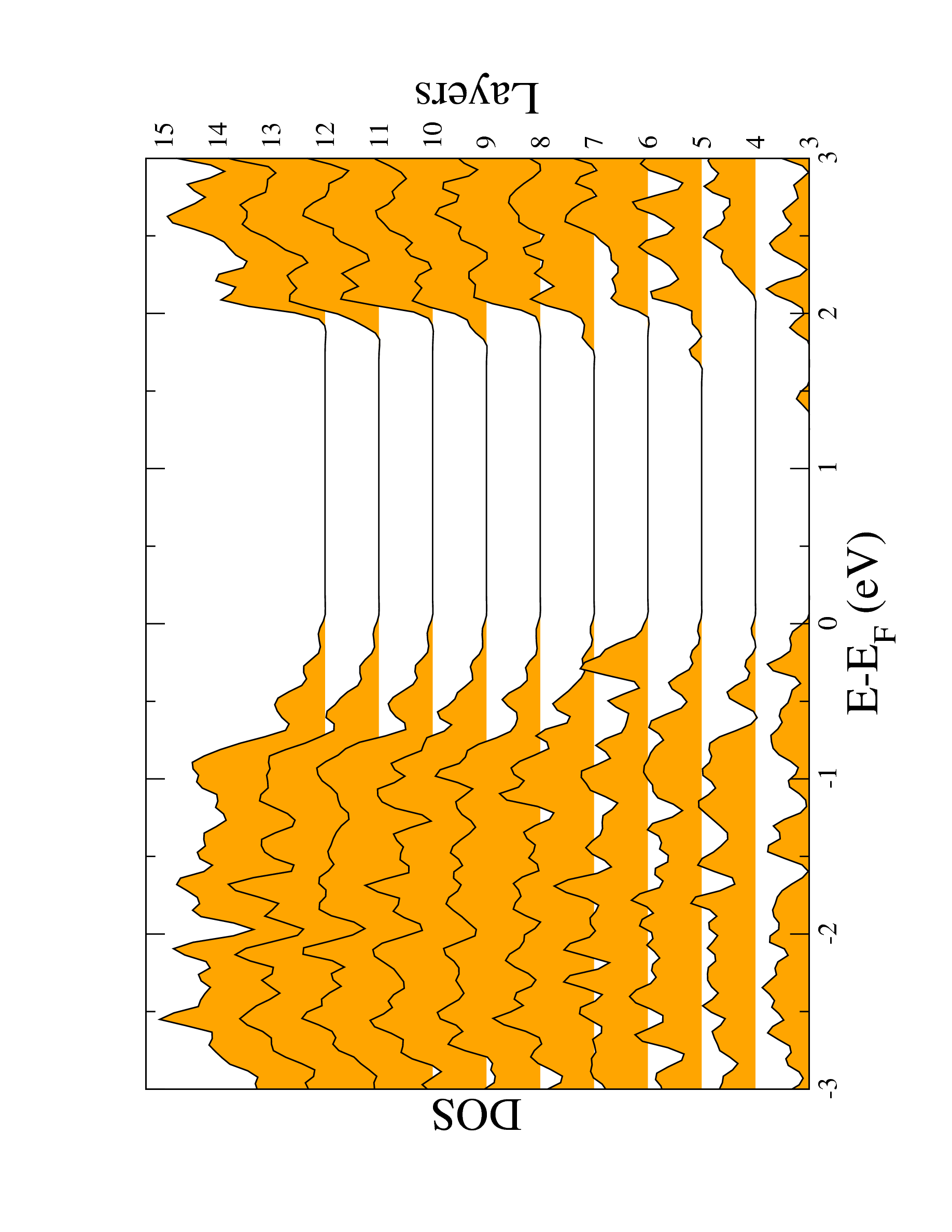}
    \end{minipage}&
    \begin{minipage}[h]{0.5\textwidth}
      \includegraphics[angle=-90,trim = 60mm 60mm 35mm 50mm, clip, width=1.\textwidth]{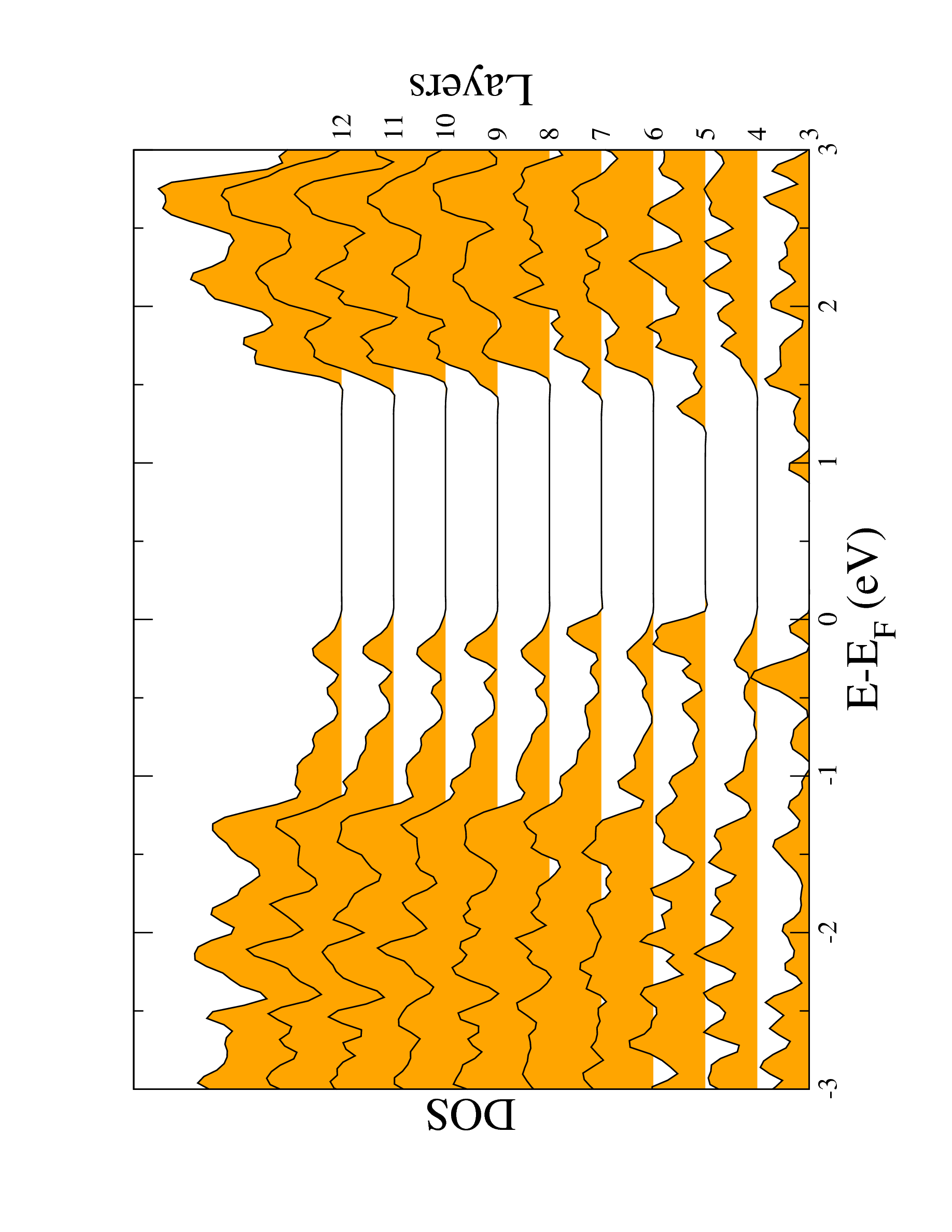}
    \end{minipage}\\
   \textbf{(i)}&\textbf{(ii)}\\
   \end{tabular}
 \caption{Rutile (110) density of states: Convergence of DOS with slab thickness for \textbf{(i)} fully relaxed (FR) and \textbf{(ii)} bottom layer 
fixed slabs (BF).}
 \label{110_dos}
 \end{center}
\end{figure}

Convergence of the density of states for the 
rutile (110) TiO\subscript{2} slabs has also been 
examined, in both the FR and BF variants,
shown in figure \ref{110_dos}. 
Wide variation in the size of the band-gap was found with the
differing number of layers present, with 
reduced band-gaps for thin slabs. 

In the case of the BF slab
this reduced band gap is maintained up to 12 layers, with
a band gap width of around 1.5 eV, while for a
12 layer FR slab it is around 2 eV, close to the reported GGA
band gap for bulk rutile. The band gap reduction
is particularly pronounced for thin slabs containing odd numbers of layers.
Examining the frontier orbitals for the 3 and 5 layer slabs we 
found that the lowest unoccupied molecular orbital is in Ti 3d orbitals
spread throughout the lattice, but with the majority localised on 
non-surface atoms. The highest occupied molecular orbital (HOMO)
is similarly spread throughout the lattice, in oxygen 2p orbitals
for the FR slab. The BF slab differs, with the HOMO localised almost
exclusively on fixed oxygens in the bottom layer.
This illustrates that the 
computational approach of fixing of the bottom layer
is responsible for the artificial shortening of the band gap. 


It is interesting that for the density of states the 
fully relaxed slab exhibited much better convergence
(towards the correct bulk GGA result) than that
with the bottom layer fixed, at variance with the convergence
of the adsorption energies. Indeed the density of states 
for a 4 layer FR slab shows good agreement with that of the 12 
layer slab, and has no states in the gap.


Taking into account the difficulty of converging the properties of 
the rutile (110) surface, we opt to use a 4 layer slab with the bottom
layer fixed for calculation of adsorption energies, as the 
the correct ordering is maintained
and the relative adsorption strengths are reasonably well converged. 
For the calculation of density of states we use a 4 layer fully 
relaxed slab, which exhibits a well converged band-gap close to the
expected GGA result.

A further point to note is that rutile (110) is also known to 
form a $1\times2$ reconstruction following
annealing at \textasciitilde 1100K\cite{chptr3_110_(1x2)}.
However, in the formation of
(110) dominated rutile nanorod DSSCs, the electrodes are annealed
at $\sim$ 400-700K \cite{chptr3_NR_110maj1,chptr3_NR_110maj} and 
thus we restrict our examination to the $1\times1$ surface.

\section{Adsorption of Acidic Anchors }

As mentioned previously the dye-semiconductor interaction
is of fundamental importance to the science of 
DSSCs. Strong binding of the sensitising dye to the TiO\subscript{2}
is essential for device stability, and this interaction 
proceeds mainly through the anchoring moiety. Similarly
for efficient charge injection good electronic overlap of the dye 
and TiO\subscript{2} electronic states is required, upon which the
anchoring group will have a significant bearing.
Typically DSSCs utilise a carboxylic acid binding strategy, and while
other anchoring groups have been used (and in some cases shown 
more desirable traits than their carboxylic acid analogues) 
the interaction of other anchoring groups with the increasingly
important rutile (110) and anatase (001) surfaces have been much 
less rigorously examined. Previous works have
examined formic acid adsorption on all three of these 
surfaces\cite{chptr3_form(101)_bb,chptr3_form_001,chptr3_form_rut_110}
and previous theoretical work has been done examining the adsorption
of phosphonic acid on anatase (101) and rutile (110)\cite{chptr3_phos_101_110_bbh,chptr3_phos(101)}. To the
best of our knowledge at the time of writing no reports have been 
made on the binding of phosphonic acid to the (001) surface, and no
reports have been made on the adsorption of the boronic acid group
on any of these three surfaces. 

\begin{figure}[h]
 \begin{center}
    \begin{minipage}[h]{0.95\textwidth}
      \includegraphics[angle = 0,trim = 0mm 15mm 0mm 15mm, clip, width=1.\textwidth]{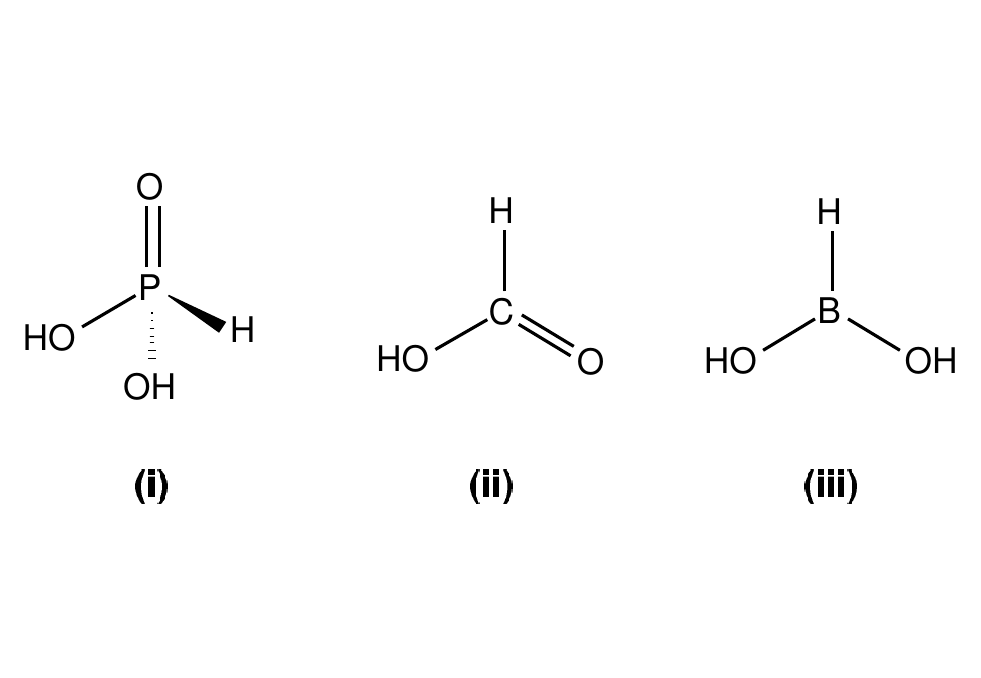}
    \end{minipage}
 \caption{Acidic binding groups examined: (i) Phosphonic acid, (ii) Formic acid and (iii) Boronic acid.}
 \label{CHPTR3_ACIDS}
 \end{center}
\end{figure}

All three of these acidic binding groups can be seen in 
figure \ref{CHPTR3_ACIDS}. While previous theoretical works have 
examined formic and phosphonic anchors on some of these surfaces, 
we repeat all of the calculations so that direct like for like
comparisons may
be made between systems that have already been examined and those
that have not. 

\begin{figure}[h]
 \begin{center}
    \begin{minipage}[h]{0.95\textwidth}
      \includegraphics[trim = 0mm 0mm 0mm 0mm, clip, width=1.\textwidth]{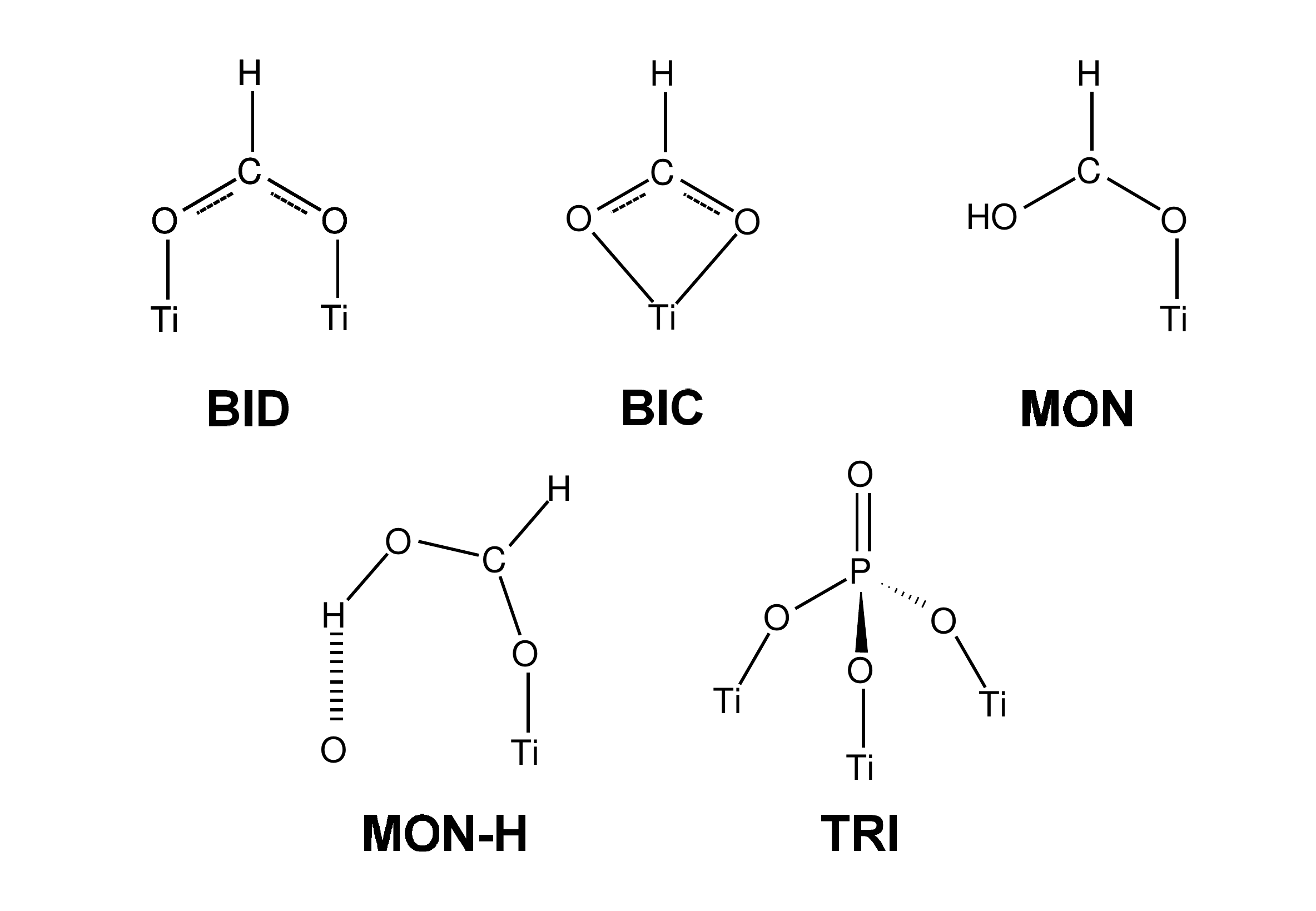}
    \end{minipage}
%
 \caption{Binding Structures for Acidic Groups: Bidentate (BID), Bidentate-chelating (BIC), Monodentate (MON), Monodentate with hydrogen bond (MON-H) and Tridentate (TRI)(Phosphonic acid is the only anchor for which the
tridentate mode is possible) }
 \label{CHPTR3_BINDING}
 \end{center}
\end{figure}

All of these anchoring groups will have several 
different possible binding motifs with the 
TiO\subscript{2} surfaces, and 
in order to gauge the strength of interaction and find the 
most stable structure we must survey several of these possiblities.
Some of the potential binding mechanisms of acidic anchors 
to TiO\subscript{2} are schematically illustrated in 
figure \ref{CHPTR3_BINDING}. Several of the binding motifs require 
the dissociation of the acid, in which case the dissociated hydrogen 
is adsorbed on a nearby undercoordinated surface oxygen atom
(not shown in the figure).

Calculation of binding energy between the acidic groups and the
TiO\subscript{2} surfaces proceeds by taking a clean, relaxed surface
slab and introducing the anchor molecule in an appropriate
binding motif. The strength of the 
interaction is then calculated by
subtracting the energy of the clean surface and the molecule from
that of the total system:

\begin{equation}
E_{ADS} = E_{TOT} - \left(E_{SURF}+E_{MOL}\right)
\end{equation}

In all instances the molecular energy is calculated for the molecule
in isolation residing in a cell of dimensions the same as that of 
the composite system. Similarly the K-point sampling and energy 
cut-off are maintained constant for each of these calculations.

\subsection{Anatase (101)}

\begin{figure}
 \begin{center}
  \begin{tabular}{c c c}
 \large
  Formic & Boronic & Phosphonic\\
 \normalsize
   \begin{minipage}[h]{0.33\textwidth}
\includegraphics[trim = 350mm 250mm 350mm 210mm, clip, width=1.\textwidth]{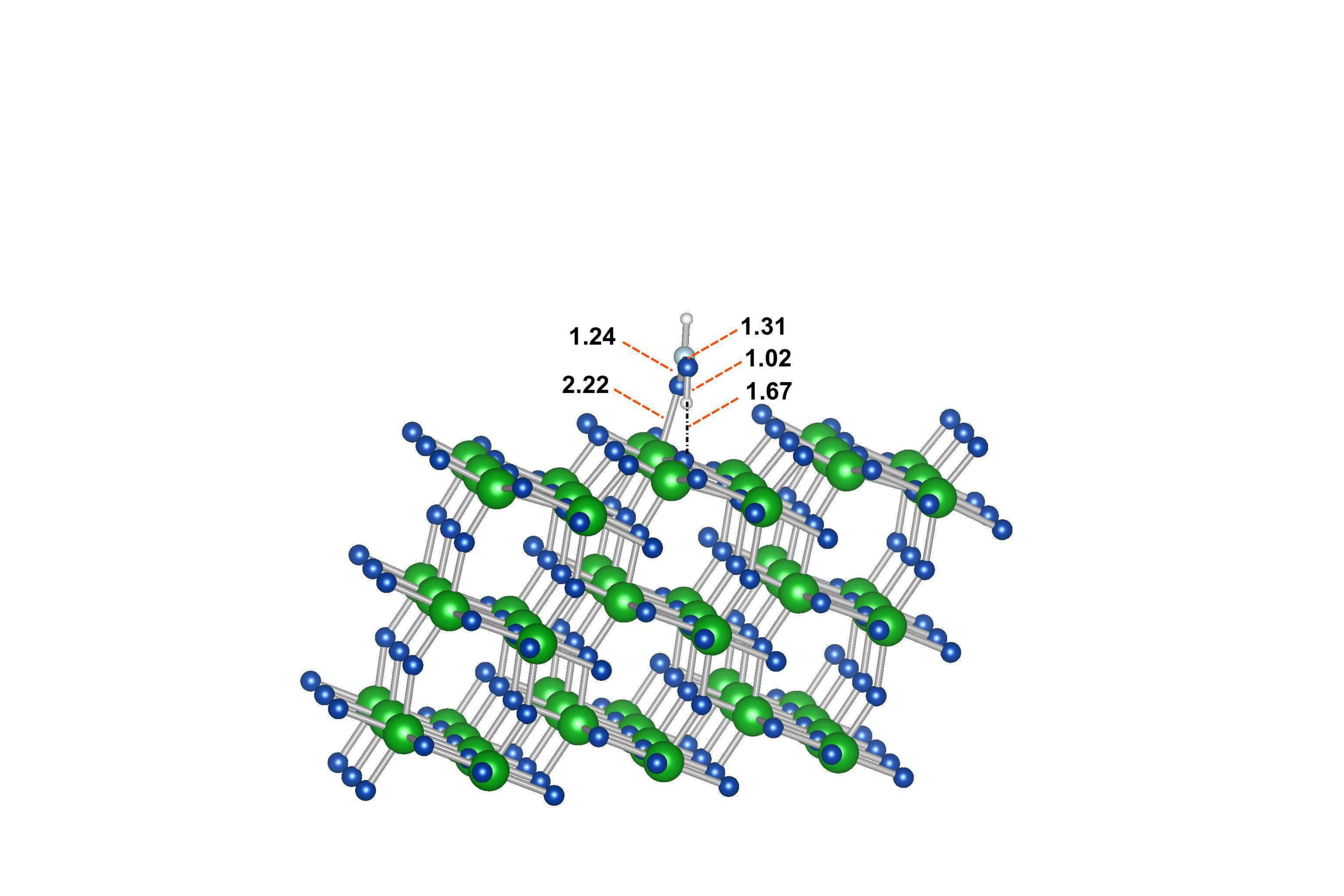}
    \end{minipage}
  &
    \begin{minipage}[h]{0.33\textwidth}
\includegraphics[trim = 350mm 250mm 350mm 210mm, clip, width=1.\textwidth]{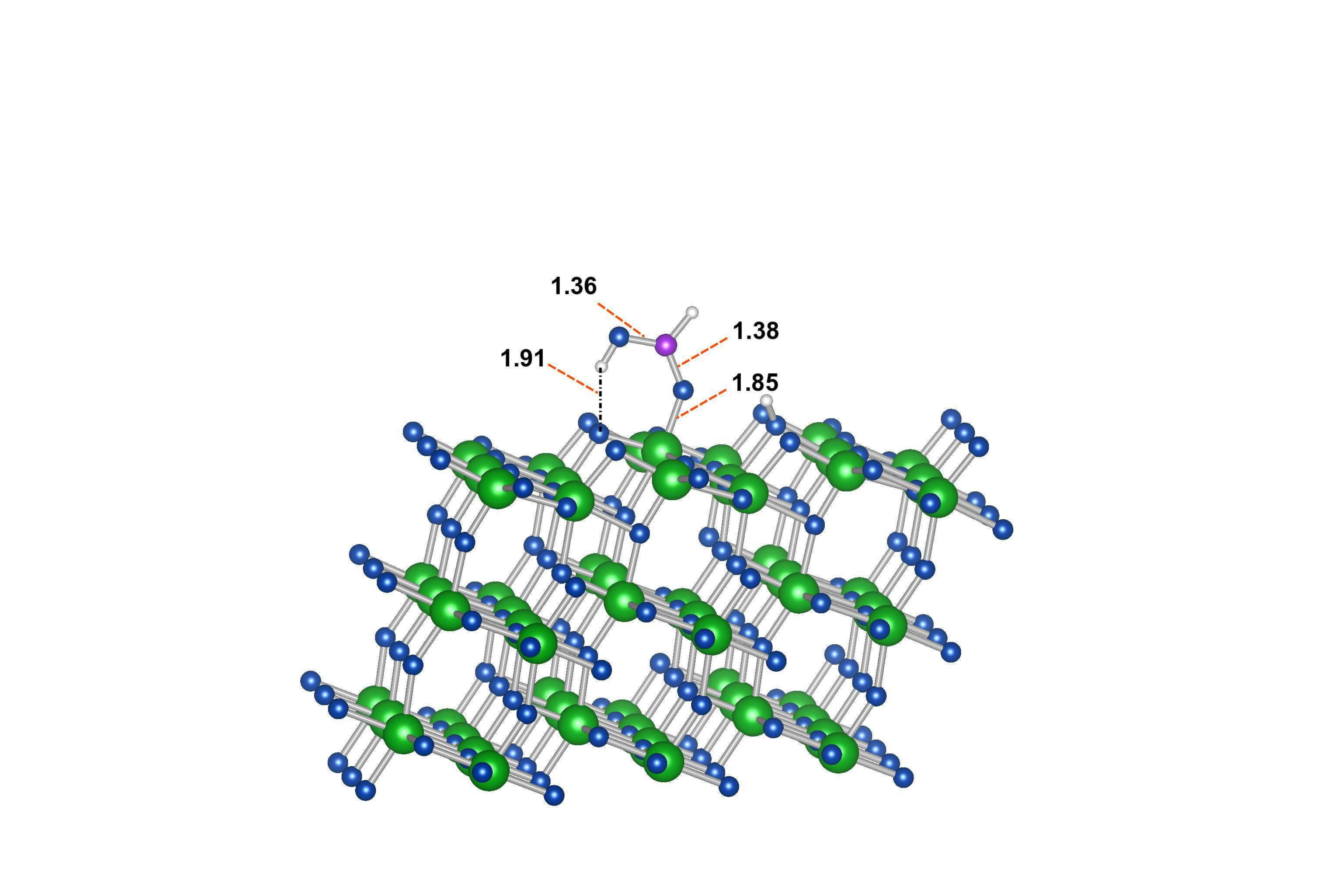}
    \end{minipage}
  & 
  \begin{minipage}[h]{0.33\textwidth}
\includegraphics[trim = 350mm 250mm 350mm 210mm, clip, width=1.\textwidth]{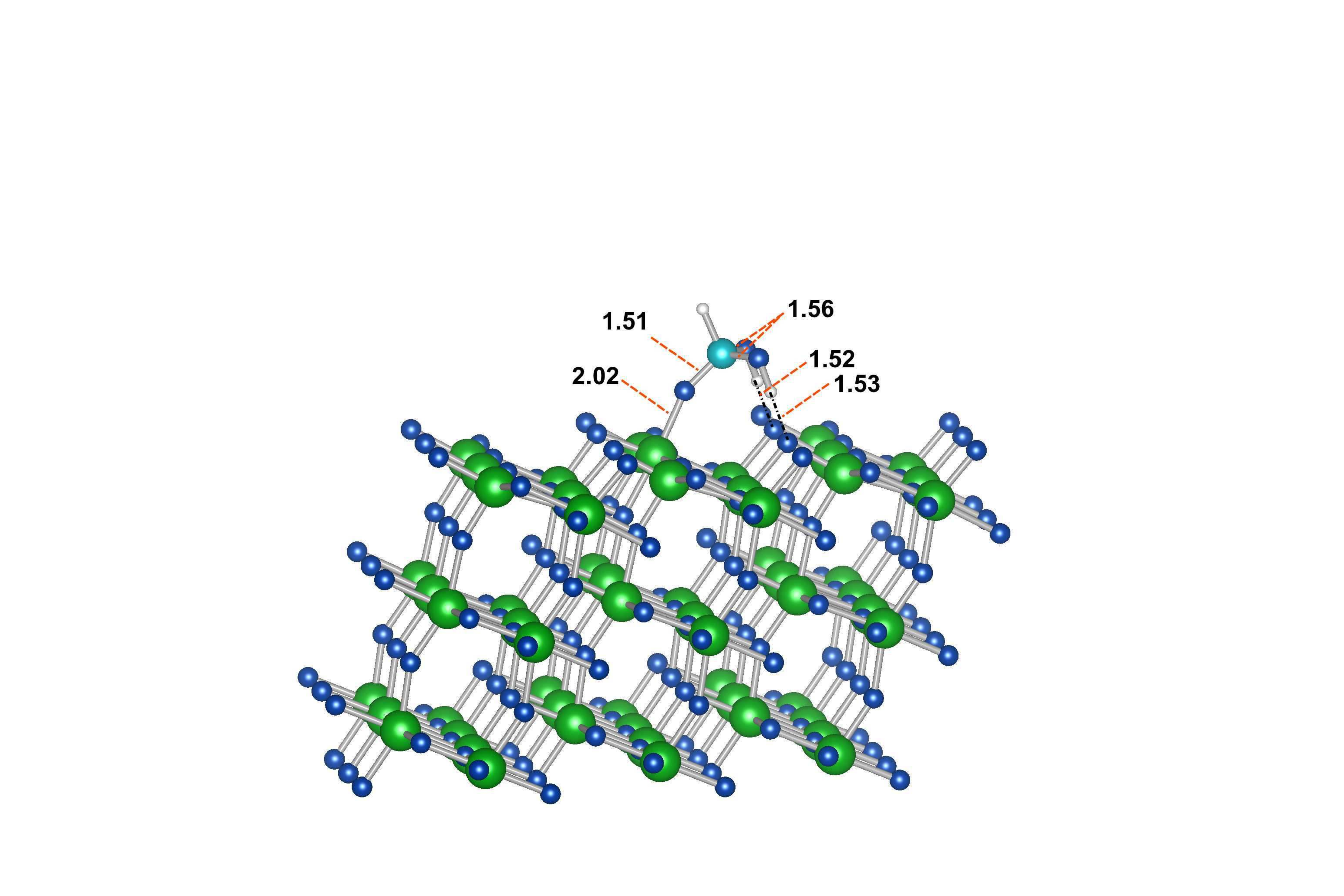}
    \end{minipage}
  \\
\textbf{(i)} MON-H&\textbf{(ii)} MON-H&\textbf{(iii)} MON-2H\\
    \begin{minipage}[h]{0.33\textwidth}
\includegraphics[trim = 350mm 250mm 350mm 210mm, clip, width=1.\textwidth]{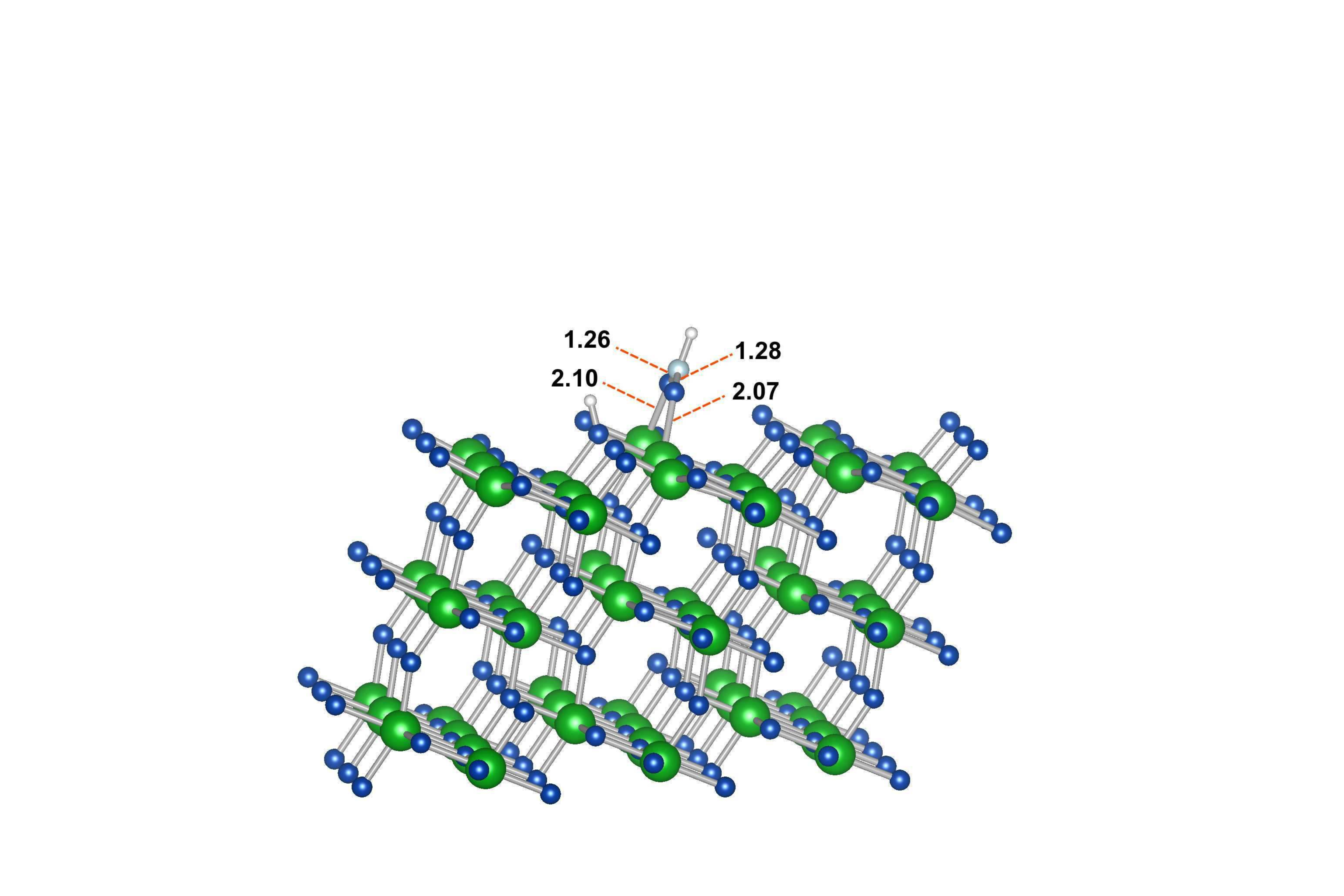}
    \end{minipage}
   &
    \begin{minipage}[h]{0.33\textwidth}
\includegraphics[trim = 350mm 250mm 350mm 210mm, clip, width=1.\textwidth]{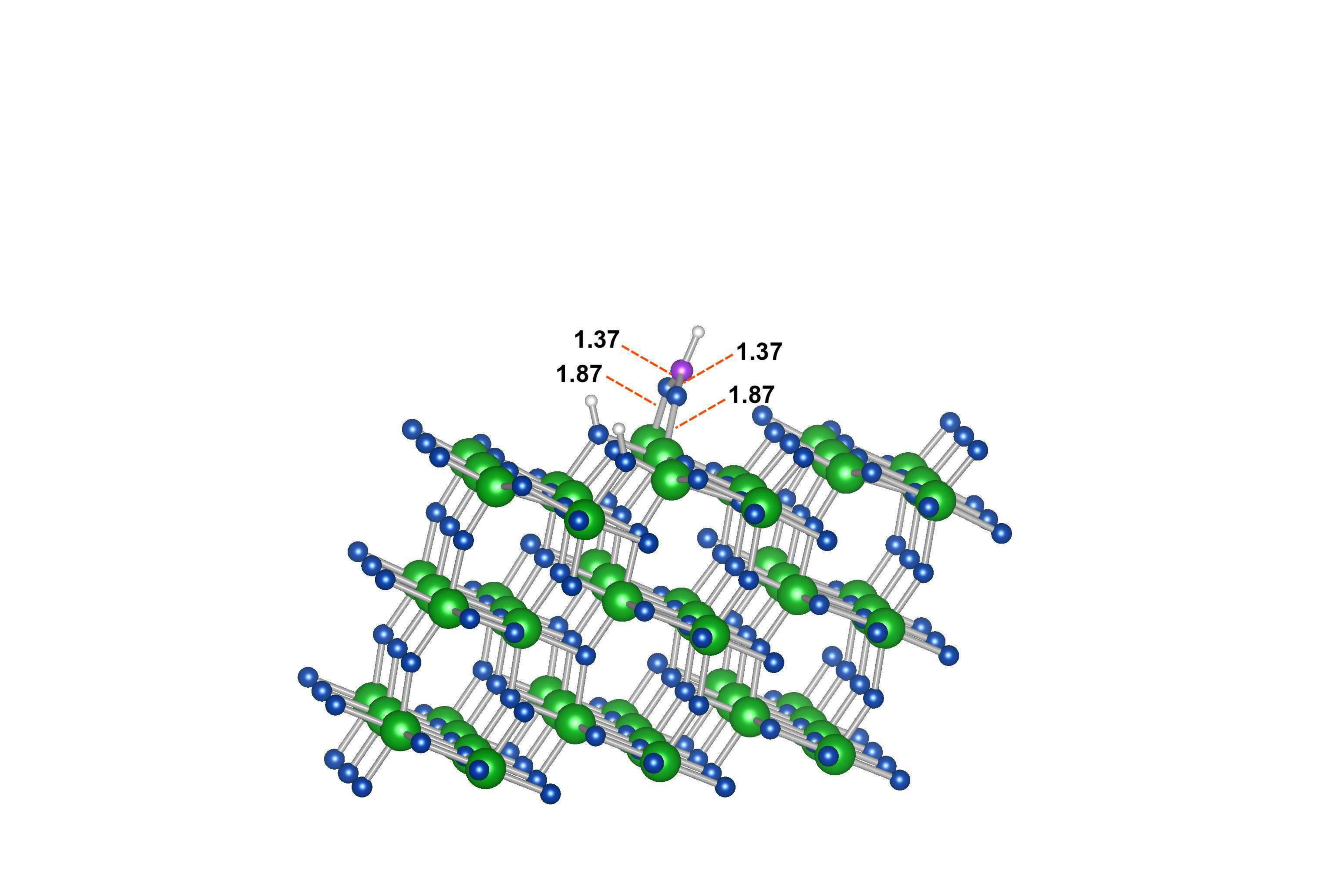}
    \end{minipage}
   &
    \begin{minipage}[h]{0.33\textwidth}
\includegraphics[trim = 350mm 250mm 350mm 210mm, clip, width=1.\textwidth]{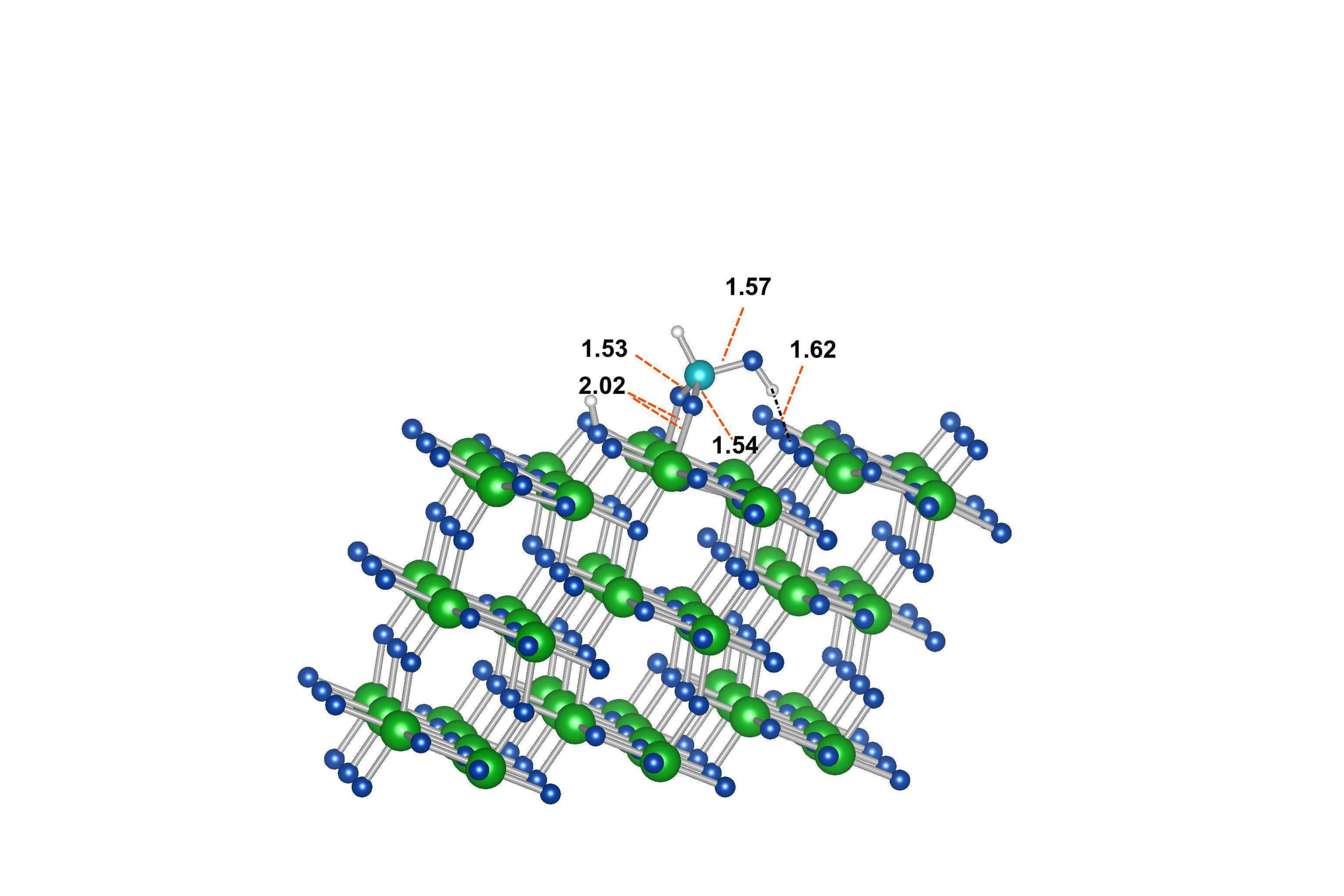}
    \end{minipage}
 \\
\textbf{(iv)} BID&\textbf{(v)} BID&\textbf{(vi)} BID-H\\
 \end{tabular}
 \caption{Anchor adsorption structures on anatase (101): Two most stable adsorption structures for formic acid (\textbf{(i)} \& \textbf{(iv)}), boronic  acid (\textbf{(ii)} \& \textbf{(v)}) and phosphonic acid (\textbf{(iii)} \& \textbf{(vi)})}
 \label{CHPTR3_FORM_101}
 \end{center}
\end{figure}

Calculated adsorption energies for all three acidic binding groups 
in several of the binding structures on the anatase (101) surface 
can be seen in table \ref{chpt3_101_bind}, and the most stable
binding structures are exhibited in figure \ref{CHPTR3_FORM_101}.

Our results illustrate that formic acid binds most strongly in the
bidentate bridging mode, with the oxygens bonding to two adjacent 
Ti(5) surface atoms. An almost equivalent binding energy is obtained 
for the monodentate binding mode with a hydrogen bond (MON-H) to the
nearest O(2) atom. A second monodentate binding mode is
reported in which the hydrogen forms a weak bond with a O(3) surface
atom (MON), with the adsorption energy being comparatively reduced. 
Finally the bidentate chelating mode (BIC) is found to be stable, 
but considerably less so than the monodentate and bridging modes.

Experimentally Fourier transform infrared spectroscopy 
report the coexistence of two different adsorption
structures, a bidentate mode and an unsymmetrical structure \cite{chptr3_form_FTIR}.
This is in agreement with
our result that the monodentate and bidentate bridging modes are 
energetically similar and likely to coexist. Previous theoretical 
work has found the most stable structure to be the bidentate
bridging mode, with the interaction of the hydrogen bonded 
monodentate mode extremely similar energetically\cite{chptr3_form(101)_bb}.

\begin{table}[h]
 \begin{center}
  \begin{tabular}{ c c c c c c c}
    \hline
     \hline
       & MON & MON-H & MON-2H & BIC & BID & BID-H  \\
     Adsorbate&&&&\\
     \hline
     Boronic    &   N/A& -0.72&   N/A& +0.12& -0.82&   N/A \\
     Formic     & -0.71& -1.02&   N/A& -0.12& -1.03&   N/A \\
     Phosphonic & -1.29&   N/A& -1.69& -0.48&   N/A& -1.82 \\
      \hline
      \hline
  \end{tabular}
 \end{center}
  \caption{Calculated adsorption energies for acids on the Anatase (101) surface (in eV.)}
\label{chpt3_101_bind}
\end{table}

The phosphonic acid group has an extra degree of freedom in terms of 
its binding modes over formic and boronic acids,
 due to the extra oxygen atom it possesses. 
Numerous adsorption structures of similar binding 
modes are therefore available to it, and we report only 
the most stable of each of these.
Examining the results for phosphonic acid we see that overall it
binds much more strongly to the surface in each of the given binding
modes than the other acids.
The most stable adsorption mode for phosphonic acid is the 
bidentate bridging mode with an additional single hydrogen 
bond (BID-H Figure~\ref{CHPTR3_FORM_101}), in which two oxygen 
atoms of the acid group form 
bonds to two adjacent Ti(5) atoms in the surface, and the final
 OH group forms a hydrogen bond to an adjacent O(2) atom.
A second comparably stable adsorption mode is found where the
phosphonic acid binds through its carbonyl group, with the two OH
groups forming hydrogen bonds to two adjacent O(2) atoms 
(MON-2H Figure~\ref{CHPTR3_FORM_101}). 
Previous studies have found similar results, 
with a DFT tight binding study finding the BID-H to be the most 
stable with the MON-2H structure being the only stable monodentate
structure found after relaxation \cite{chptr3_phos_101_110_bbh}. 
A second study found the two 
to be comparable in energy, but with the monodentate structure slightly more stable
by 0.13 eV \cite{chptr3_phos(101)}. 
Our calculations do obtain further stable monodentate
structures, all of which contain a single hydrogen bond irrespective
of whether the hydrogen dissociates to reside on the nearest O(2) of the
surface or not. Similarly to the previous works no stable tridentate 
binding mode was obtained.


Chemical adsorption of the boronic acid requires the 
dissociation of at least one hydrogen atom from either of its
OH groups. From the results reported in table \ref{chpt3_101_bind}
we can see that its most stable
structure is the bidentate bridging mode, similar to that of
both the formic and phosphonic acid groups. However the strength
of the interaction for this mode, and for all others, is found to 
be significantly weaker than that of the formic acid and phosphonic
acid groups. 
While it is still possible for the
boronic acid group to form a stable bond with the (101) surface, 
and therefore anchor
dyes to the TiO\subscript{2} electrode, it will 
have poorer device stability and dye take up
over dyes anchoring through phosphonic and formic acid groups. 
Given that the typical nanoparticle DSSC electrode is dominated 
by the (101) anatase surface this result 
explains the experimentally observed trend \cite{chptr3_bor_dye}, 
that TiO\subscript{2} sensitised with boronic acid groups gave 
low surface coverage, and resulting low IPCE values. Increasing the
number of boronic anchoring moieties had the effect of increasing the
surface coverage, and consequently the IPCE. 

Stabilisation of the monodentate binding modes through hydrogen
bond formation is another interesting feature, with the 
monodentate modes of both the formic and phosphonic groups 
being stabilised in such a way. Boronic acid also tended 
towards the formation of these hydrogen bonds when adsorbed in 
each of the monodentate modes
 investigated (again only the most stable of these is reported).

\subsection{Rutile (110)}

\begin{figure}
 \begin{center}
  \begin{tabular}{c c c}
    \begin{minipage}[h]{0.33\textwidth}
\includegraphics[trim = 350mm 250mm 350mm 210mm, clip, width=1.\textwidth]{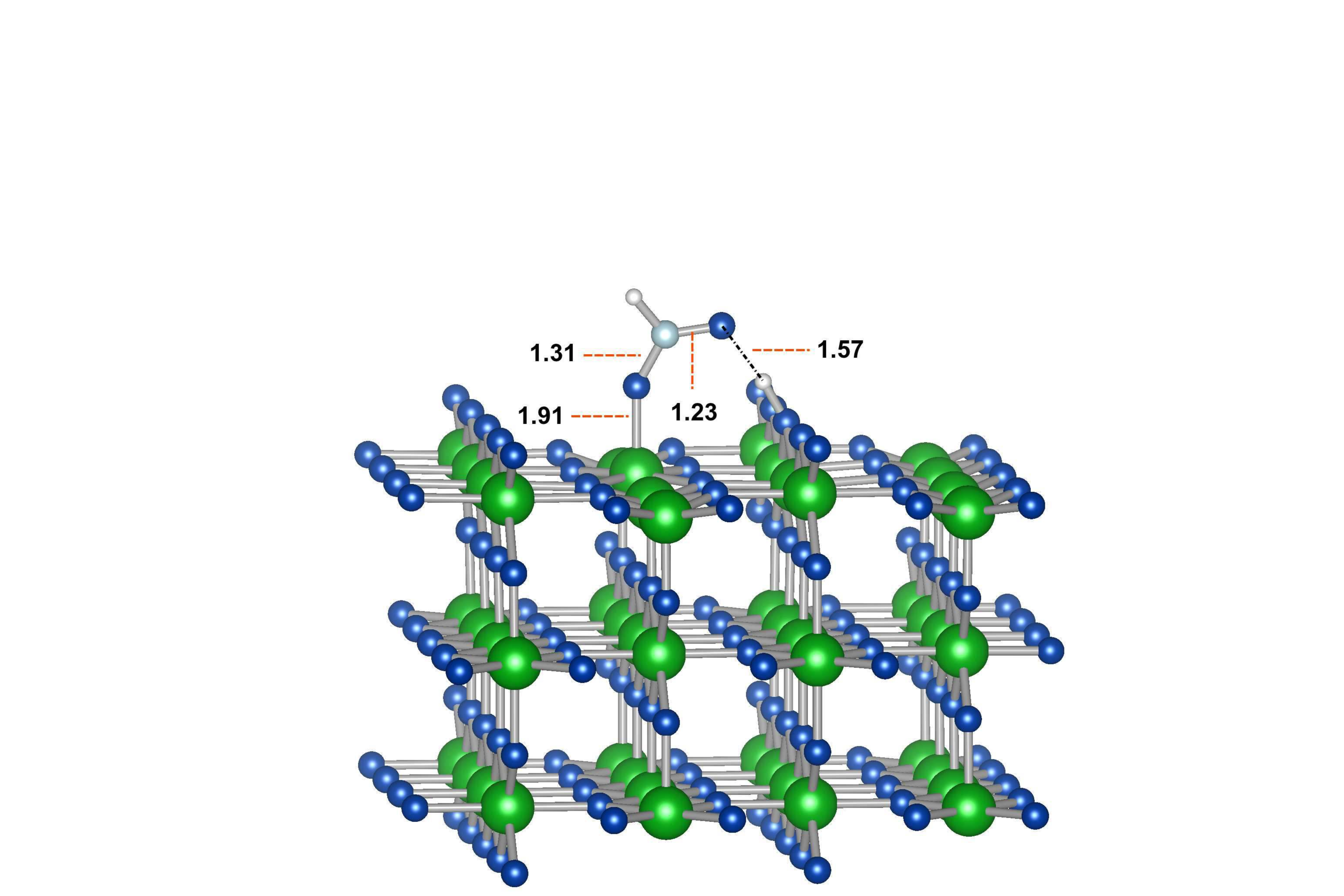}
    \end{minipage}
  &
    \begin{minipage}[h]{0.33\textwidth}
\includegraphics[trim = 350mm 250mm 350mm 210mm, clip, width=1.\textwidth]{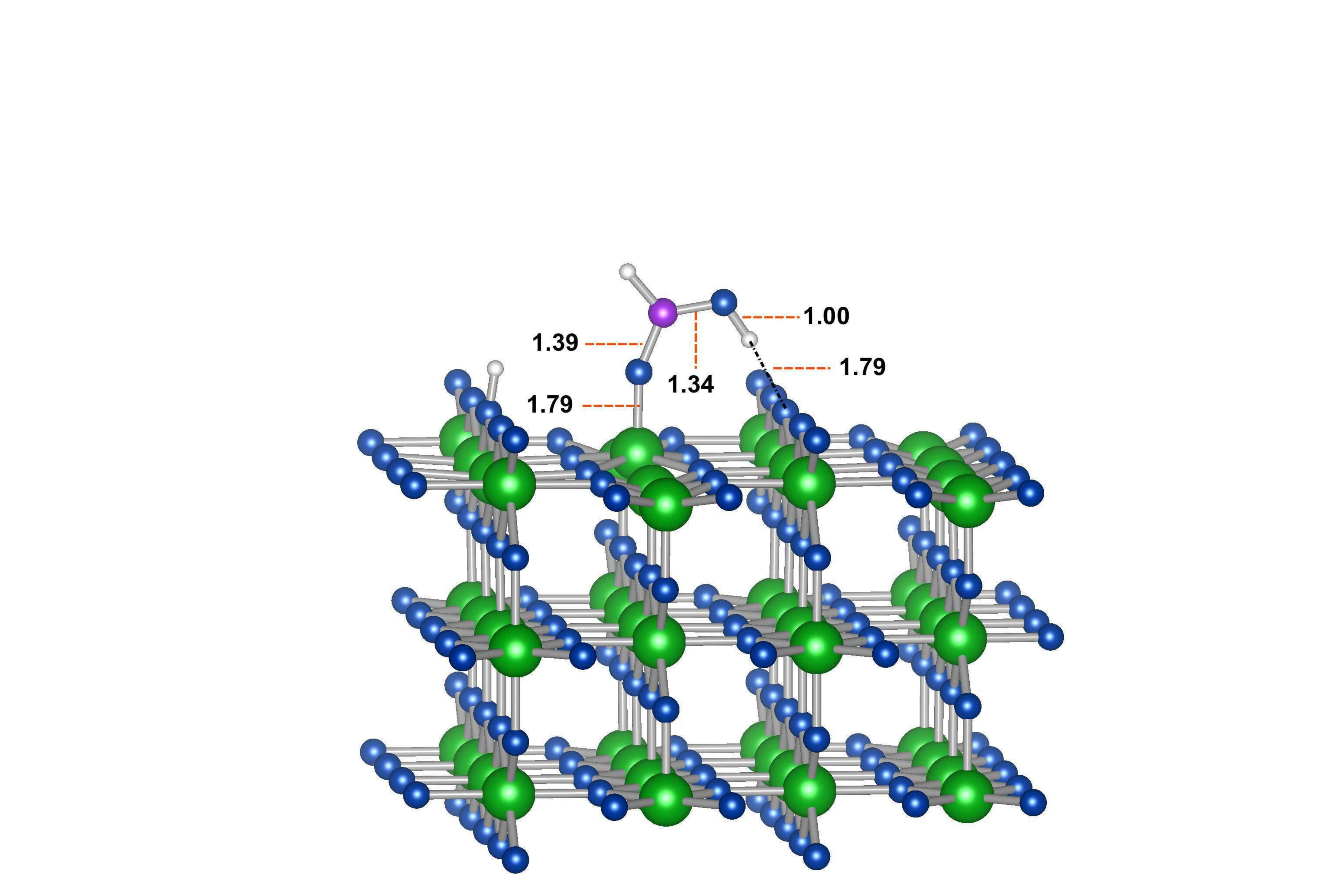}
    \end{minipage}
  &
  \begin{minipage}[h]{0.33\textwidth}
\includegraphics[trim = 350mm 250mm 350mm 210mm, clip, width=1.\textwidth]{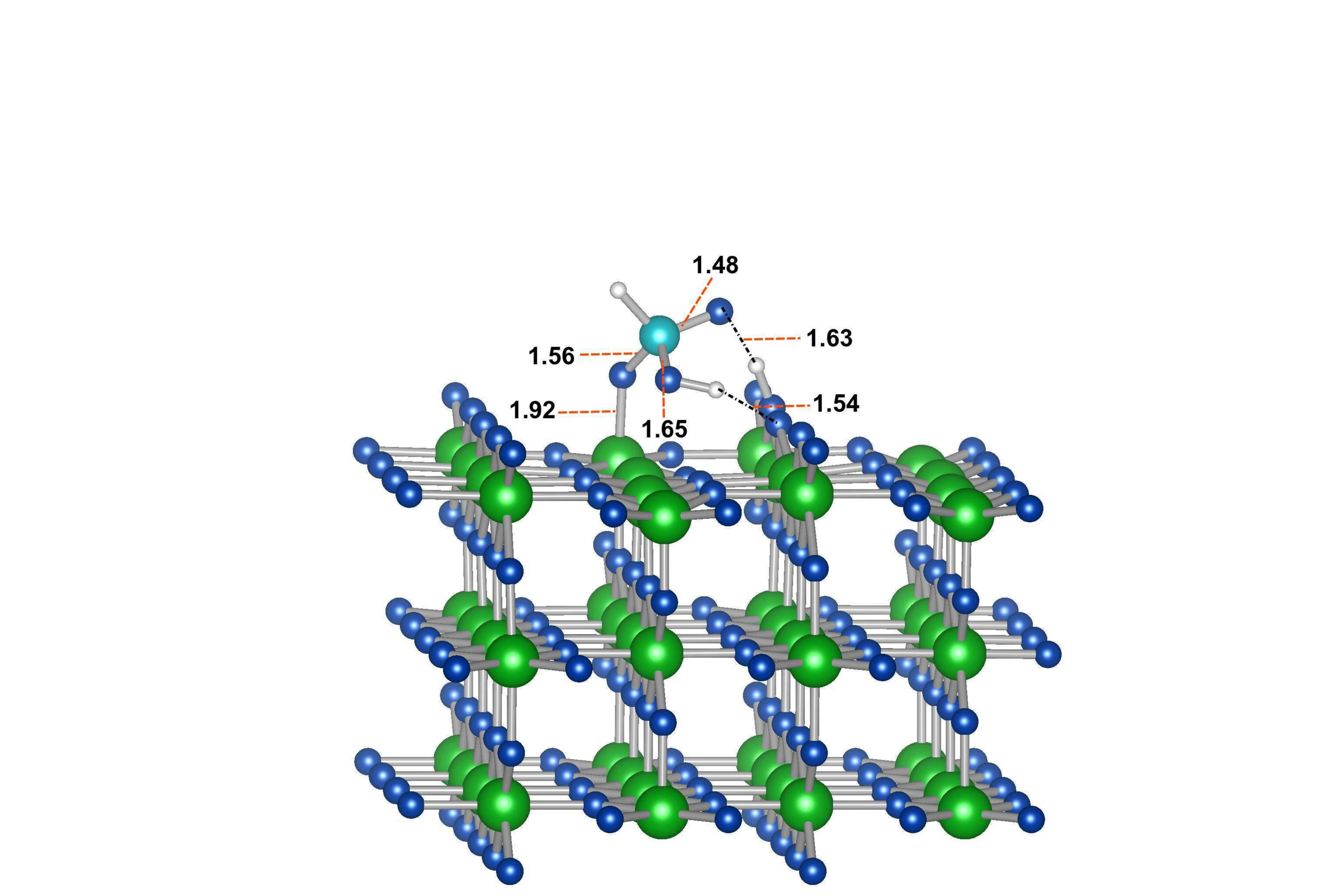}
    \end{minipage}
  \\
\textbf{(i)} MON-H&\textbf{(ii)} MON-H&\textbf{(iii)} MON-H\\
    \begin{minipage}[h]{0.33\textwidth}
\includegraphics[trim = 350mm 250mm 350mm 210mm, clip, width=1.\textwidth]{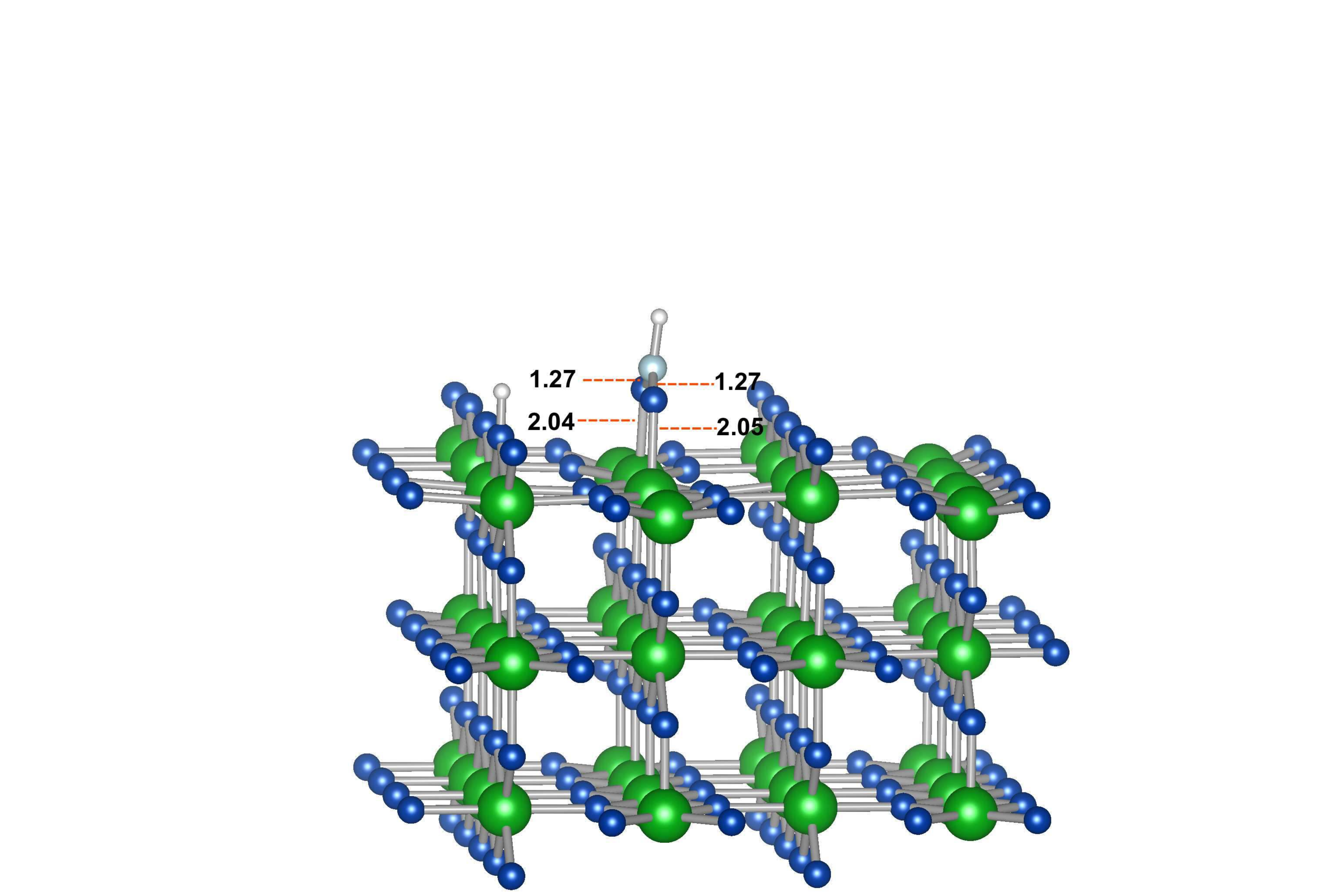}
    \end{minipage}
   &
    \begin{minipage}[h]{0.33\textwidth}
\includegraphics[trim = 350mm 250mm 350mm 210mm, clip, width=1.\textwidth]{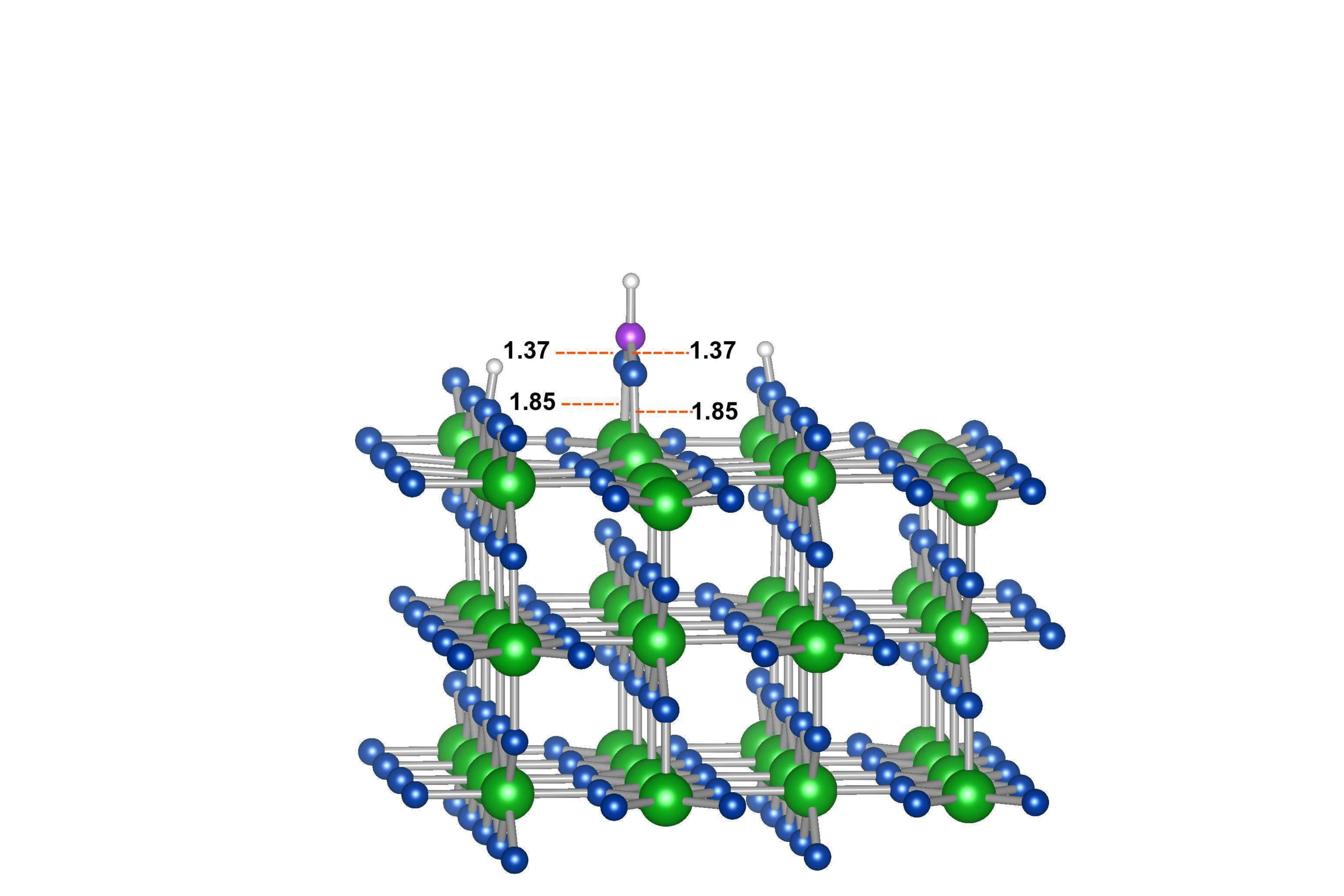}
    \end{minipage}
   &
    \begin{minipage}[h]{0.33\textwidth}
\includegraphics[trim = 350mm 250mm 350mm 210mm, clip, width=1.\textwidth]{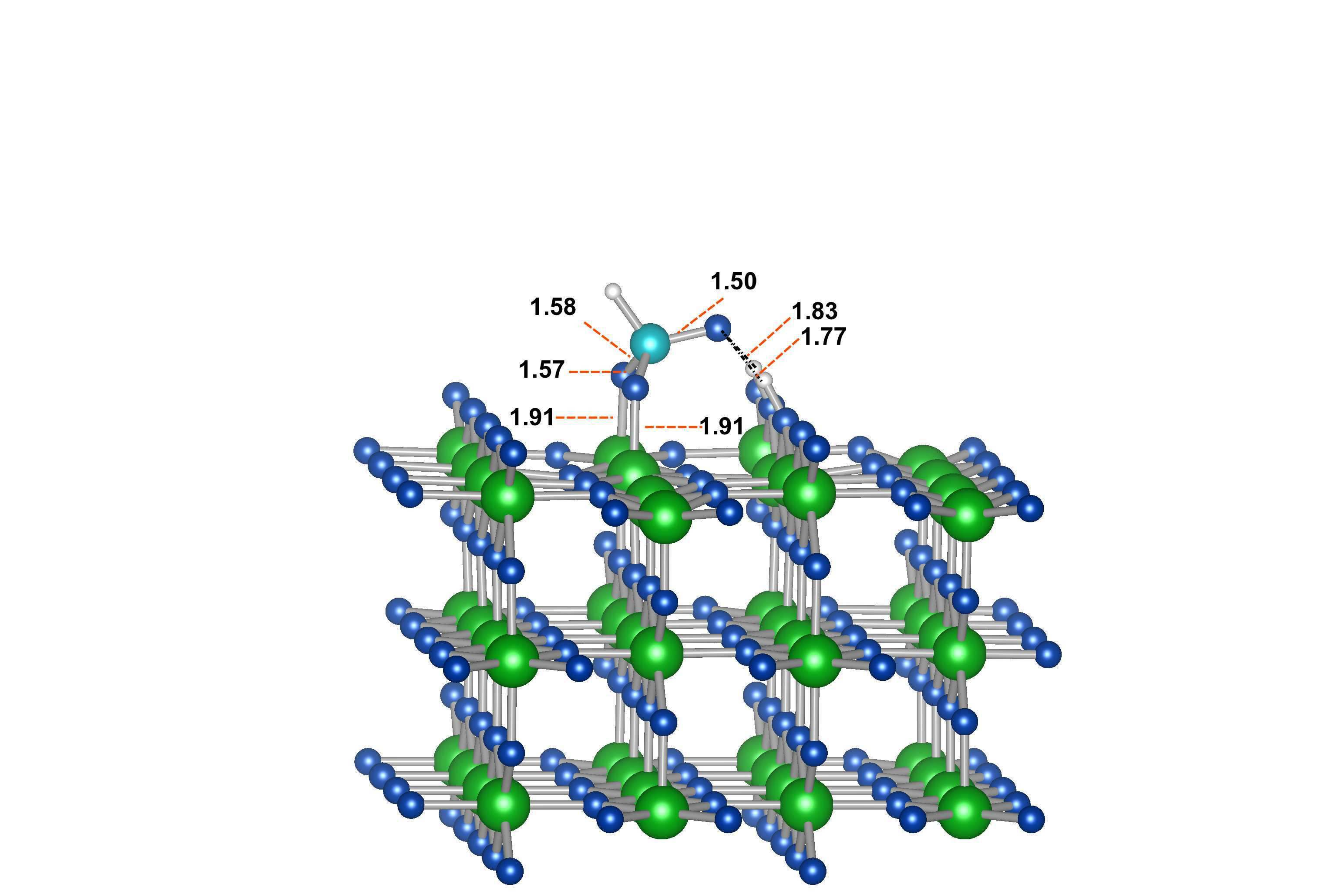}
    \end{minipage}
 \\
\textbf{(iv)} BID&\textbf{(v)} BID&\textbf{(vi)} BID-H\\
 \end{tabular}
 \caption{Anchor adsorption structures on rutile (110): Bidentate and monodentate structures for formic acid (\textbf{(i)} \& \textbf{(iv)}), boronic  acid (\textbf{(ii)} \& \textbf{(v)}) and phosphonic acid (\textbf{(iii)} \& \textbf{(vi)})}
 \label{CHPTR3_FORM_110}
 \end{center}
\end{figure}

Examining the interaction of our binding groups with the rutile (110)
surface (table \ref{chpt3_Rutile_bind}) a different trend with respect to 
the interactions on the anatase (101) surface is immediately observed;
in the most stable binding mode the interaction of all 
three anchoring moieties with rutile (110) is considerably stronger 
than that with 
anatase (101). Our direct comparison of adsorption energies illustrates
this trend, which in some sense has been reported by proxy when comparing 
previous studies for formic acid adsorption on the (110) and (101) 
surfaces \cite{chptr3_form2_rut_110,chptr3_form(101)_bb} and
for phosphonic acid \cite{chptr3_phos(101),chptr3_phos_110}.
The two most stable binding structures
for each of the anchors on the (110) surface
can be seen in
figure \ref{CHPTR3_FORM_110}.  
 
\begin{table}
 \begin{center}
  \begin{tabular}{c c c c c c}
    \hline
     \hline
       & MON & MON-H & BIC & BID & BID-H\\
     Adsorbate&&&&&\\
    \hline
     Boronic     & -0.73& -1.01& +0.53& -1.13& N/A\\
     Formic      & -0.71& -1.03& -0.22& -1.36& N/A\\
     Phosphonic & N/A& -1.54& -0.60& N/A& -2.05\\
      \hline
      \hline
  \end{tabular}
 \end{center}
  \caption{Calculated adsorption energies for acids on the Rutile (110) surface (in eV.)}
\label{chpt3_Rutile_bind}
\end{table}

As with the anatase (101) surface, for each anchoring group
the bidentate bridging mode is again found to be the most stable
structure. However there is a contrast when comparing the relative
stabilities of each mode, as on the (110) rutile surface the
bidentate bridging mode is considerably more stable than the
most stable monodentate modes, with perhaps the exception of 
the boronic anchor for which the bidentate mode is more stable 
by around 0.15 eV. Experimentally FTIR studies also find one
binding mechanism for formate on rutile (110), which is identified
as being the bidentate bridging mode by Hartree-Fock studies 
\cite{chptr3_110_form_FTIR}.

We note that the relative binding stabilities maintain the same order as
the anatase (101) surface,
such that $\text{phosphonic} > 
\text{formic} > \text{boronic}$. 
This is a significant result, suggesting that dyes containing
phosphonic acids will again have the ability to
bind more strongly than carboxylic anchored dyes when adhering
to the majority (110) surface exposed in rutile nanorod electrodes.

Both the formic and boronic acids bind in a very similar manner
in their most stable modes, as can seen in figure \ref{CHPTR3_FORM_110}.
Addition of a hydrogen bond is found to stabilise the adsorption structure
for the boronic and formic anchor. The phosphonic acid again differs in its
 binding modes slightly, due to its different molecular structure, and
has more scope for forming hydrogen bonds than the other two anchors. 
In both of the phosphonic adsorption modes shown in
\ref{CHPTR3_FORM_110}, two hydrogen bonds are formed.  Two dissociated
hydrogen atoms, attached to O(2) surface atoms, coordinate to 
the single carbonyl group on the phosphonic acid in
BID-H. In MON-H the hydrogen on the phosphonic OH group coordinates
to a single O(2) atom, along with the carbonyl group coordinating to  
the dissociated hydrogen. 
Similarly to the (101) surface no stable tridentate mode is found 
rather, when attemping to bind
the phosphonic group in a tridentate structure, relaxation
returns the mode to a BID-H structure.

\subsection{Anatase (001)}

Similarly to the rutile (110) surface, the anatase (001) surface 
reconstructs to form a ($1\times4$) termination in order to 
minimise the surface
energy \cite{chptr3_001_recon}, and thereby becoming less reactive. 
Capping with hydrofluoric acid results in a
fluorine terminated ($1\times1$) surface which is more stable than the
(101) surface. Using this capping agent it is possible obtain single
crystals with extremely high percentages of the (001) surface exposed. 
There is some debate over whether the (001) surface retains
its ($1\times1$) termination after removal of the 
fluorine capping agent by thermal processing. Experimentally 
the crystals
are reported as remaining unchanged\cite{chptr3_nature_001},
with a recent 
theoretical work showing that the process of removing the fluorine 
capping agent will result in the formation of the ($1\times4$) 
reconstruction\cite{chptr3_HF-001_reconstructs}. 
We report here only results on the ($1\times1$) surface
termination, and view the study on the ($1\times4$) reconstruction
as an important future extension of the work.

\begin{table}
 \begin{center}
  \begin{tabular}{c c c c c c}
    \hline
     \hline
       & MON-H & BIC & BID & BID2 & TRI \\
     Adsorbate&&&&&\\
     \hline
     Boronic     & -1.69& -0.65& -0.66 & -4.08 & N/A\\
     Formic      & -1.96& -1.70& -0.73 & -1.72 & N/A\\
     Phosphonic & -2.70& -1.82& -2.85 & -3.21 & -2.99\\
      \hline
      \hline
  \end{tabular}
 \end{center}
  \caption{Calculated adsorption energies for acids on the Anatase (001) surface (in eV.)}
\label{chpt3_001_bind}
\end{table}

\begin{figure}
 \begin{center}
  \begin{tabular}{c c c}
    \begin{minipage}[h]{0.33\textwidth}
\includegraphics[trim = 350mm 250mm 350mm 210mm, clip, width=1.\textwidth]{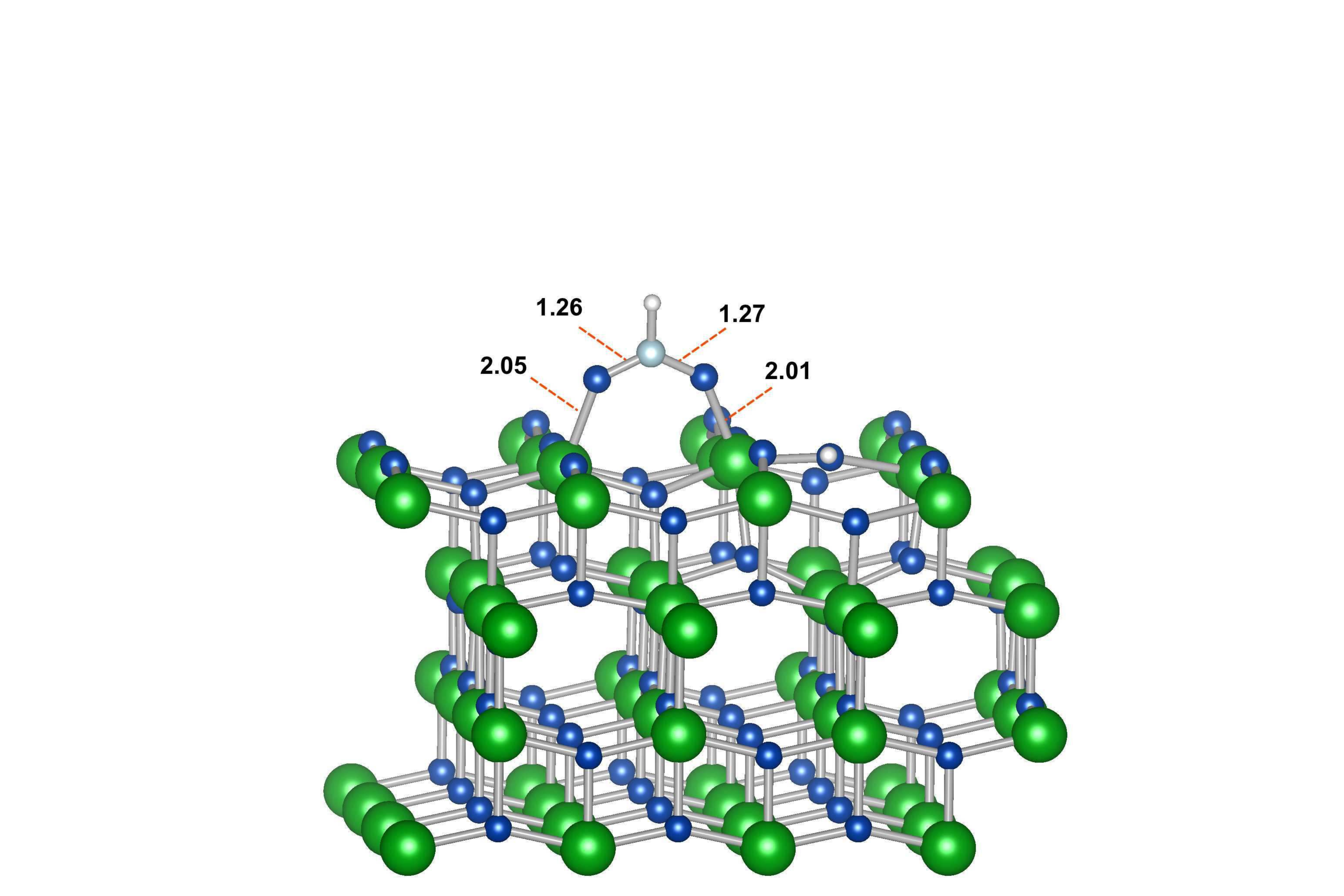}
    \end{minipage}
  &
    \begin{minipage}[h]{0.33\textwidth}
\includegraphics[trim = 350mm 250mm 350mm 210mm, clip, width=1.\textwidth]{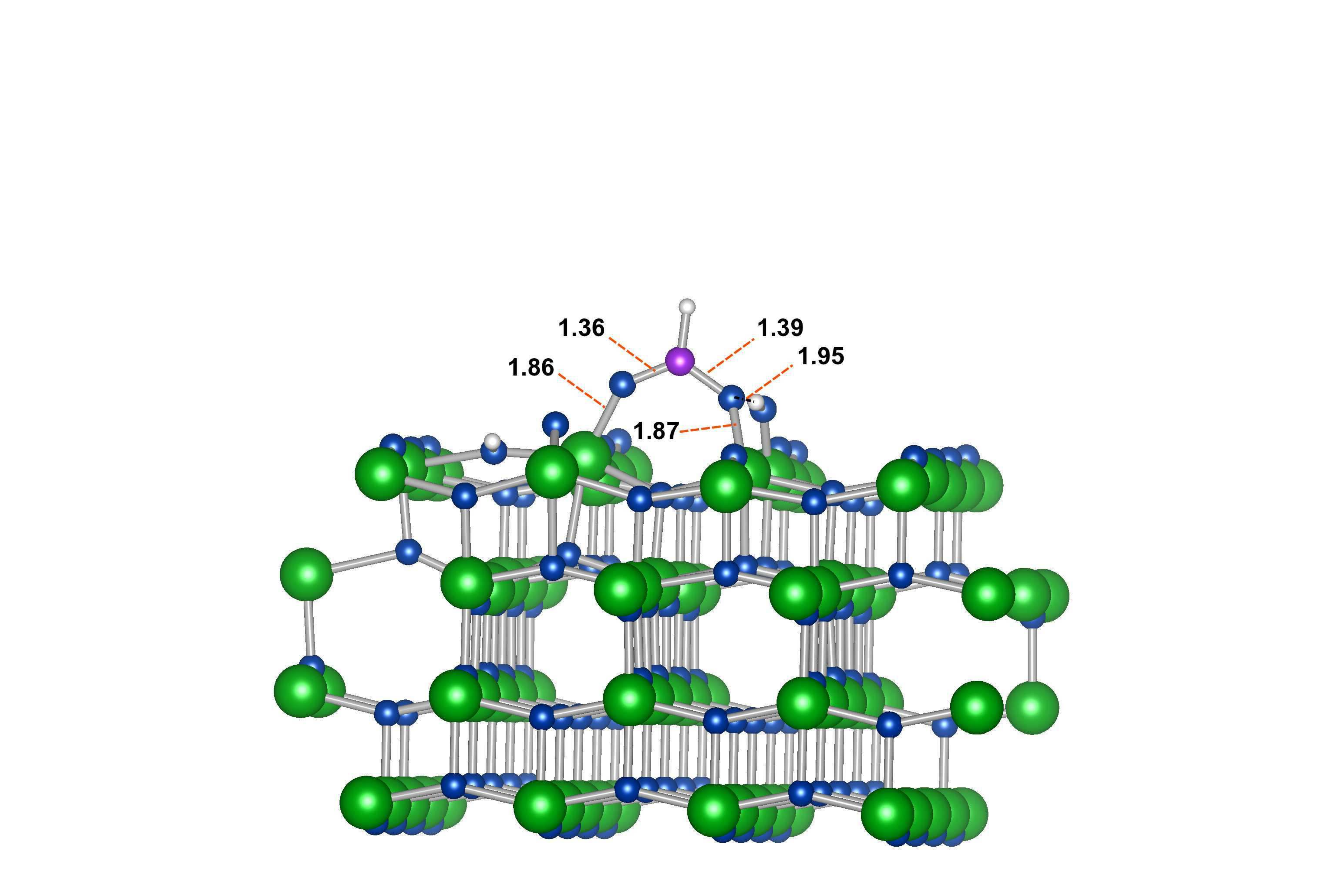}
    \end{minipage}
  &
  \begin{minipage}[h]{0.33\textwidth}
\includegraphics[trim = 325mm 250mm 375mm 210mm, clip, width=1.\textwidth]{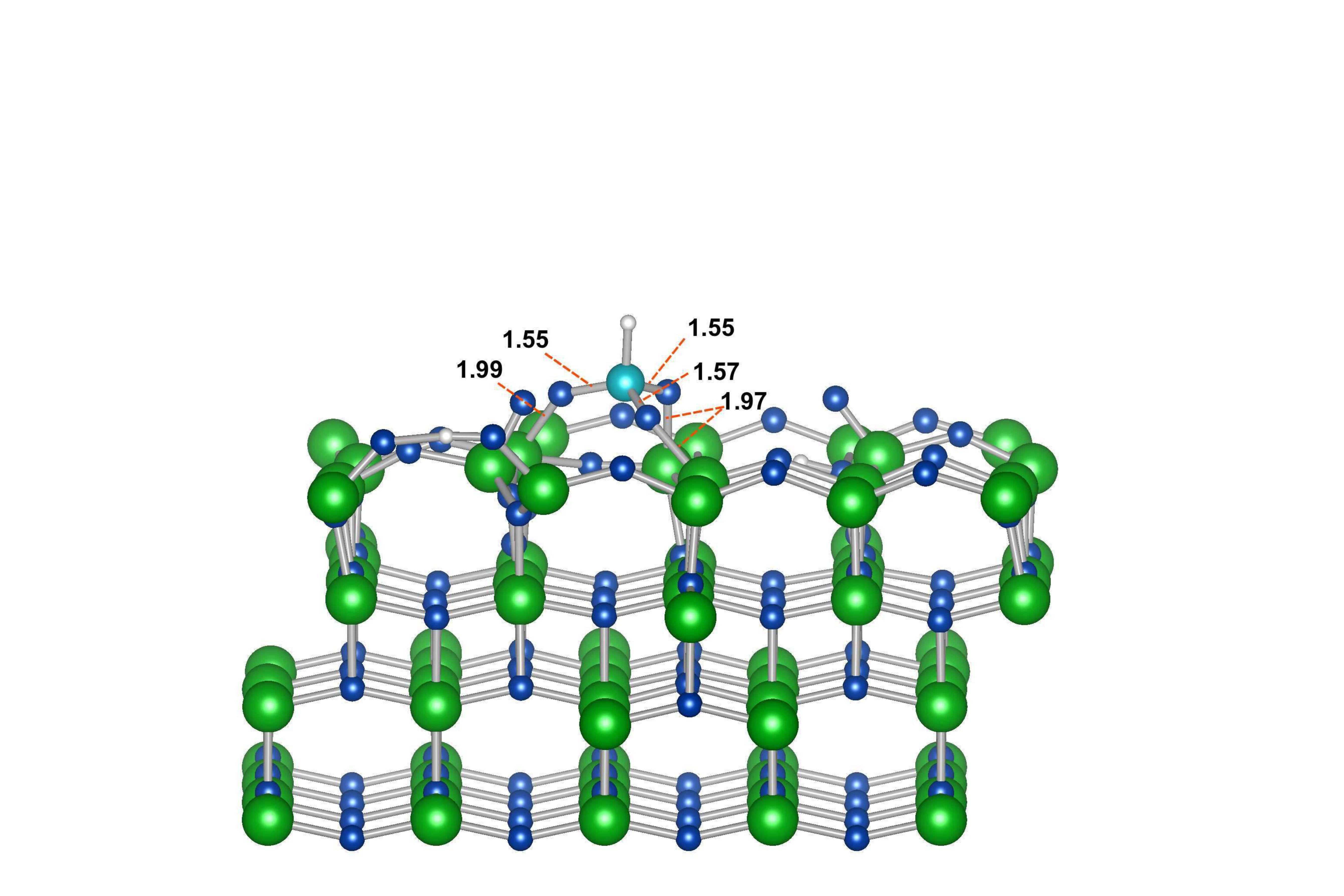}
    \end{minipage}
  \\
\textbf{(i)} BID&\textbf{(ii)} BID&\textbf{(iii)} TRI\\
    \begin{minipage}[h]{0.33\textwidth}
\includegraphics[trim = 350mm 250mm 350mm 210mm, clip, width=1.\textwidth]{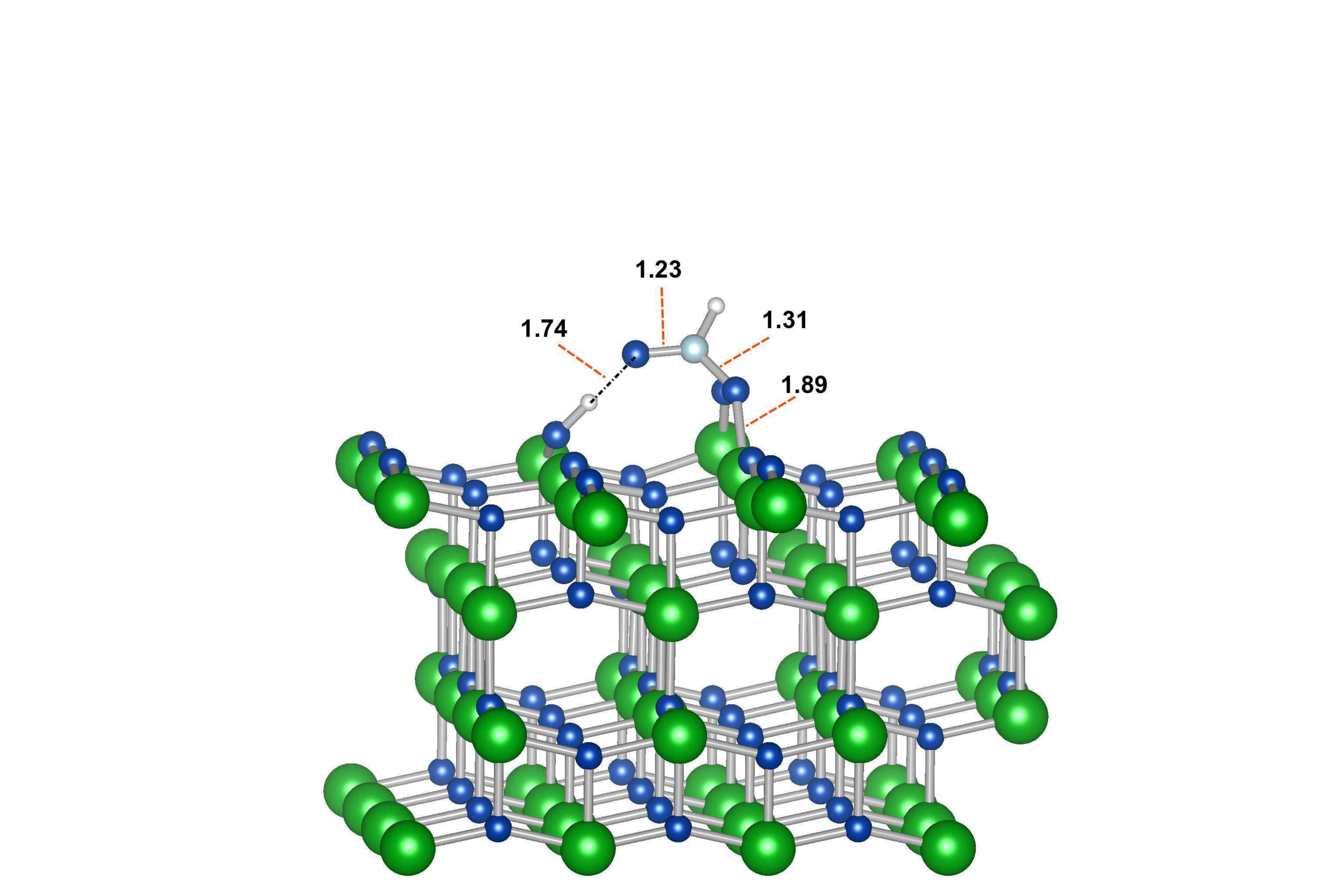}
    \end{minipage}
   &
    \begin{minipage}[h]{0.33\textwidth}
\includegraphics[trim = 275mm 250mm 425mm 210mm, clip, width=1.\textwidth]{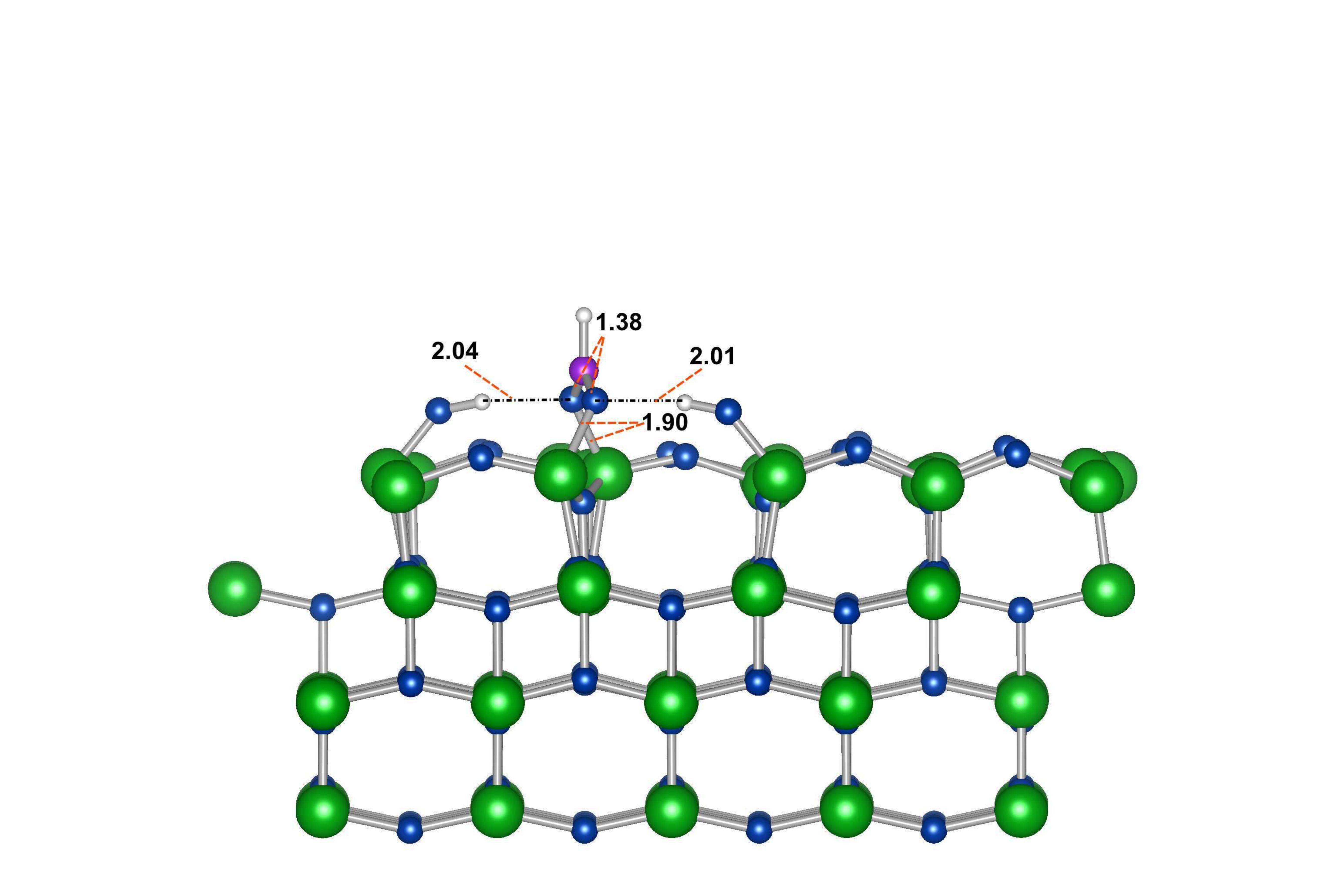}
    \end{minipage}
   &
    \begin{minipage}[h]{0.33\textwidth}
\includegraphics[trim = 350mm 250mm 350mm 210mm, clip, width=1.\textwidth]{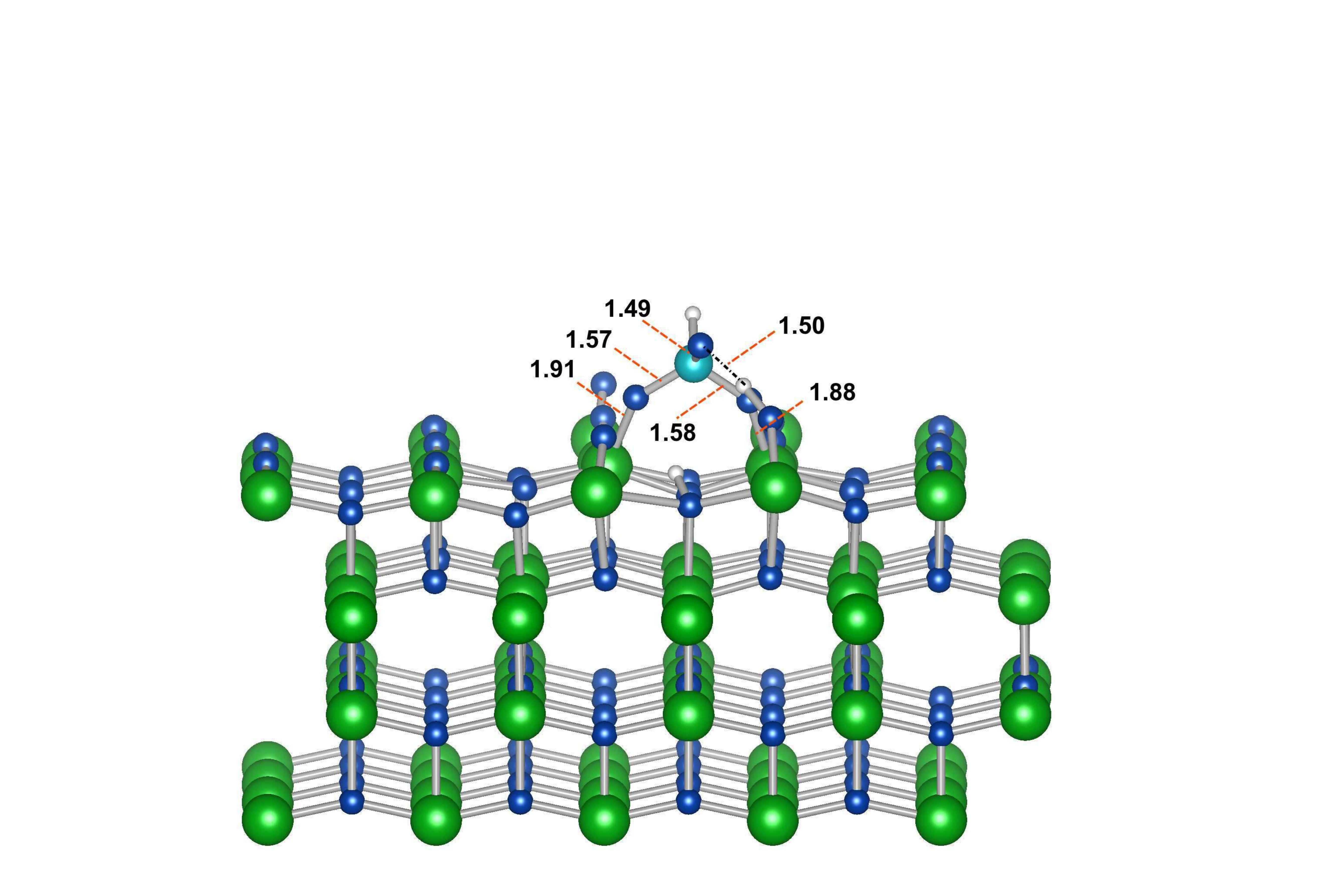}
    \end{minipage}
 \\
\textbf{(iv)} MON-H&\textbf{(v)} BID2&\textbf{(vi)} BID2\\
 \end{tabular}
 \caption{Anchor adsorption structures on anatase (001): Most stable adsorption structures for formic acid (\textbf{(i)} \& \textbf{(iv)}), boronic  acid (\textbf{(ii)} \& \textbf{(v)}) and phosphonic acid (\textbf{(iii)} \& \textbf{(vi)})}
 \label{CHPTR3_FORM_001}
 \end{center}
\end{figure}

Calculated adsorption energies for all three anchors on the (001)
surface can be seen in table ~\ref{chpt3_001_bind}, along with selected
adsorption structures in figure ~\ref{CHPTR3_FORM_001}. 
Bidentate bridging can occur in two ways when the anchor binds to
two Ti(5) atoms in the corrugated surface, with the anchor either bound
 above the O(2) atom (labelled BID in table ~\ref{chpt3_001_bind}) , 
or above the O(3) atom in the surface (labelled BID2 in table
~\ref{chpt3_001_bind}). 
Binding above the O(3) atom results in a more stable structure than when
doing so above the O(2) surface atom,
this is as a result of the associated stress when the 
outward corrugation of the O(2) atom is 
reversed as a result of repulsion from the anchor binding above it.

High adsorption energies are indicative of the high reactivity 
of the (001) surface. This can also be seen in the tendency of the
O(2) atoms to form surface -OH groups on addition of dissociated hydrogen
atoms (see for example the boronic binding motifs in 
figure \ref{CHPTR3_FORM_001}), a feature not present 
in either rutile (110)
or anatase (101) surfaces. Another noticeable trait is the 
stabilisation of the bidentate chelating mode.

For the formic acid adsorption the monodentate structure is now found to
be the most stable, with the bidentate bridging mode again reasonably
similar in 
energy, as it was for the (101) surface. 
However the chelating mode is now also 
found to be similar energetically. All three modes are more stable than
on both the the anatase (101) and rutile (110) surfaces.

In the case of the phosphonic and boronic acids the
most stable structures are again found to be the bidentate bridging 
modes (BID2).
The phosphonic acid has numerous significantly stable binding 
structures of similar adsorption energy.
The boronic acid group also has several stable binding structures,
however the most stable bidentate mode (BID2) is significantly more 
energetically favoured than the next most stable (MON-H)

Looking at figure ~\ref{CHPTR3_FORM_001}(v) the BID2 mode for boronic anchor has, in addition to the
usual bidenate O-Ti bond, the two dissociated hydrogen atoms,
which have formed surface OH groups, also coordinating 
to the binding oxygen atoms. This goes 
some way to explain the stability of the system; moving one of 
these co-ordinating H-atoms to another O(2) atom further from 
the adsorbate, and thereby preventing it from forming a hydrogen bond
with the binding oxygen atom, results in a 0.88 eV less stable structure.
 
For the (001) surface we have the important result that a 
reorganisation of the most stable 
binding anchors occurs for this surface, giving us 
$\text{boronic} > \text{phosphonic} >> \text{formic}$, 
illustrating that \emph{most} used carboxylic anchor is
 therefore predicted to have the \emph{least} stable 
binding structure for the (001) surface and that the \emph{least} used 
boronic anchor is predicted to have a \emph{significantly
stronger} binding structure than both phosphonic and formic acids.

We recognise the importance of performing these calculations on the
reconstructed surface, in order to verify the impressive stability of
the boronic-(001) surface coupling. However, irrespective of this,
our results highlight a significant point when designing dyes
for DSSCs; the anchor choice should necessarily depend on the
majority surface exposed in the electrode. 
This is a particularly important point to consider
at present, given the current trend of exploring and exploiting the
properties of other electrode TiO\subscript{2} morphologies 
for which the (101) is no longer 
the dominant exposed surface. As a final point, 
the impressive binding of the boronic and phosphonic anchors to 
the unreconstructed (001) surface can be seen as making a strong
case for experimentalists to find a way of exploiting its 
reactivity by functionalising the surface before any 
reconstruction can occur. 


\section{Dye Adsorption}

Electronic structure in the composite dye-TiO\subscript{2} system
forming a DSSC will underpin the device efficiency. 
Adsorption of carefully 
selected dyes on a mesoporous TiO\subscript{2} electrode
will introduce occupied states
in the band-gap of TiO\subscript{2}. Reducing the effective 
band-gap of the system in this manner thereby provides a 
photoexcitation pathway for photons which fall outside of the UV 
range of TiO\subscript{2}. Among other factors, the relative 
postition of the dye localised frontier orbitals with respect to 
those of the TiO\subscript{2} electrode will have a 
significant role in defining the open circuit voltage and 
therefore the device efficiency. 
Altering the anchoring group will necessarily change the electronic
structure of the dye, and may therefore affect the 
postitioning of its electronic levels relative to that of the 
TiO\subscript{2}, thereby directly affecting the
device efficieny.

\begin{figure}[h]
  \begin{center}
   \begin{tabular}{c c c}
\begin{minipage}[h]{0.3\textwidth}
      \includegraphics[trim = 0mm 0mm 0mm 0mm, clip, width=1.\textwidth]{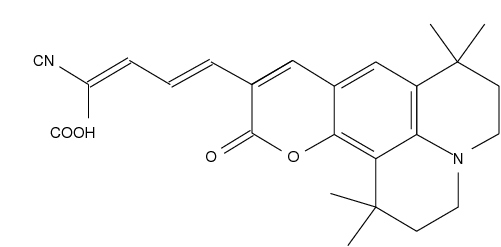}
      \end{minipage}&
        \begin{minipage}[h]{0.3\textwidth}
        \includegraphics[trim = 0mm 0mm 0mm 0mm, clip, width=1.\textwidth]{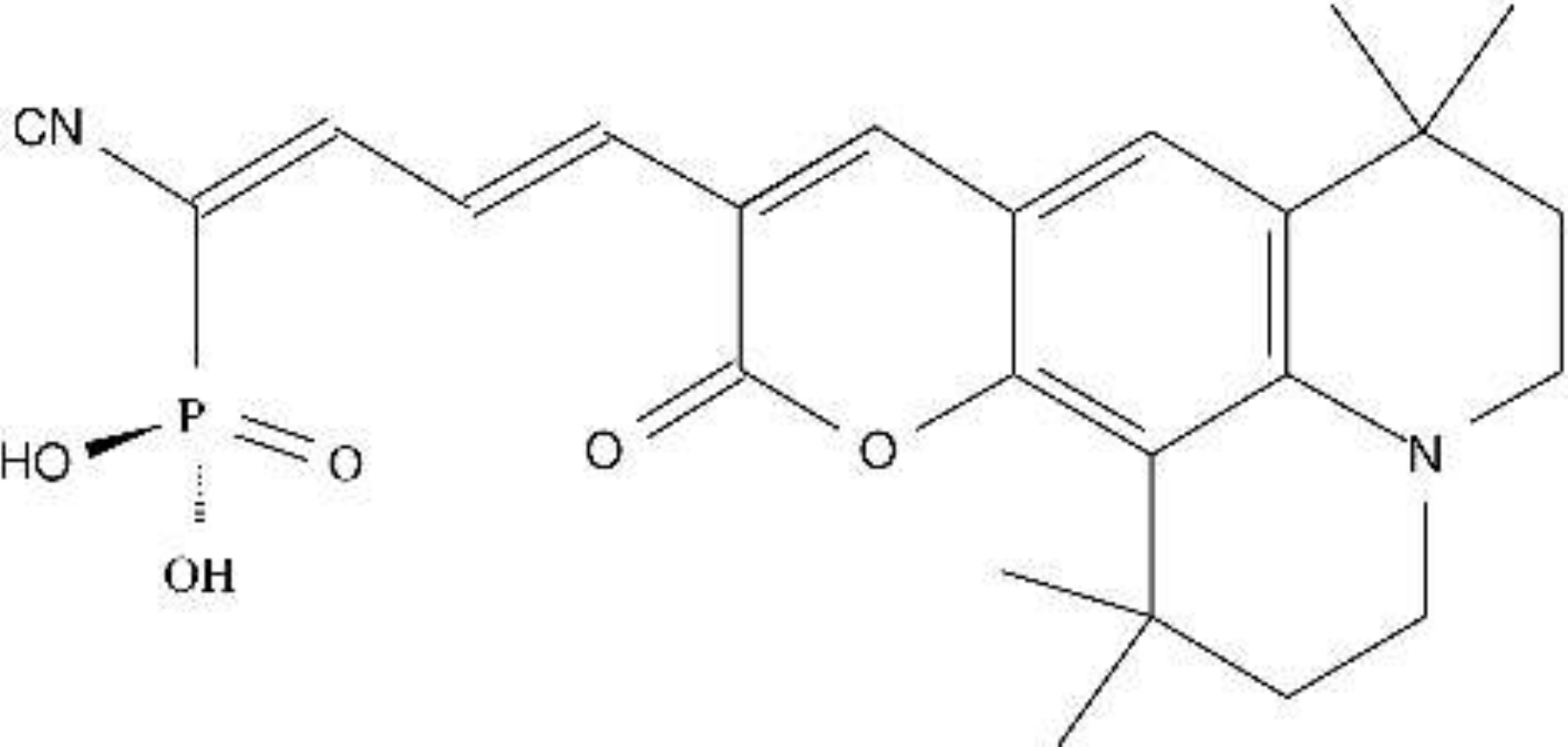}
         \end{minipage}&
      \begin{minipage}[h]{0.3\textwidth}
         \includegraphics[trim = 0mm 0mm 0mm 0mm, clip, width=1.\textwidth]{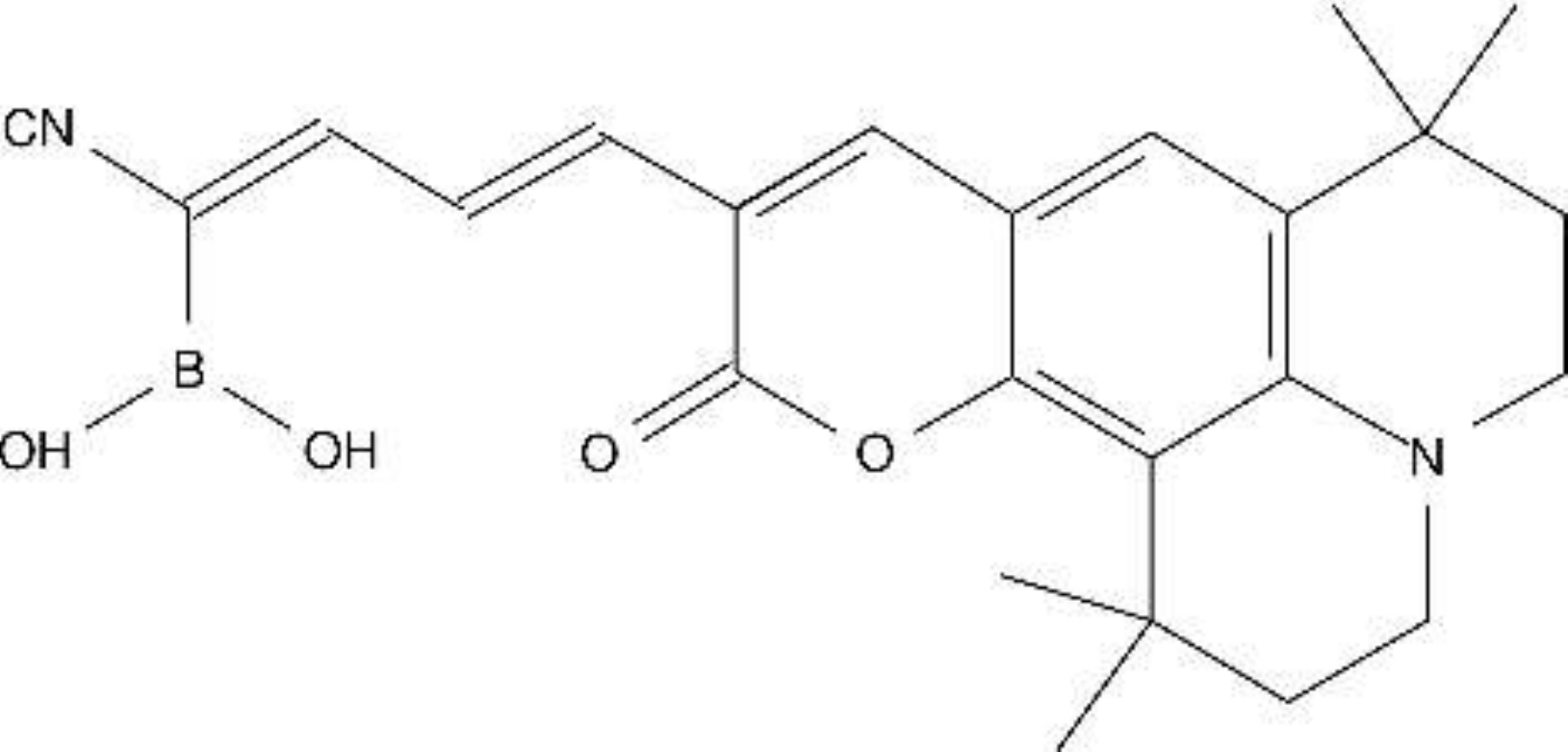}
      \end{minipage}\\
        \textbf{(i)} & \textbf{(ii)}&\textbf{(iii)}\\
     \end{tabular}
     \end{center}
     \caption{Chemical structures for the adsorbed dye molecules:(i) NKX-2311, (ii) NKX-2311P, (iii) NKX-2311B}
    \label{dye_struct}
\end{figure}

In order to gauge the extent that changing 
the anchoring group in a DSSC will affect the electronic structure,
we have performed calculations on a well documented coumarin based
dye, named NKX-2311\cite{Hara_c343,c343_femto,c343_tddft,c343_QMD}, 
adsorbed on each of our 
three TiO\subscript{2}
surfaces. NKX-2311 utilises the carboxylic acid binding moiety, 
allowing us to exmaine the effect of using a 
differing anchor by simply
replacing the anchoring group with either the phosphoric or boronic
anchors. The chemical structures for NKX-2311 and the same 
coumarin dye 
utilising the phosphoric and boronic anchors, which we name 
NKX-2311P and NKX-2311B respectively, can be 
seen in figure ~\ref{dye_struct}. 

\subsection{Anatase (101)}

\begin{figure}
 \begin{center}
  \begin{tabular}{c c c}
    \begin{minipage}[h]{0.33\textwidth}
\includegraphics[trim = 180mm 0mm 200mm 0mm, clip, width=1.\textwidth]{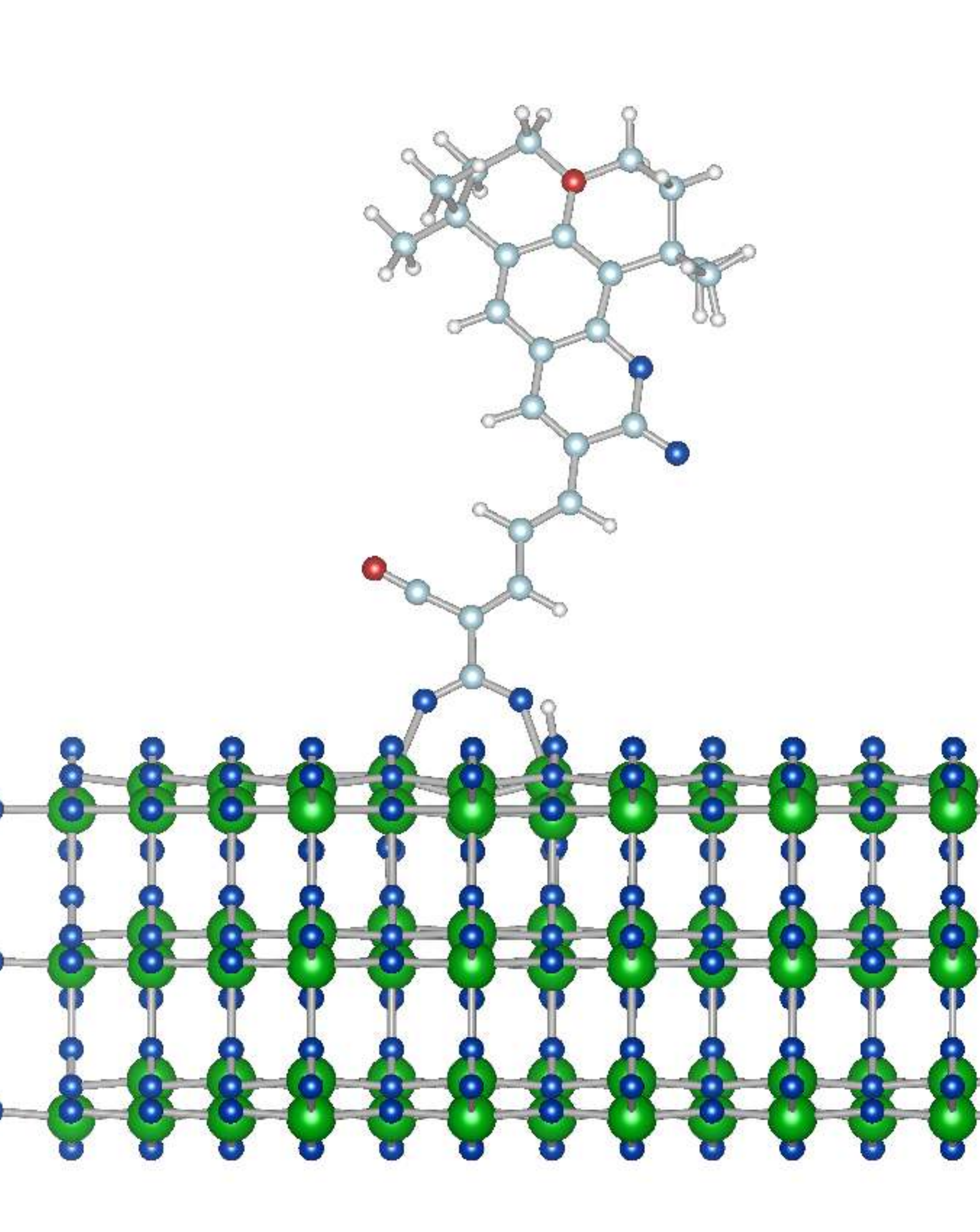}
    \end{minipage}
  &
    \begin{minipage}[h]{0.33\textwidth}
\includegraphics[trim = 180mm 0mm 200mm 0mm, clip, width=1.\textwidth]{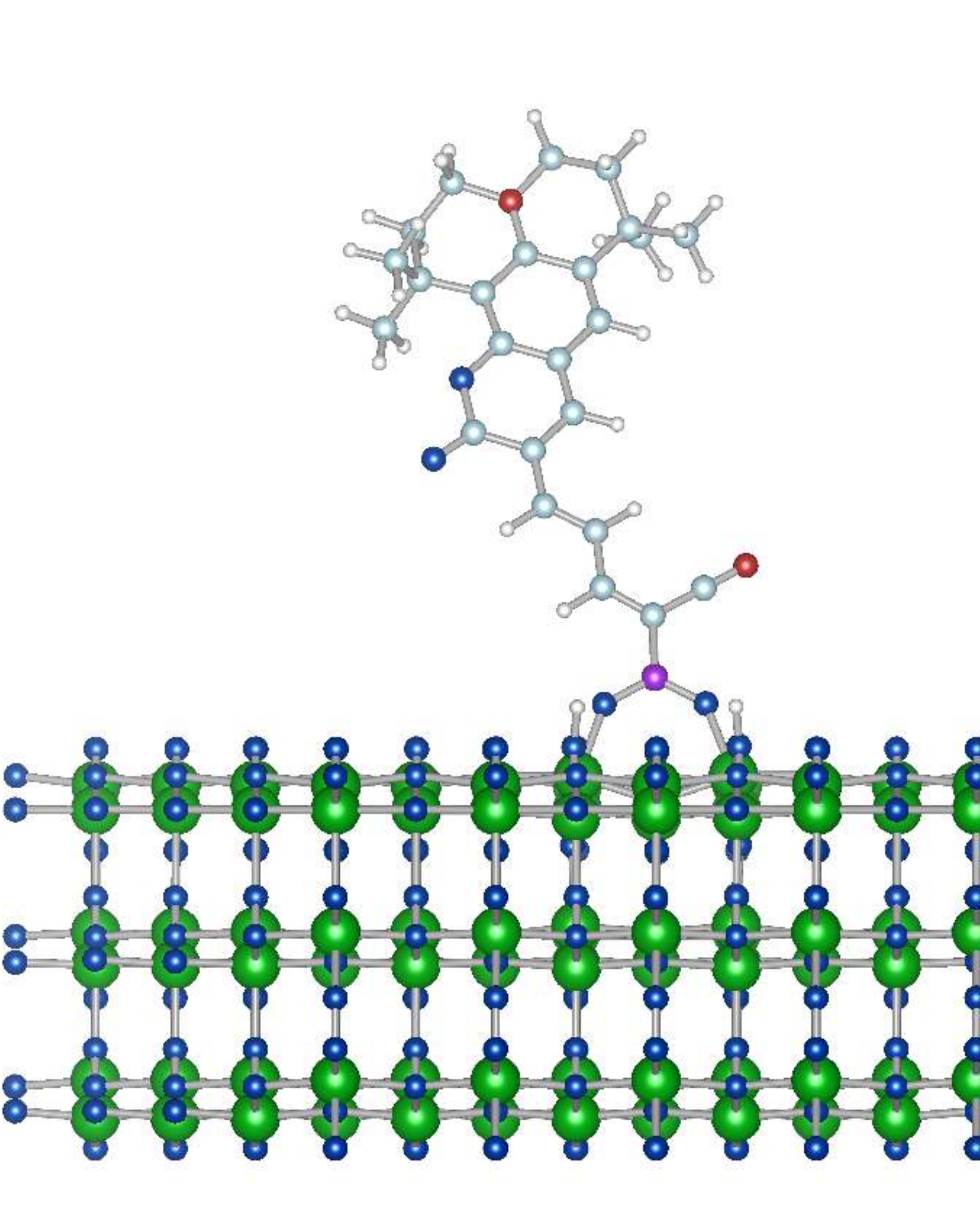}
    \end{minipage}
  & 
  \begin{minipage}[h]{0.33\textwidth}
\includegraphics[trim = 180mm 0mm 200mm 0mm, clip, width=1.\textwidth]{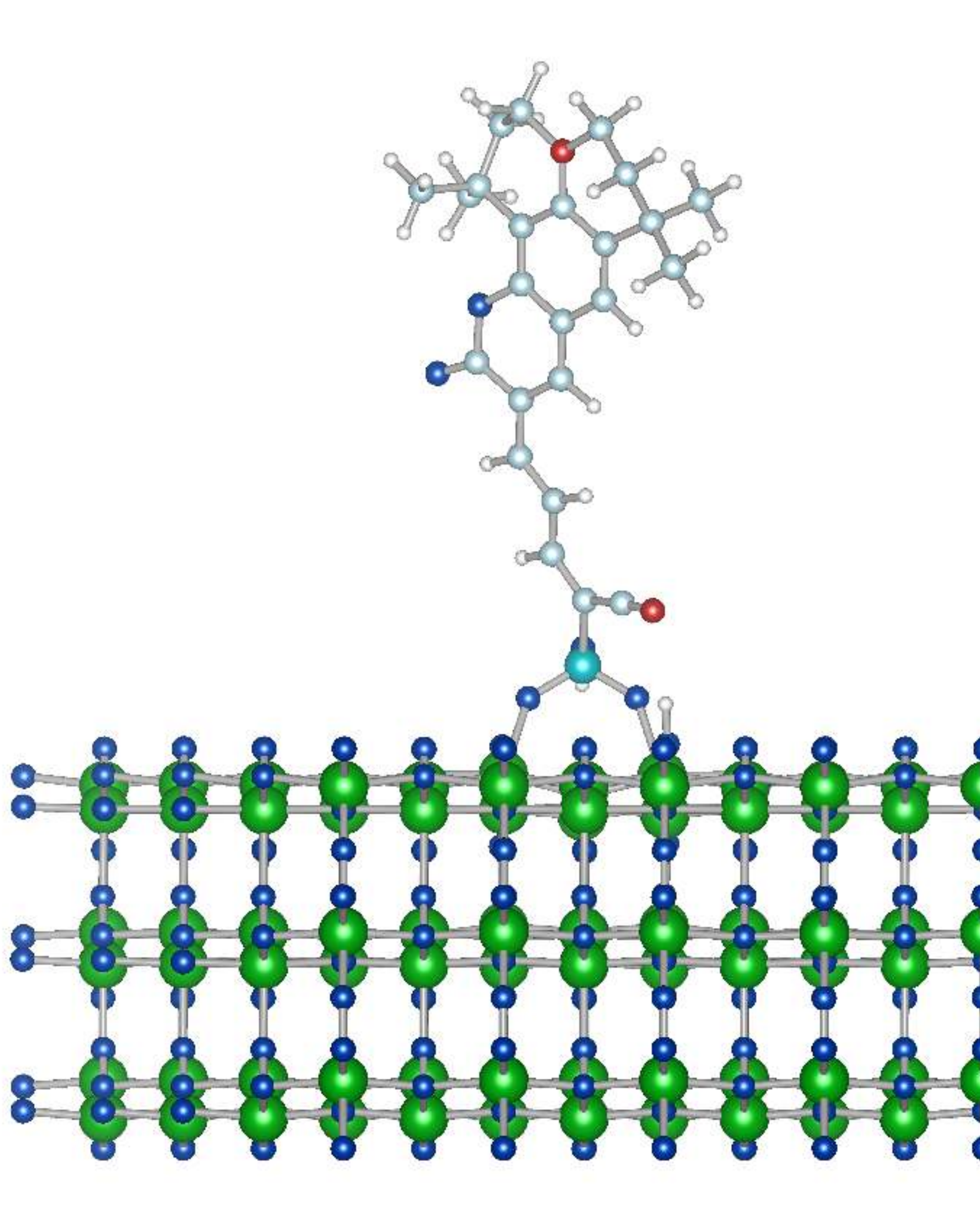}
    \end{minipage}
  \\
 
 $E_{Ads} = -1.11$ eV& $E_{Ads} = -0.91$ eV  & $E_{Ads}= -1.88$ eV \\
   \begin{minipage}[h]{0.33\textwidth}
\includegraphics[trim = 180mm 0mm 200mm 0mm, clip, width=1.\textwidth]{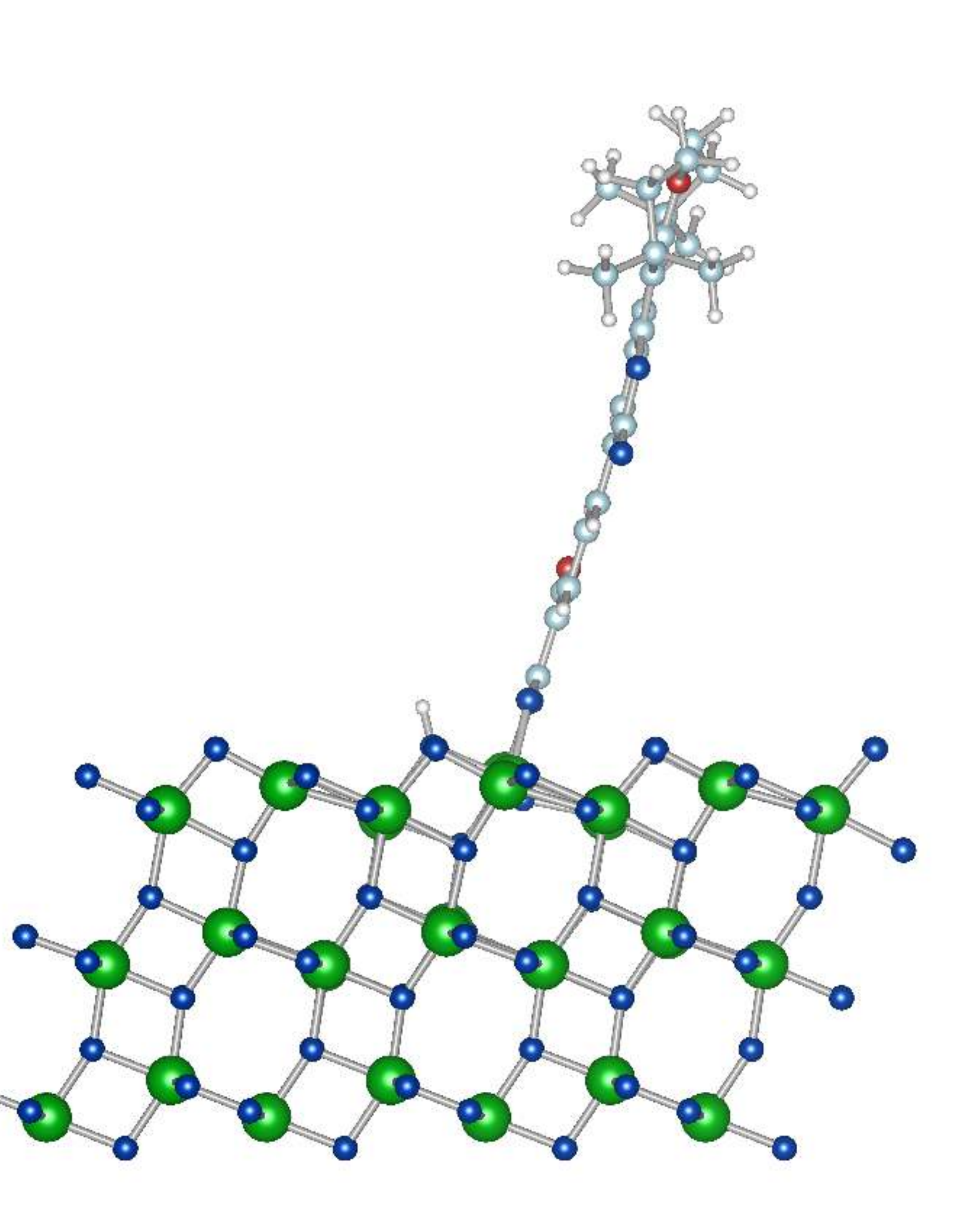}
    \end{minipage}
   &
    \begin{minipage}[h]{0.33\textwidth}
\includegraphics[trim = 180mm 0mm 200mm 0mm, clip, width=1.\textwidth]{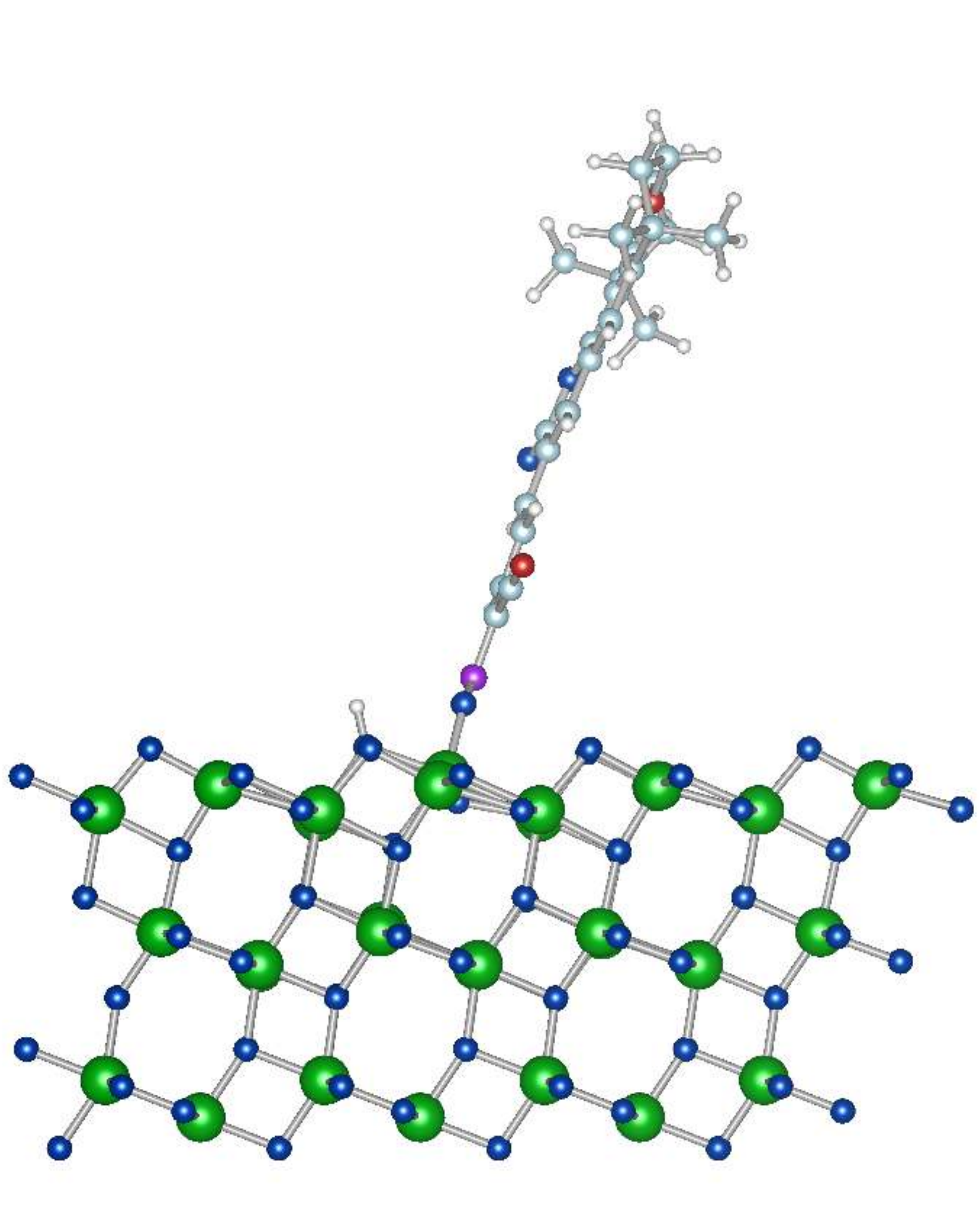}
    \end{minipage}
   &
    \begin{minipage}[h]{0.33\textwidth}
\includegraphics[trim = 180mm 0mm 200mm 0mm, clip, width=1.\textwidth]{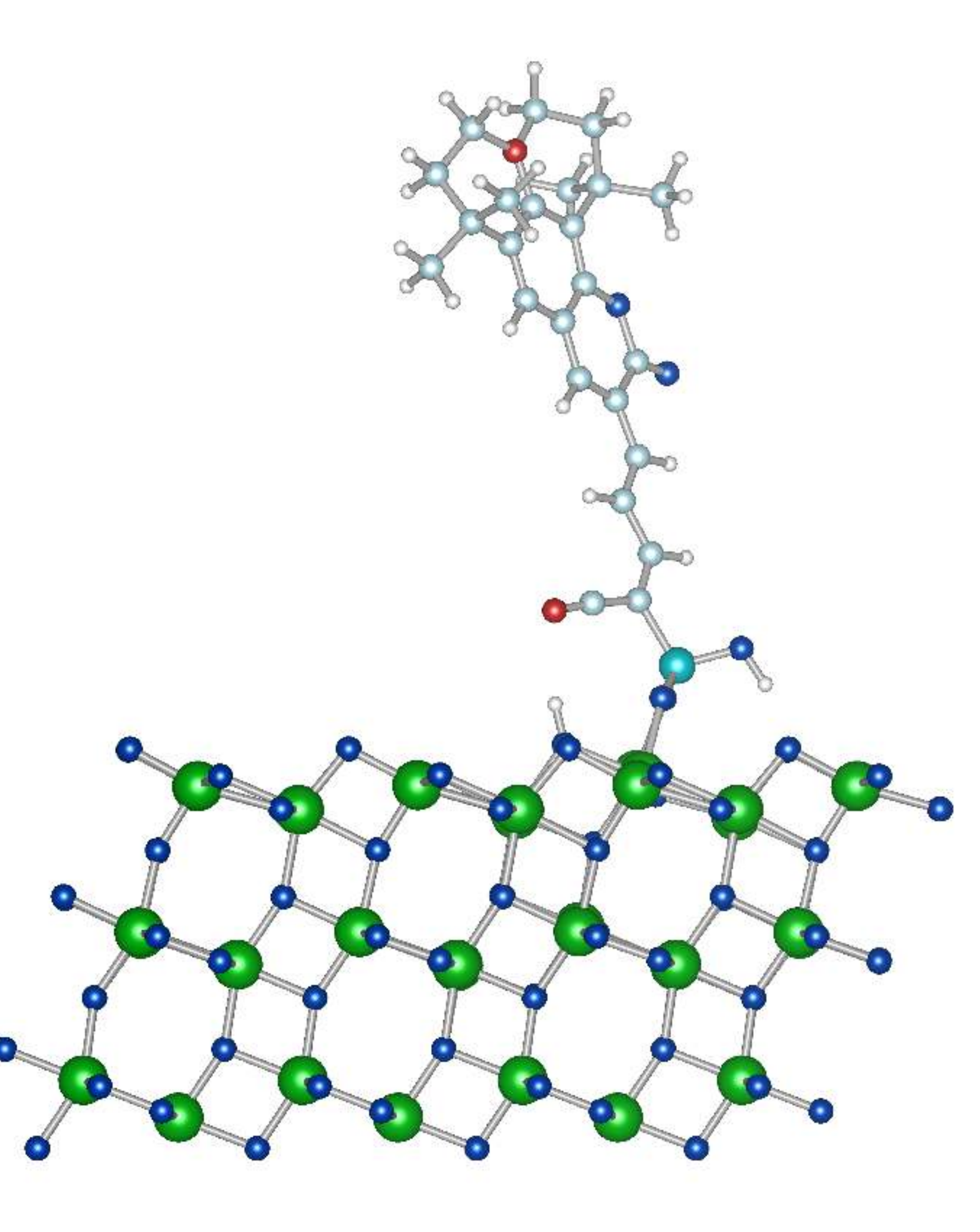} 
    \end{minipage}
 \\ 
\textbf{(i)}&\textbf{(ii)}&\textbf{(iii)}\\
 \end{tabular}
 \caption{Relaxed dye adsorption geometries on Anatase (101): \textbf{(i)} NKX-2311, \textbf{(ii)} NKX-2311B and \textbf{(iii)} NKX-2311P. Front (top) and side (bottom) perspectives are shown. }
 \label{DYE_ADS_101}
 \end{center}
\end{figure}

For each surface we have opted 
to examine the dyes bound in the strongest adsorption
configuration found for each of the isolated anchoring groups from 
the previous section. In the case of the anatase (101) surface
all three anchors were found to bind preferentially in a bidentate 
bridging mode. Relaxed structures for each of our three dye variants 
on the (101) surface can be seen in figure \ref{DYE_ADS_101}, 
along with the calculated adsorption energies.

Examining the adosoption geometries we see that very little 
difference is found in comparison to the geometry of the 
respective anchor groups in isolation,  and we find that the bond 
lengths between the anchor and the TiO\subscript{2}
changing only very slightly (the largest change is $\sim$ 2\%). 
Each of the dyes is found to bind more strongly to the
surface than the anchors in isolation by $\sim 0.1-0.2$ eV. 
The relative strength of the adsorption interaction
remains the same as that of the
isolated anchors, with NKX-2311P $>$ NKX2311 $>$ NKX-2311B, 
illustrating that in terms of binding properties alone the
phosphonic anchor is most effective on the (101) surface.

\begin{figure}
\begin{center}
 \begin{minipage}[c]{0.7\textwidth}
  \includegraphics[angle=-90,origin=c,trim = 30mm 0mm 30mm 0mm, clip,width=1.0\textwidth]{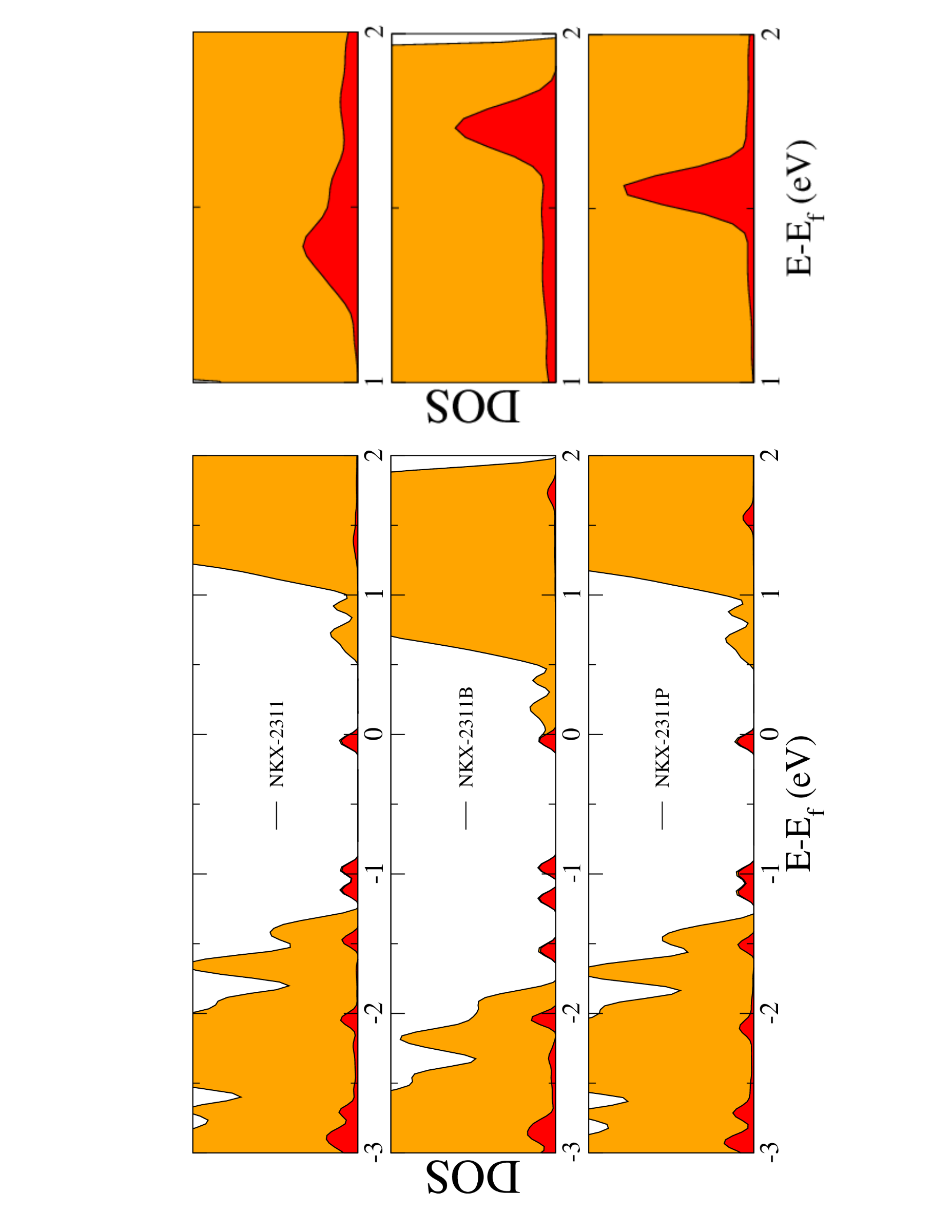}
  \end{minipage}

\caption{Projected density of states for dyes adsorbed on the anatase (101) surface. Top: NKX-2311,  Middle: NKX-2311B, Bottom: NKX-2311P. Total DOS in orange, with the dye localised PDOS in red. On the right hand side the 
LUMO\superscript{*} is enlarged for clarity.}
 \label{101_PDOS}
\end{center}
\end{figure}

In order to examine the electronic overlap between the adsorbed
dye and the TiO\subscript{2} lattice we have 
generated partial density of states plots (PDOS), along with 
the projection on the adsorbed dyes, which can be seen in 
figure \ref{101_PDOS}. Examining the PDOS we see that
all three dyes introduce dye localised states in the gap, as well
as in the conduction band. 
Previous theoretical work on the NKX-2311 dye adsorbed on a
TiO\subscript{2} cluster calculated the
HOMO to dye localised LUMO energy difference as being 1.425eV \cite{c343_tddft}.
 Here we obtain a simliar value of $\sim$ 1.4 eV.
Comparing the PDOS for NKX-2311 and NKX-2311P
we can see that for this surface there is only a marginal change
in the electronic structure as a result of using a phosphonic 
anchoring group. The highest occupied molecular orbital (HOMO)
is dye localised in both cases, and resides around 0.5 eV below
the TiO\subscript{2} conduction band. Further dye localised
states are introduced at around the same energy ($\sim 0.1-0.3$ eV) 
above the valence band in both cases. 

In the case of the boronic
anchor we have a markedly different electronic structure; relative 
shifts of the dye localised states occur both within 
the TiO\subscript{2} 
band-gap and with respect to each other, with the result that
a further dye localised state is introduced into the gap.
It is also worth noting that the HOMO is found to reside essentially 
at the bottom of the band-gap, with the
composite system ostenibly acting as a conductor. 
However, semi-local GGA density
functional theory is known to underestimate the band-gap in 
semiconductors, as a result of a lack of the discontinuity in the
exchange-correlation potential \cite{band-gap}. This results in 
our calculated band-gap being $\sim$ 2 eV as opposed to 
the experimentally reported value of 3.2 eV. 

Bearing this point in mind, and given that DFT is a ground state
theory we now discuss the possible consequences of the
electronic structure on DSSC efficiency. At face value the 
remarkably similar electronic structure of the NKX-2311P and NKX-2311
dyes would suggest that they would be likely to have similar
device properties. However examining the dye localised LUMO
states (LDOS\superscript{*})
in the right hand side of figure \ref{101_PDOS} we see that the
NKX-2311 LDOS\superscript{*} is considerably more
extended throughout the TiO\subscript{2} conduction band. Charge
injection times will depend upon the electronic overlap
between the TiO\subscript{2} and dye orbitals, suggesting that
the carboxylic anchor bound dye will have a more efficient injection
mechanism that the NKX-2311P and NKX-2311B dyes. This
result agrees with previous work examining the effect of the
anchoring group on the adsorption of a pyridine molecule\cite{FWHM}, in
which the carboxylic acid bound molecule was found to have a larger
full width at half maximum value for the dye localised LUMO.
Experiments have also been found this to be the case,
with more efficient charge injection through the carboxylic
anchoring group\cite{cvp_elect,cvp2}. While this
more efficient injection does not necessarily automatically result in
higher efficiencies for all carboxylic bound dyes, it has certainly shown
to be the case for some dye groups\cite{Carb_vs_phos,cvp2}.

For NKX-2311B we can note that despite the lower binding energy
to the (101) surface, 
and similar LUMO\superscript{*} overlap as NKX-2311P, the extra 
introduced dye localised state in the band-gap is significant. 
Increasing the number of excitation
pathways, by this introduction of a new gap state, may lead to
an increase in the short-circuit current, J\subscript{SC}, and
to an increase in device efficiency over NKX-2311P.


\subsection{Rutile (110)}

\begin{figure}
 \begin{center}
  \begin{tabular}{c c c}
    \begin{minipage}[h]{0.33\textwidth}
\includegraphics[trim = 380mm 0mm 300mm 0mm, clip, width=1.\textwidth]{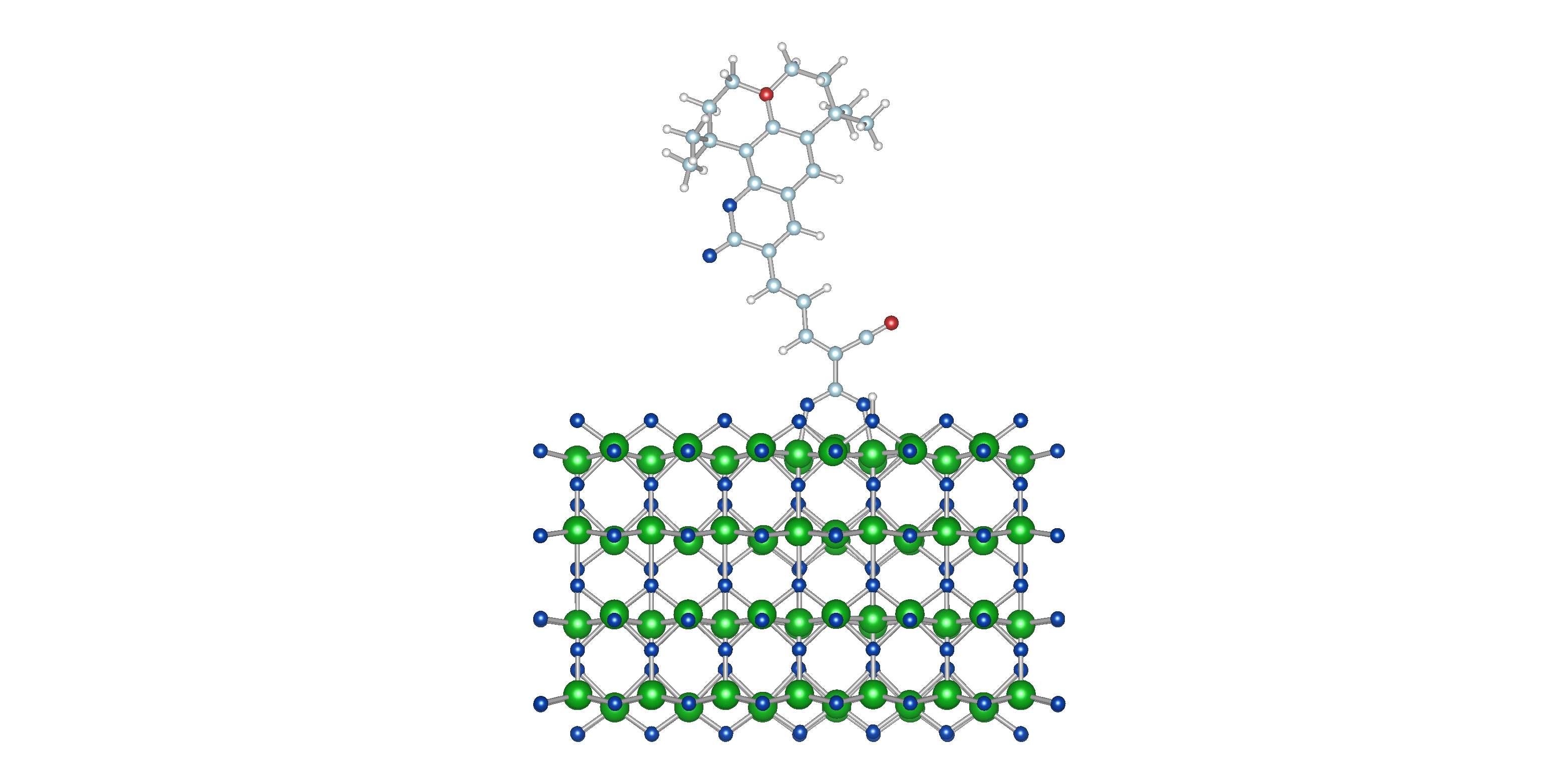}
    \end{minipage}
  &
    \begin{minipage}[h]{0.33\textwidth}
\includegraphics[trim = 380mm 0mm 300mm 0mm, clip, width=1.\textwidth]{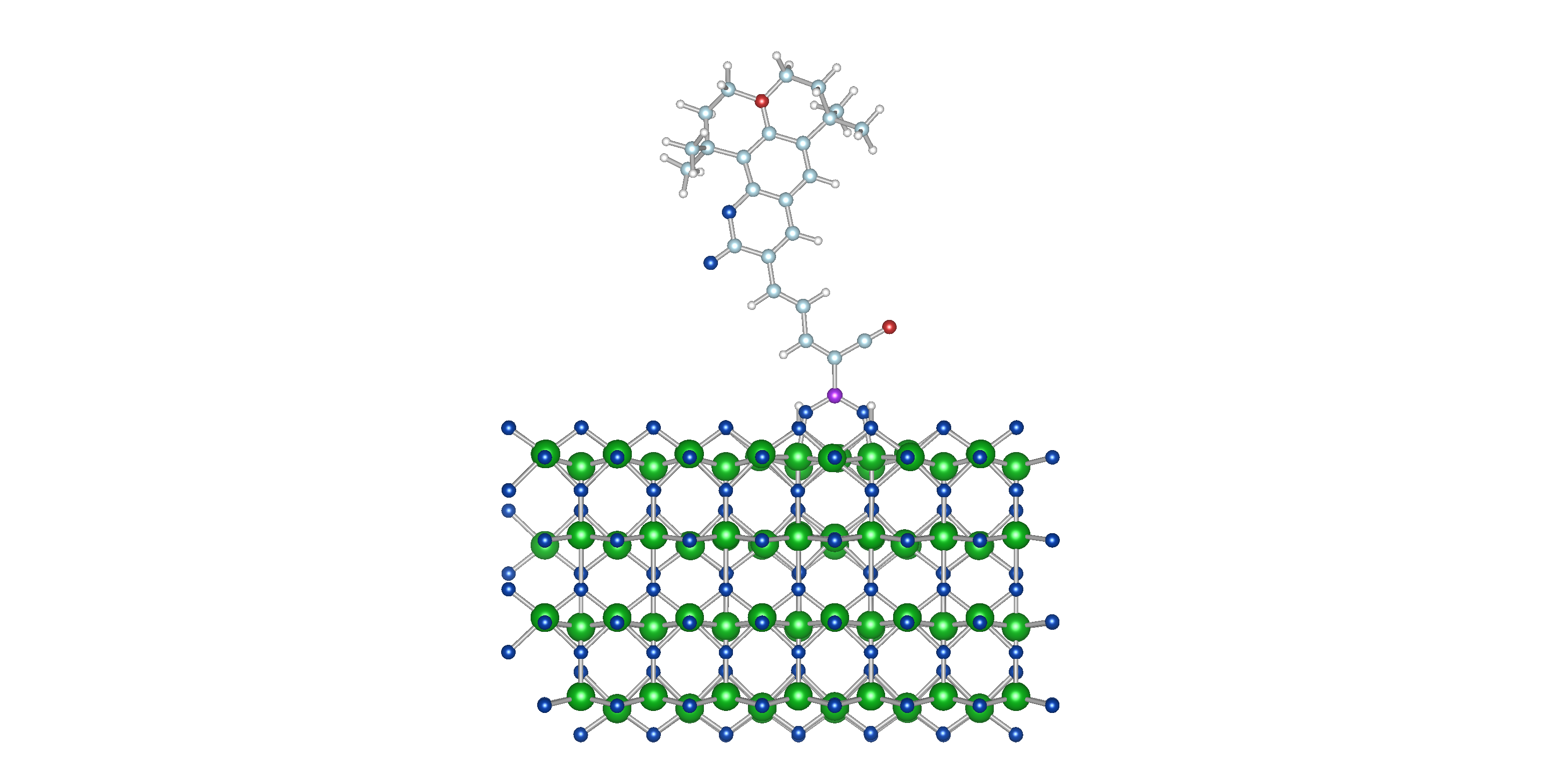}
    \end{minipage}
  & 
  \begin{minipage}[h]{0.33\textwidth}
\includegraphics[trim = 380mm 0mm 300mm 0mm, clip, width=1.\textwidth]{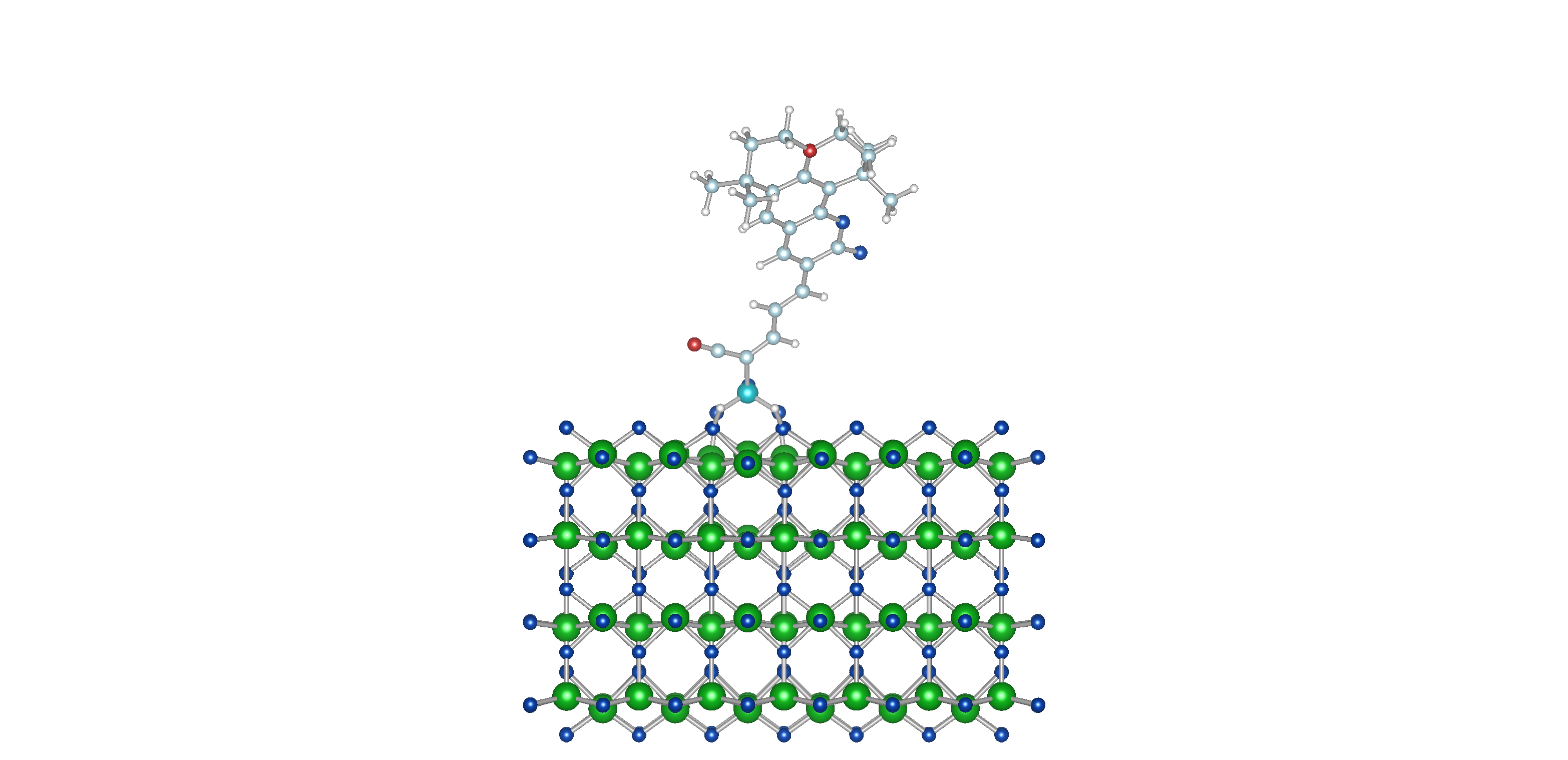}
    \end{minipage}
  \\
  $E_{Ads} = -1.74$ eV& $E_{Ads} = -1.53$ eV  & $E_{Ads}= -2.26$ eV \\
   \begin{minipage}[h]{0.33\textwidth}
\includegraphics[trim = 380mm 0mm 300mm 0mm, clip, width=1.\textwidth]{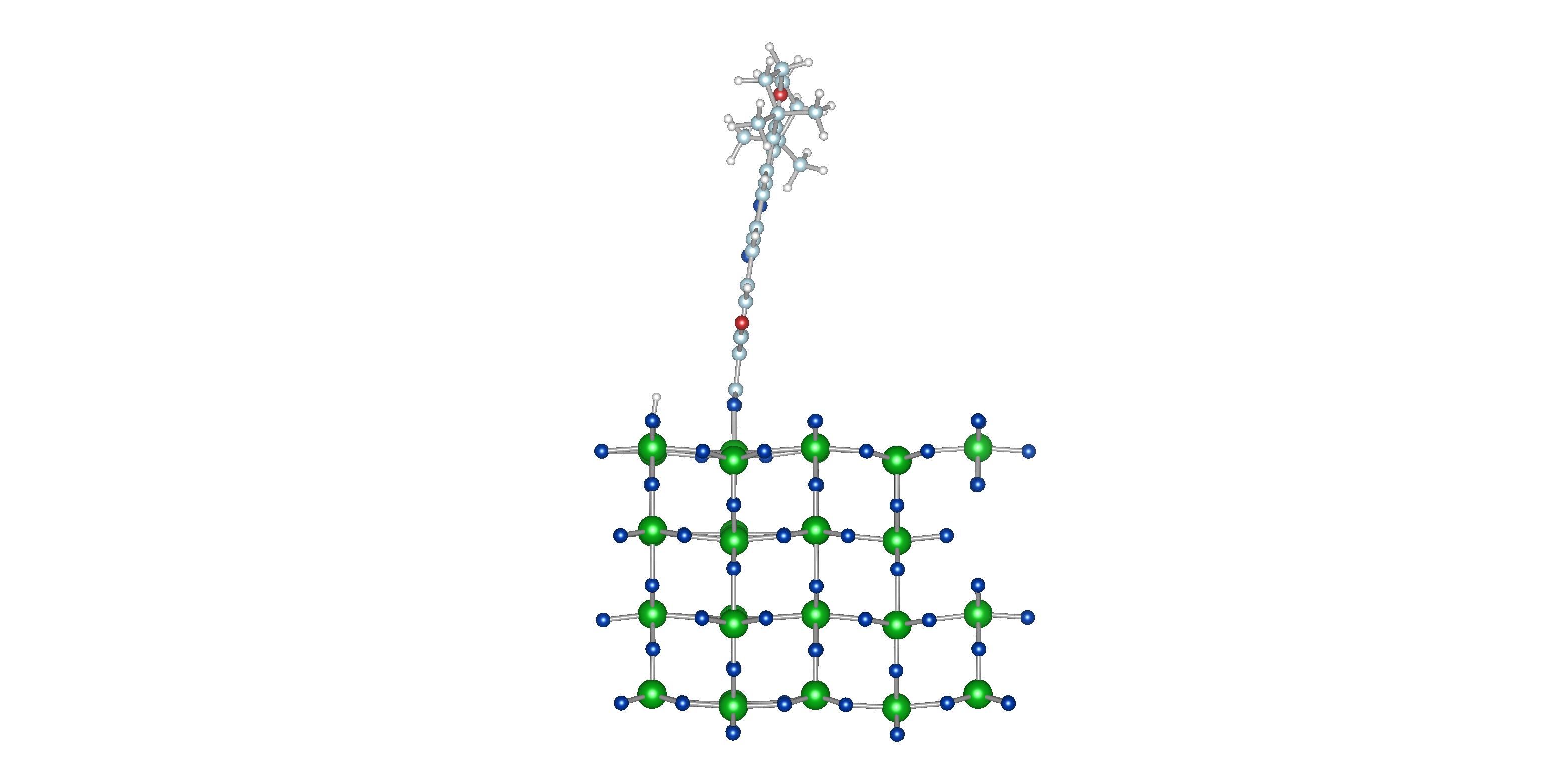}
    \end{minipage}
   &
    \begin{minipage}[h]{0.33\textwidth}
\includegraphics[trim = 380mm 0mm 300mm 0mm, clip, width=1.\textwidth]{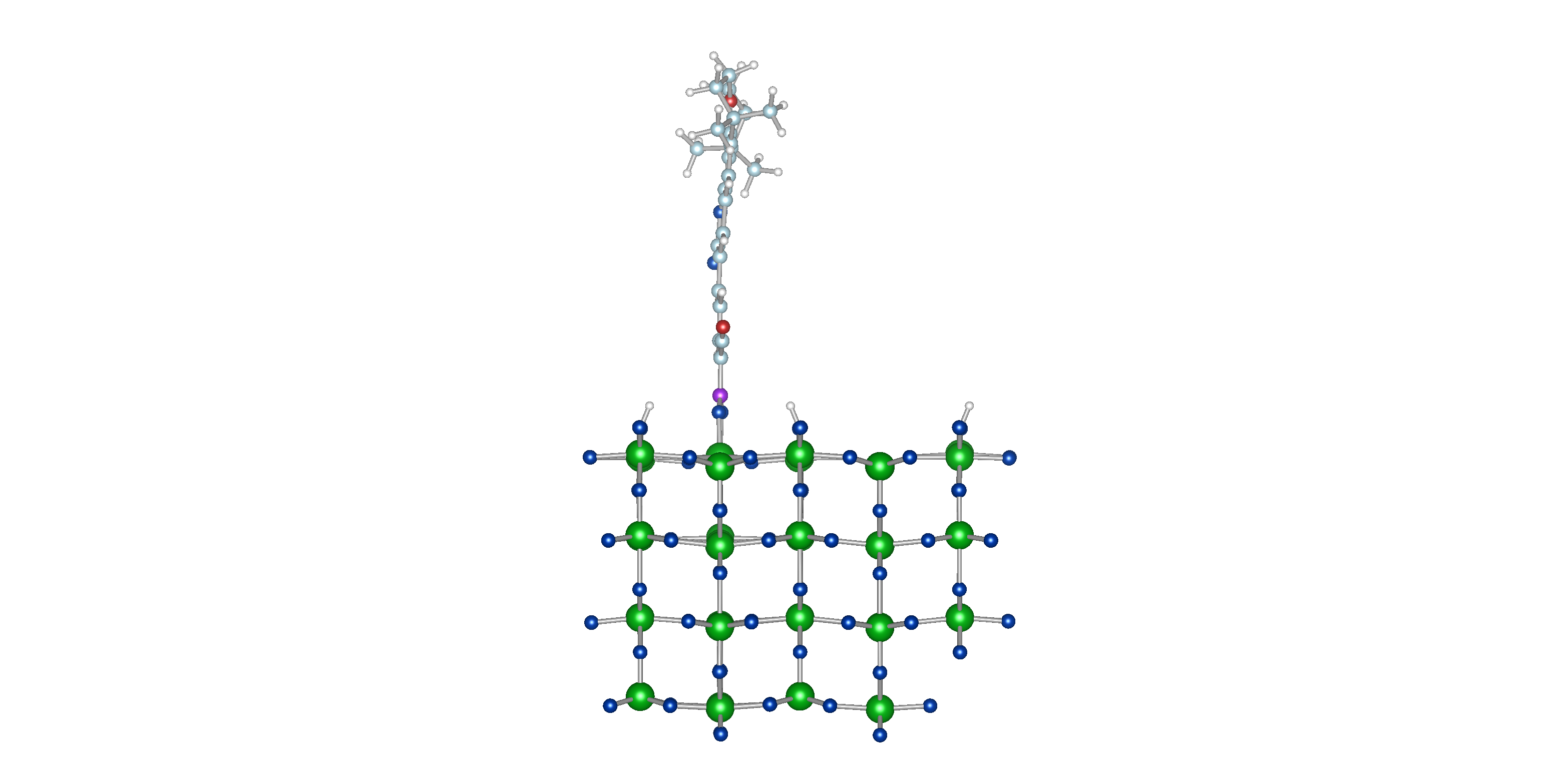}
    \end{minipage}
   &
    \begin{minipage}[h]{0.33\textwidth}
\includegraphics[trim = 380mm 0mm 300mm 0mm, clip, width=1.\textwidth]{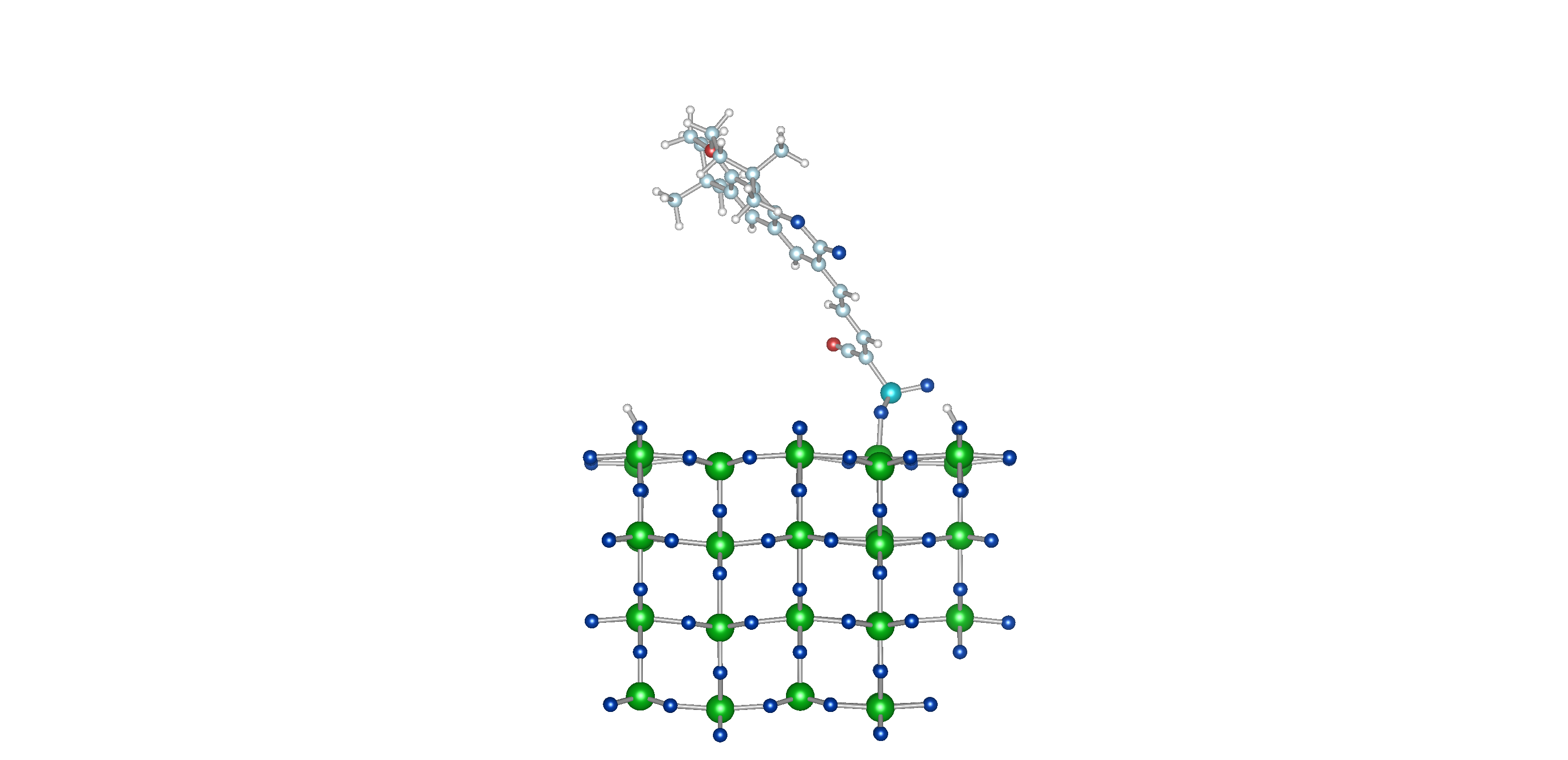} 
    \end{minipage}
 \\ 
\textbf{(i)}&\textbf{(ii)}&\textbf{(iii)}\\
 \end{tabular}
 \caption{Relaxed dye adsorption geometries on Rutile (110): \textbf{(i)} NKX-2311, \textbf{(ii)} NKX-2311B and \textbf{(iii)} NKX-2311P. Front (top) and side (bottom) perspectives are shown. }
 \label{DYE_ADS_110}
 \end{center}
\end{figure}

For the rutile (110) surface, the dye binding
geometries are again very similar to that
of the isolated anchor groups, as seen in the relaxed 
adsorption structures shown
in figure \ref{DYE_ADS_110}. The ordering of the
relative strengths of the interaction is maintained (also shown in 
figure \ref{DYE_ADS_110}) with 
NKX-2311P $>$ NKX-2311 $>$ NKX-2311B, which again illustrates
that the phosphonic acid anchor is the most attractive in terms of
binding properties alone. More generally we note that each of the 
dyes bind more strongly to the (110) rutile surface than the anatase
(101). 

\begin{figure}
\begin{center}
 \begin{minipage}[c]{0.7\textwidth}
  \includegraphics[angle=-90,origin=c,trim = 30mm 0mm 30mm 0mm, clip,width=1.0\textwidth]{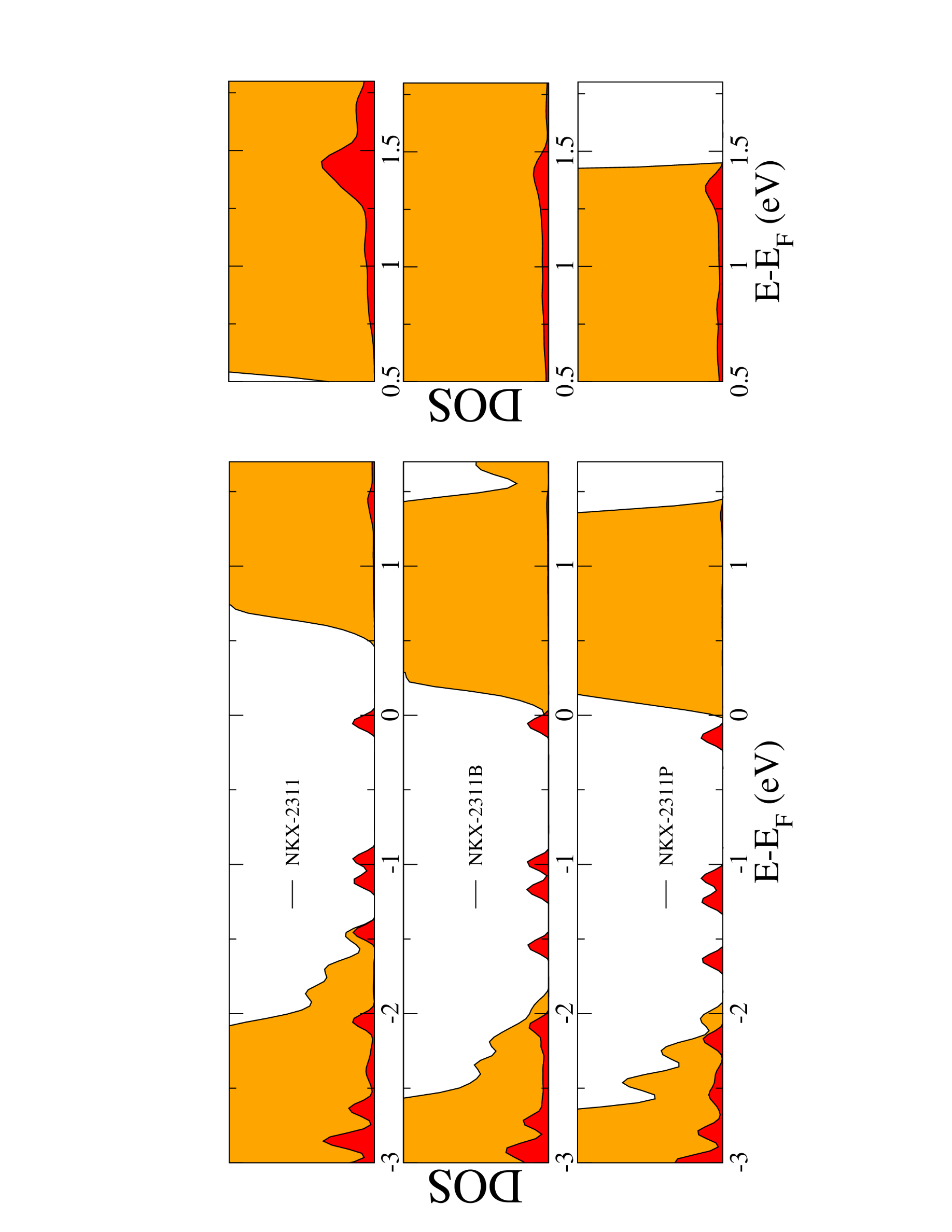}
  \end{minipage}

\caption{Projected density of states for dyes adsorbed on the rutile (110) surface. Top: NKX-2311,  Middle: NKX-2311B, Bottom: NKX-2311P. Total DOS in orange, with the dye localised PDOS in red. On the right hand side the
LUMO\superscript{*} is enlarged for clarity.}
 \label{110_PDOS}
\end{center}
\end{figure}

Examining the electronic properties in figure \ref{110_PDOS} we can see
that as opposed to the anatase (101) surface the electronic
structre for the NKX-2311P dye now follows that of the NKX-2311B 
dye, 
with shifted states in the gap relative to the NKX-2311 dye. This again
has the effect of introducing an extra state in the gap about
$0.3$eV above the 
valence band for both of these dyes, with the HOMO just below the 
calculated conduction band. For the NKX-2311 dye the HOMO is 
further from the conduction band, and there are fewer states in the gap.
As with the boronic anchored dye on the (101) surface
 this may again have the effect of increasing the number of excitation 
pathways for the boronic and phosphonic anchored dyes over
the formic anchor, and lead to higher J\subscript{SC} values. 
This different behaviour of the phosphonic anchored dye on the rutile (110)
as opposed to the anatase (101) surface
 is an interesting result illustrating that, as with the isolated anchor 
groups, 
there are considerable differences between the results on alternative
TiO\subscript{2} surfaces.

Examining the LUMO\superscript{*} states we again see that 
the NKX-2311 dye has a larger number of dye localised states well 
spread out within the TiO\subscript{2} conduction band. As with
these dyes on the (101) surface this significant overlap would suggest
that the carboxylic acid anchor may have the most efficient charge 
injection mechanism. 

\subsection{Anatase (001)}

\begin{figure}
 \begin{center}
  \begin{tabular}{c c c}
    \begin{minipage}[h]{0.33\textwidth}
\includegraphics[trim = 180mm 0mm 200mm 0mm, clip, width=1.\textwidth]{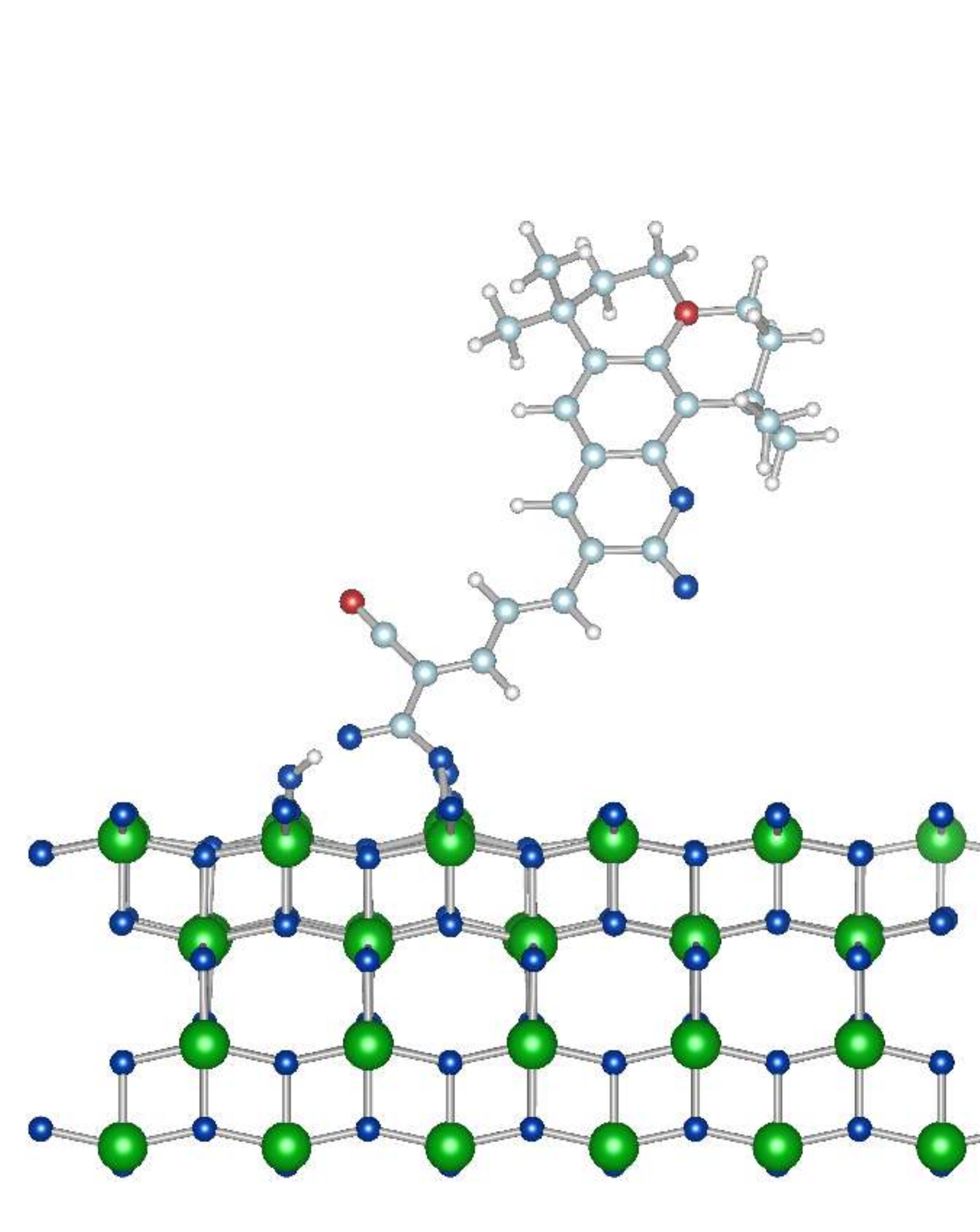}
    \end{minipage}
  &
    \begin{minipage}[h]{0.33\textwidth}
\includegraphics[trim = 180mm 0mm 200mm 0mm, clip, width=1.\textwidth]{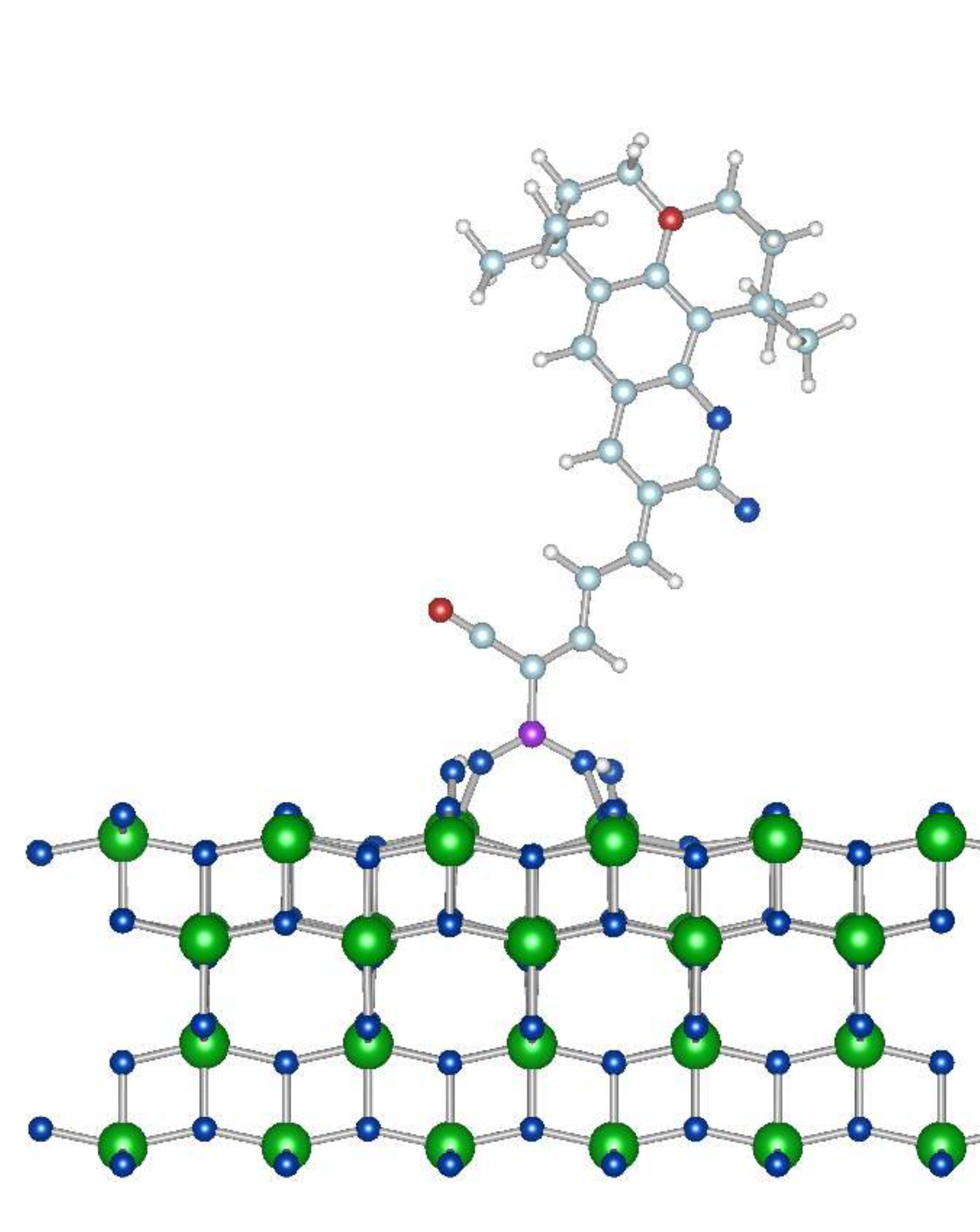}
    \end{minipage}
  & 
  \begin{minipage}[h]{0.33\textwidth}
\includegraphics[trim = 180mm 0mm 200mm 0mm, clip, width=1.\textwidth]{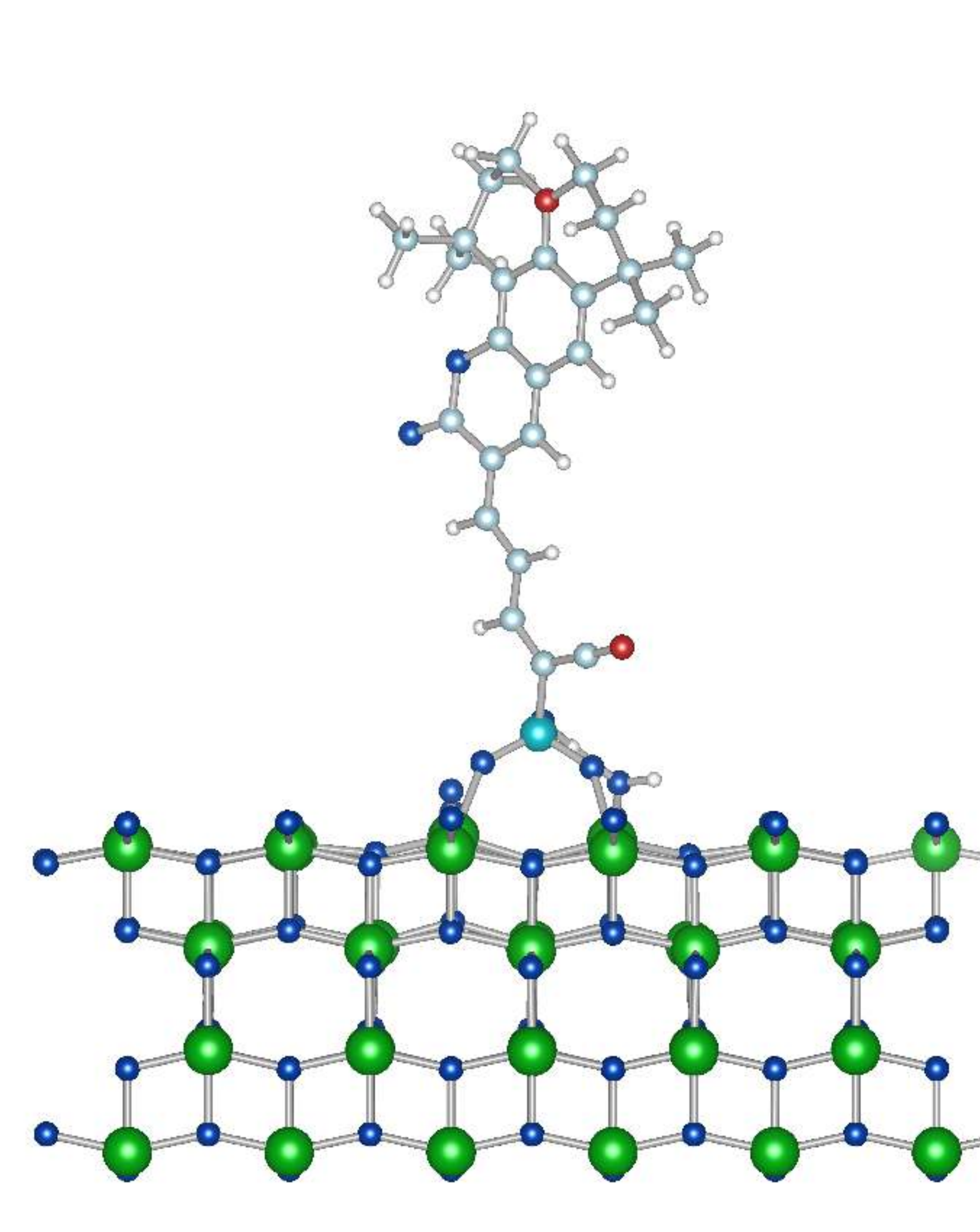}
    \end{minipage}
  \\

 $E_{Ads} = -2.80$ eV& $E_{Ads} = -4.35$ eV  & $E_{Ads}= -4.26$ eV \\

    \begin{minipage}[h]{0.33\textwidth}
\includegraphics[trim = 180mm 0mm 200mm 0mm, clip, width=1.\textwidth]{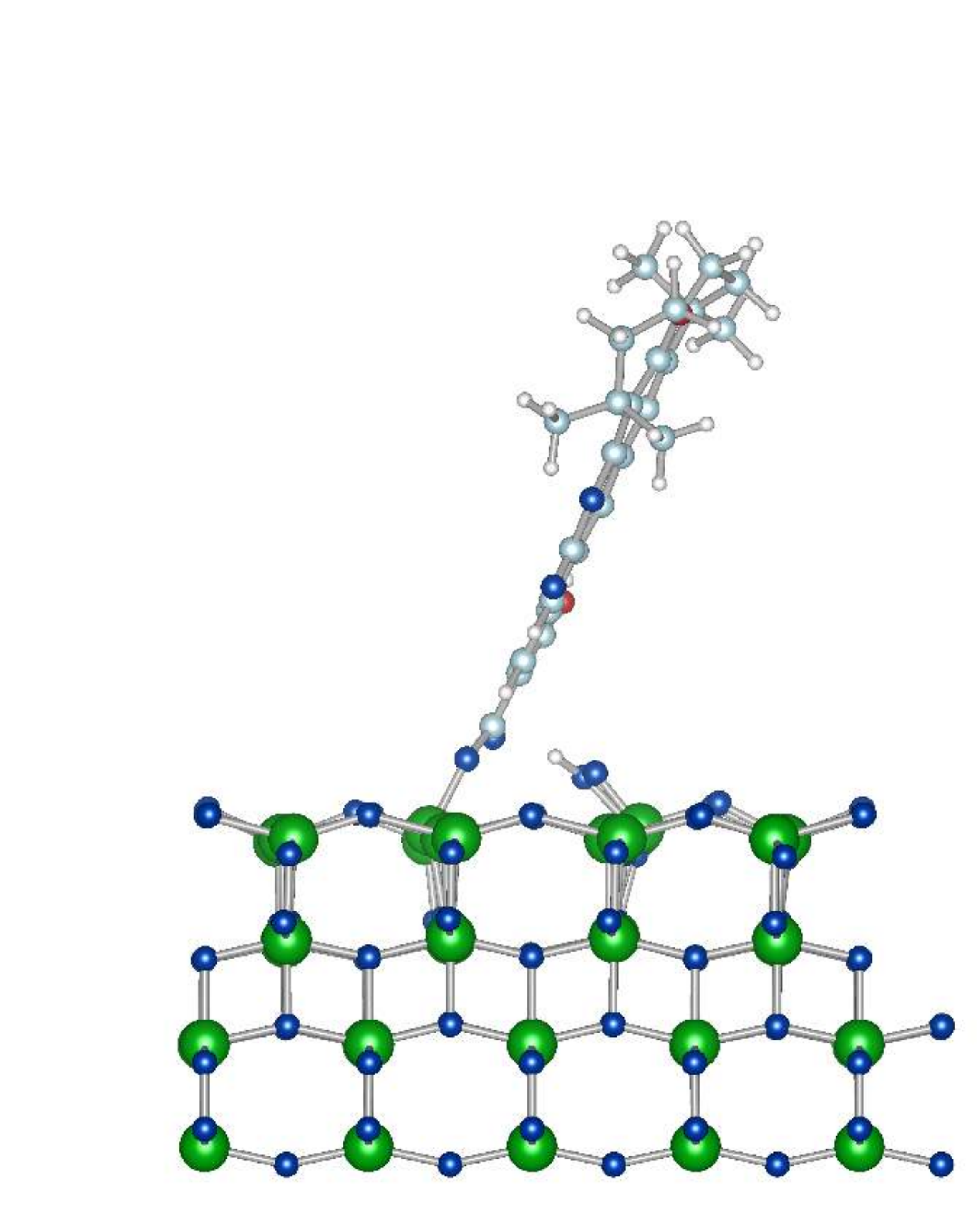}
    \end{minipage}
   &
    \begin{minipage}[h]{0.33\textwidth}
\includegraphics[trim = 180mm 0mm 200mm 0mm, clip, width=1.\textwidth]{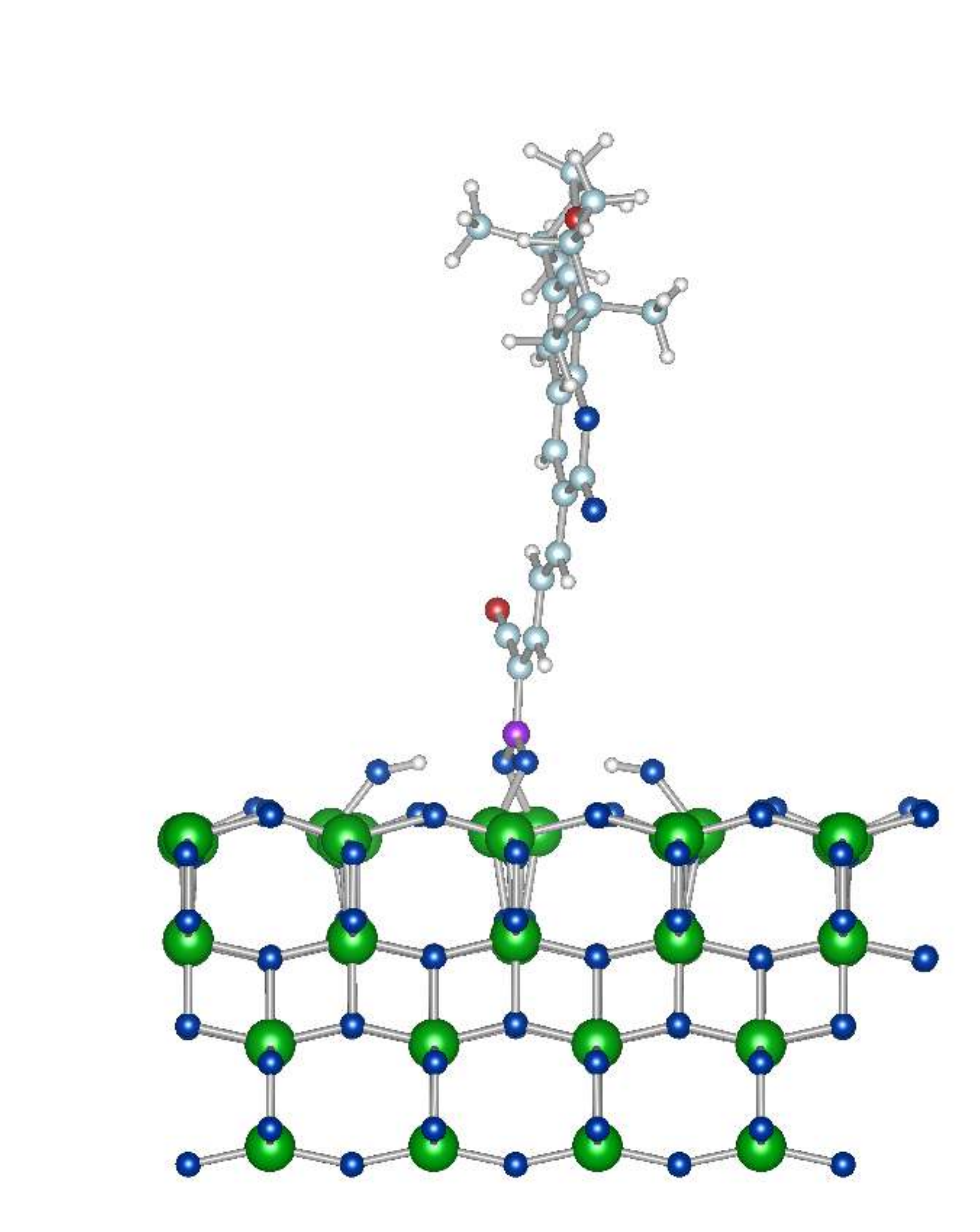}
    \end{minipage}
   &
    \begin{minipage}[h]{0.33\textwidth}
\includegraphics[trim = 180mm 0mm 200mm 0mm, clip, width=1.\textwidth]{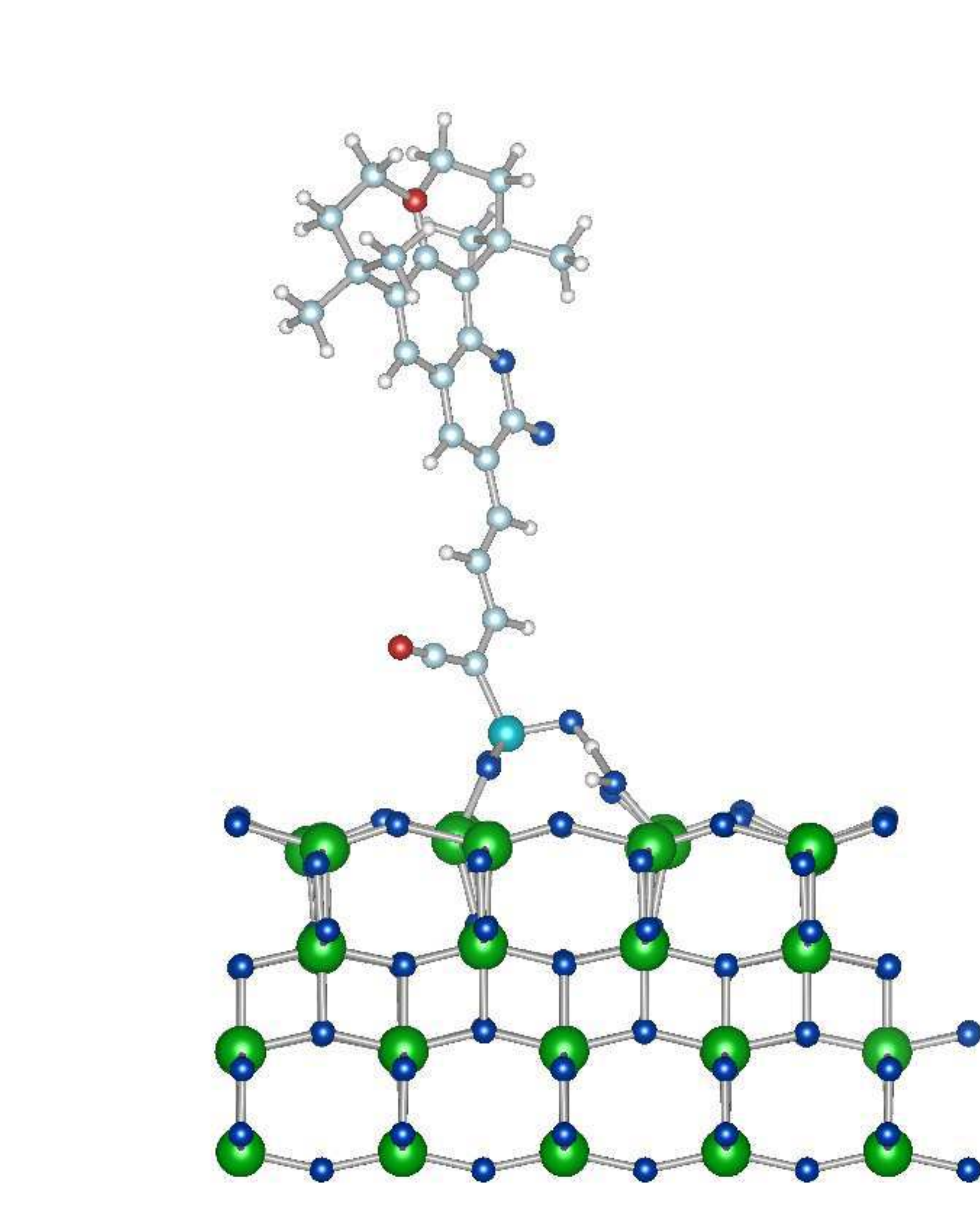} 
    \end{minipage}
 \\ 
\textbf{(i)}&\textbf{(ii)}&\textbf{(iii)}\\
 \end{tabular}
 \caption{Relaxed dye adsorption geometries on Anatase (001): \textbf{(i)} NKX-2311, \textbf{(ii)} NKX-2311B and \textbf{(iii)} NKX-2311P. Front (top) and side (bottom) perspectives are shown. }
 \label{DYE_ADS_001}
 \end{center}
\end{figure}

Relaxed structures for the NKX-2311 dyes on the anatase (001) surface, 
along with the calculated adsorption energies, may be seen in figure
\ref{DYE_ADS_001}. Very similar structures to those of the 
isolated anchors are again found for the formic and boronic bound dyes,
with relevant bond lengths for the anchors to the surface 
changing by at most a few hundredths of 
an Angstrom. However the binding structure of the NKX-2311P dye
has a couple of considerable differences with respect to that of the
isolated anchor. While for the isolated anchor the carbonyl group
formed a hydrogen bond to one of the dissociated hydrogen atoms, 
in this case this hydrogen atom has reassociated 
to the carbonyl group. Similarly this reassociation of surface hydrogen has
been shown to provide a stabilising effect for dyes binding to the
anatase (101) surface\cite{prot_stab}. The second dissociated hydrogen atom
has been transferred on the surface from a O(3) atom to an O(2) atom, 
breaking a Ti-O bond, with this hydroxyl group in turn forming a
hydrogen bond with the OH group of the anchor.

Examining the adsorption energies we see that there is a considerable
stabilisation of both the NKX-2311 and NKX-2311P dyes over the
isolated anchor groups. It is likely that this stabilisation for 
the NKX-2311P dye may be attributed in part to this reorganisation and 
reassociation of the hydrogen groups on the surface. 
On closer examination of the geometry for the NKX-2311 dye we find that,
although the bonds between the anchor and the surface change only very 
slightly, the Ti atom bonded to the surface OH group has shifted
causing the Ti-OH bond length to decrease from 1.94 \AA\ for the isolated
dye to 1.74 \AA. Subsequent changes in bond lengths occur 
between this Ti atom and the oxygen atoms of the lattice, which may 
be responsible for this stabilisation.

While the NKX-2311 and NKX-2311P dyes are stabilised so too is the 
NKX-2311B dye, albeit much less so, and it remains the most
stable anchoring moeity. Again we have the reorganisation of the
relative binding strengths with respect to the rutile (110) and anatase (101) surfaces, with NKX-2311B $>$ NKX-2311P $>>$ NKX-2311, highlighting
the fact that in terms of binding to the surface the carboxylic acid
group is at a considerable disadvantage compared to the boronic and 
phosphonic anchors.

\begin{figure}
\begin{center}
 \begin{minipage}[c]{0.7\textwidth}
  \includegraphics[angle=-90,origin=c,trim = 30mm 0mm 30mm 0mm, clip,width=1.0\textwidth]{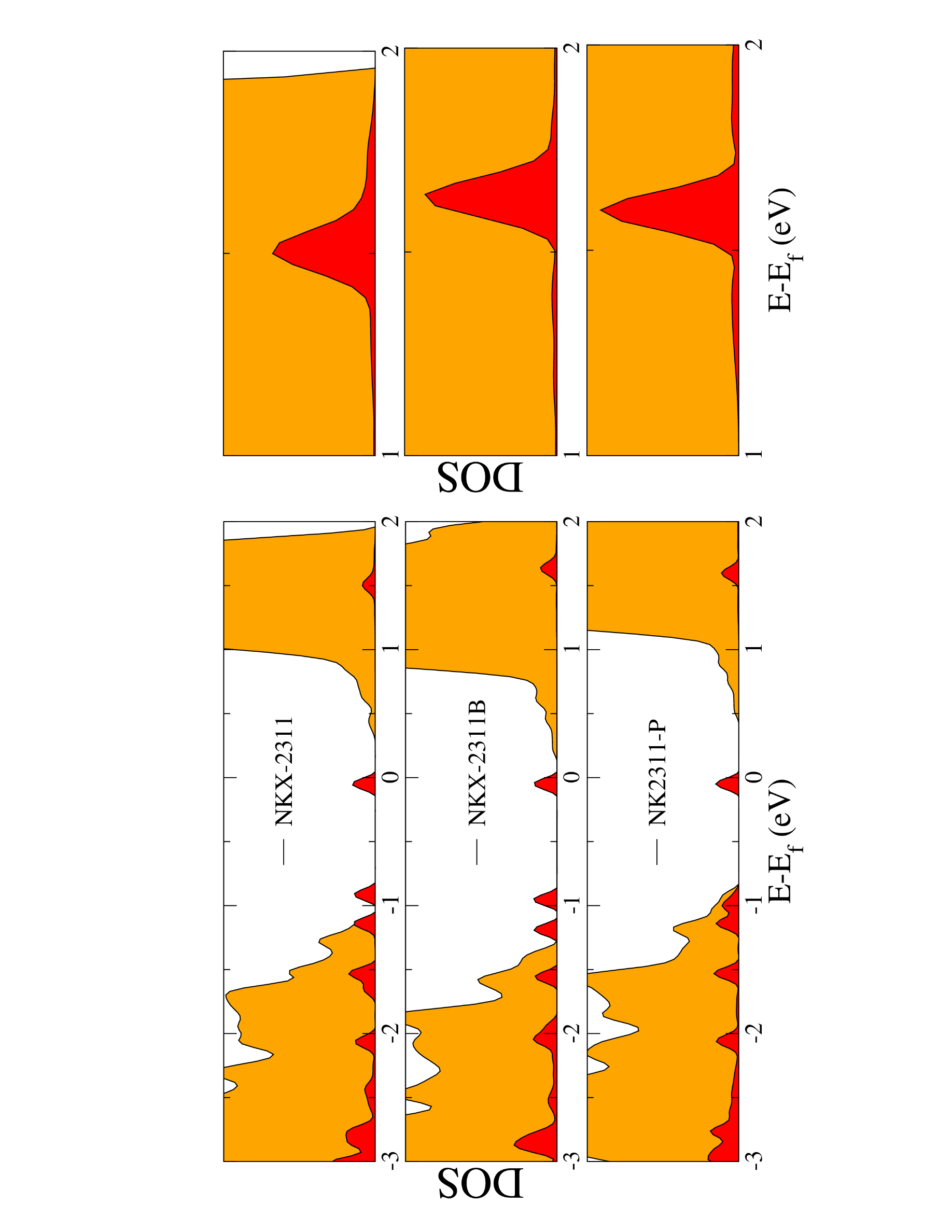}
  \end{minipage}

\caption{Projected density of states for dyes adsorbed on the anatase (001) surface. Top: NKX-2311,  Middle: NKX-2311B, Bottom: NKX-2311P. Total DOS in orange, with the dye localised PDOS in red. On the right hand side the
LUMO\superscript{*} is enlarged for clarity.}
 \label{001_PDOS}
\end{center}
\end{figure}

Partial density of states, along with 
the projection on the adsorbed dyes, can be seen in 
figure \ref{001_PDOS}. Once again dye localised states are 
introduced into the band-gap for all three systems. However
these dye localised states are shifted relative to the 
TiO\subscript{2} bands depending on the anchoring group used.
For the NKX-2311P dye only one state is located within the gap, 
with two further dye levels
located at the very top of the valence band. On changing the anchor
to the carboxylic group for NKX-2311 we see that a shift forces a 
further state to appear in the gap. Similarly changing to the boronic
anchor we see a similar shift introducing a third dye localised state
in the gap. Introduction of the
 extra gap states for the NKX-2311B dye could potentially increase 
the number permitted photo-excitation pathways in comparison to
NKX-2311 and NKX-2311P.

Examining the LUMO\superscript{*} states from our PDOS plots
we can see that for the (001) surface the carboxylic acid
bound dye now has a very similar overlap with the TiO\subscript{2}
states as the NKX-2311B and NKX-2311P dyes. This would suggest that
the electron injection may be similar for all three on 
the (001) surface. This is an interesting result, which when coupled
with the extra states in the gap for the NKX-2311B dye, suggests
that NKX-2311B would have a higher J\subscript{SC} than
both NKX-2311 and NKX-2311P. Coupling this point with the 
excellent binding of the NKX-2311B dye 
suggests that the boronic anchor could be an impressive 
alternative to the carboxylic anchor for the (001) surface.

\section{Conclusions}

We have introduced the two main TiO\subscript{2}
polymorphs of interest for use as electrodes in dye 
sensitised solar cells and characterised three of the most 
important TiO\subscript{2} surfaces for use in DSSCs.

Examining the interaction of three binding anchors 
with each of these surfaces allowed us to compare and contrast the 
benefits of the most used carboxylic acid group with under-used
phosphonic and boronic anchors. 
Our results show that significant differences exist for the
anchors when binding to different surfaces, 
with considerable stabilisation and reorganisation of the 
relative binding strengths depending on the surface the anchors
adsorb to. 

Extending this investigation to examine the effect anchors have
when binding dyes to these surfaces, we found similar results. 
Variation of the anchor group was found to produce considerable
differences on the binding strengths and electronic structure
of the composite system depending on the surface chosen.
  
This highlights the important, but often neglected, 
point; that anchoring groups perform \emph{differently} on different 
surfaces and that the anchor choice should necessarily
depend on the majority surface exposed in the elctrode.
 Although somewhat obvious, this conclusion
is a useful pointer to experimentalists working in the field and 
is something that should be taken into consideration when designing 
sensitising dyes. This is a point that is particularly important to 
consider at present, given the current interest in 
other TiO\subscript{2} electrode morphologies for 
which the anatase (101) surface is no longer the dominant exposed surface.

Finally the reactivity, and impressive binding 
strength of the anchors when attaching to unreconstructed anatase (001) 
surface is found to make a compelling case for experimentalists to 
exploit this reactivity by functionalising before surface reconstruction, 
with the boronic and phosphonic anchoring groups in particular showing
great promise as anchors when adhereing dyes to this surface.

\section*{Acknowledgements}
C.O'R. is supported by the MANA-WPI project and D.R.B. was 
funded by the Royal Society.
We thank Umberto Terranova for useful
discussions. This work made use of the facilities of HECToR, the UK's national high-performance computing service, which is provided by UoE HPCx Ltd at the University of Edinburgh, Cray Inc and NAG Ltd, and funded by the Office of Science and Technology through EPSRC's High End Computing Programme. Calculations were performed at HECToR through the UKCP Consortium. The authors acknowledge the use of the UCL Legion High Performance Computing Facility, and associated support services, in the completion of this work.
\section{Supporting Information}
Most stable adsorption structures are available online \cite{figshare}.

\bibliography{Anchors}

\ifx\mcitethebibliography\mciteundefinedmacro
\PackageError{achemso.bst}{mciteplus.sty has not been loaded}
{This bibstyle requires the use of the mciteplus package.}\fi
\begin{mcitethebibliography}{68}
\providecommand*{\natexlab}[1]{#1}
\mciteSetBstSublistMode{f}
\mciteSetBstMaxWidthForm{subitem}{(\alph{mcitesubitemcount})}
\mciteSetBstSublistLabelBeginEnd{\mcitemaxwidthsubitemform\space}
{\relax}{\relax}

\bibitem[Linsebigler et~al.(1995)Linsebigler, Lu, and Yates]{chptr3_photoc}
Linsebigler,~A.~L.; Lu,~G.; Yates,~J.~T. \emph{Chemical Reviews} \textbf{1995},
  \emph{95}, 735--758\relax
\mciteBstWouldAddEndPuncttrue
\mciteSetBstMidEndSepPunct{\mcitedefaultmidpunct}
{\mcitedefaultendpunct}{\mcitedefaultseppunct}\relax
\EndOfBibitem
\bibitem[Nakata and Fujishima(2012)]{chptr3_photoc2}
Nakata,~K.; Fujishima,~A. \emph{Journal of Photochemistry and Photobiology C:
  Photochemistry Reviews} \textbf{2012}, \emph{13}, 169 -- 189\relax
\mciteBstWouldAddEndPuncttrue
\mciteSetBstMidEndSepPunct{\mcitedefaultmidpunct}
{\mcitedefaultendpunct}{\mcitedefaultseppunct}\relax
\EndOfBibitem
\bibitem[Wang et~al.(2011)Wang, Watson, Sung, Tseng, Bouis, and
  Fernando]{chptr3_whitepig}
Wang,~D.~L.; Watson,~S.~S.; Sung,~L.-P.; Tseng,~I.-H.; Bouis,~C.~J.;
  Fernando,~R. \emph{Journal of Coatings Technology and Research}
  \textbf{2011}, \emph{8}, 19--33\relax
\mciteBstWouldAddEndPuncttrue
\mciteSetBstMidEndSepPunct{\mcitedefaultmidpunct}
{\mcitedefaultendpunct}{\mcitedefaultseppunct}\relax
\EndOfBibitem
\bibitem[Maness et~al.(1999)Maness, Smolinski, Blake, Huang, Wolfrum, and
  Jacoby]{chptr3_bacter}
Maness,~P.-C.; Smolinski,~S.; Blake,~D.~M.; Huang,~Z.; Wolfrum,~E.~J.;
  Jacoby,~W.~A. \emph{Applied and environmental microbiology} \textbf{1999},
  \emph{65}, 4094--4098\relax
\mciteBstWouldAddEndPuncttrue
\mciteSetBstMidEndSepPunct{\mcitedefaultmidpunct}
{\mcitedefaultendpunct}{\mcitedefaultseppunct}\relax
\EndOfBibitem
\bibitem[Mills et~al.(1993)Mills, Davies, and Worsley]{chptr3_waterp}
Mills,~A.; Davies,~R.~H.; Worsley,~D. \emph{Chem. Soc. Rev.} \textbf{1993},
  \emph{22}, 417--425\relax
\mciteBstWouldAddEndPuncttrue
\mciteSetBstMidEndSepPunct{\mcitedefaultmidpunct}
{\mcitedefaultendpunct}{\mcitedefaultseppunct}\relax
\EndOfBibitem
\bibitem[amd Michael~Gr\"{a}tzel(1991)]{OReagan_Grat}
amd Michael~Gr\"{a}tzel,~B.~O. \emph{Nature} \textbf{1991}, \emph{353}, 737 --
  740\relax
\mciteBstWouldAddEndPuncttrue
\mciteSetBstMidEndSepPunct{\mcitedefaultmidpunct}
{\mcitedefaultendpunct}{\mcitedefaultseppunct}\relax
\EndOfBibitem
\bibitem[Diebold(2003)]{chptr3_diebold}
Diebold,~U. \emph{Surface Science Reports} \textbf{2003}, \emph{48},
  53--229\relax
\mciteBstWouldAddEndPuncttrue
\mciteSetBstMidEndSepPunct{\mcitedefaultmidpunct}
{\mcitedefaultendpunct}{\mcitedefaultseppunct}\relax
\EndOfBibitem
\bibitem[Listorti et~al.(2011)Listorti, O'Regan, and
  Durrant]{chpt3_durrant_charge}
Listorti,~A.; O'Regan,~B.; Durrant,~J.~R. \emph{Chemistry of Materials}
  \textbf{2011}, \emph{23}, 3381--3399\relax
\mciteBstWouldAddEndPuncttrue
\mciteSetBstMidEndSepPunct{\mcitedefaultmidpunct}
{\mcitedefaultendpunct}{\mcitedefaultseppunct}\relax
\EndOfBibitem
\bibitem[K.~Nazeeruddin et~al.(1997)K.~Nazeeruddin, Pechy, and
  Gr\"{a}tzel]{chpt3_black_dye}
K.~Nazeeruddin,~M.; Pechy,~P.; Gr\"{a}tzel,~M. \emph{Chem. Commun.}
  \textbf{1997},  1705--1706\relax
\mciteBstWouldAddEndPuncttrue
\mciteSetBstMidEndSepPunct{\mcitedefaultmidpunct}
{\mcitedefaultendpunct}{\mcitedefaultseppunct}\relax
\EndOfBibitem
\bibitem[Nazeeruddin et~al.(2001)Nazeeruddin, P\'{e}chy, Renouard, Zakeeruddin,
  Humphry-Baker, Comte, Liska, Cevey, Costa, Shklover, Spiccia, Deacon,
  Bignozzi, and Gr\"{a}tzel]{chpt3_N719}
Nazeeruddin,~M.~K. et~al. \emph{Journal of the American Chemical Society}
  \textbf{2001}, \emph{123}, 1613--1624, PMID: 11456760\relax
\mciteBstWouldAddEndPuncttrue
\mciteSetBstMidEndSepPunct{\mcitedefaultmidpunct}
{\mcitedefaultendpunct}{\mcitedefaultseppunct}\relax
\EndOfBibitem
\bibitem[Nazeeruddin et~al.(1993)Nazeeruddin, Kay, Rodicio, Humphry-Baker,
  Mueller, Liska, Vlachopoulos, and Gr\"{a}tzel]{chpt3_N3}
Nazeeruddin,~M.~K.; Kay,~A.; Rodicio,~I.; Humphry-Baker,~R.; Mueller,~E.;
  Liska,~P.; Vlachopoulos,~N.; Gr\"{a}tzel,~M. \emph{Journal of the American
  Chemical Society} \textbf{1993}, \emph{115}, 6382--6390\relax
\mciteBstWouldAddEndPuncttrue
\mciteSetBstMidEndSepPunct{\mcitedefaultmidpunct}
{\mcitedefaultendpunct}{\mcitedefaultseppunct}\relax
\EndOfBibitem
\bibitem[Zhang and F.~Banfield(1998)]{chpt3_ana}
Zhang,~H.; F.~Banfield,~J. \emph{J. Mater. Chem.} \textbf{1998}, \emph{8},
  2073--2076\relax
\mciteBstWouldAddEndPuncttrue
\mciteSetBstMidEndSepPunct{\mcitedefaultmidpunct}
{\mcitedefaultendpunct}{\mcitedefaultseppunct}\relax
\EndOfBibitem
\bibitem[Lazzeri et~al.(2001)Lazzeri, Vittadini, and Selloni]{chptr3_selloni}
Lazzeri,~M.; Vittadini,~A.; Selloni,~A. \emph{Phys. Rev. B} \textbf{2001},
  \emph{63}, 155409\relax
\mciteBstWouldAddEndPuncttrue
\mciteSetBstMidEndSepPunct{\mcitedefaultmidpunct}
{\mcitedefaultendpunct}{\mcitedefaultseppunct}\relax
\EndOfBibitem
\bibitem[Vittadini et~al.(2000)Vittadini, Selloni, Rotzinger, and
  Gr\"{a}tzel]{chptr3_formic_101}
Vittadini,~A.; Selloni,~A.; Rotzinger,~F.~P.; Gr\"{a}tzel,~M. \emph{The Journal
  of Physical Chemistry B} \textbf{2000}, \emph{104}, 1300--1306\relax
\mciteBstWouldAddEndPuncttrue
\mciteSetBstMidEndSepPunct{\mcitedefaultmidpunct}
{\mcitedefaultendpunct}{\mcitedefaultseppunct}\relax
\EndOfBibitem
\bibitem[Xu et~al.(2012)Xu, Noei, Buchholz, Muhler, Wöll, and
  Wang]{chptr3_formic_101_exp}
Xu,~M.; Noei,~H.; Buchholz,~M.; Muhler,~M.; Wöll,~C.; Wang,~Y. \emph{Catalysis
  Today} \textbf{2012}, \emph{182}, 12 -- 15\relax
\mciteBstWouldAddEndPuncttrue
\mciteSetBstMidEndSepPunct{\mcitedefaultmidpunct}
{\mcitedefaultendpunct}{\mcitedefaultseppunct}\relax
\EndOfBibitem
\bibitem[Zakeeruddin et~al.(1997)Zakeeruddin, Nazeeruddin, Pechy, Rotzinger,
  Humphry-Baker, Kalyanasundaram, Gr\"{a}tzel, Shklover, and
  Haibach]{chptr3_phos_dye}
Zakeeruddin,~S.~M.; Nazeeruddin,~M.~K.; Pechy,~P.; Rotzinger,~F.~P.;
  Humphry-Baker,~R.; Kalyanasundaram,~K.; Gr\"{a}tzel,~M.; Shklover,~V.;
  Haibach,~T. \emph{Inorganic Chemistry} \textbf{1997}, \emph{36},
  5937--5946\relax
\mciteBstWouldAddEndPuncttrue
\mciteSetBstMidEndSepPunct{\mcitedefaultmidpunct}
{\mcitedefaultendpunct}{\mcitedefaultseppunct}\relax
\EndOfBibitem
\bibitem[Pechy et~al.(1995)Pechy, Rotzinger, Nazeeruddin, Kohle, Zakeeruddin,
  Humphry-Baker, and Gratzel]{chptr3_phos_expt}
Pechy,~P.; Rotzinger,~F.~P.; Nazeeruddin,~M.~K.; Kohle,~O.; Zakeeruddin,~S.~M.;
  Humphry-Baker,~R.; Gratzel,~M. \emph{J. Chem. Soc.{,} Chem. Commun.}
  \textbf{1995}, \emph{0}, 65--66\relax
\mciteBstWouldAddEndPuncttrue
\mciteSetBstMidEndSepPunct{\mcitedefaultmidpunct}
{\mcitedefaultendpunct}{\mcitedefaultseppunct}\relax
\EndOfBibitem
\bibitem[Altobello et~al.(2004)Altobello, Bignozzi, Caramori, Larramona, Quici,
  Marzanni, and Lakhmiri]{chptr3_bor_dye}
Altobello,~S.; Bignozzi,~C.; Caramori,~S.; Larramona,~G.; Quici,~S.;
  Marzanni,~G.; Lakhmiri,~R. \emph{Journal of Photochemistry and Photobiology
  A: Chemistry} \textbf{2004}, \emph{166}, 91 -- 98\relax
\mciteBstWouldAddEndPuncttrue
\mciteSetBstMidEndSepPunct{\mcitedefaultmidpunct}
{\mcitedefaultendpunct}{\mcitedefaultseppunct}\relax
\EndOfBibitem
\bibitem[Katono et~al.(2011)Katono, Bessho, Meng, Humphry-Baker, Rothenberger,
  Zakeeruddin, Kaxiras, and Gr\"{a}tzel]{chptr3_cyano_dye}
Katono,~M.; Bessho,~T.; Meng,~S.; Humphry-Baker,~R.; Rothenberger,~G.;
  Zakeeruddin,~S.~M.; Kaxiras,~E.; Gr\"{a}tzel,~M. \emph{Langmuir}
  \textbf{2011}, \emph{27}, 14248--14252\relax
\mciteBstWouldAddEndPuncttrue
\mciteSetBstMidEndSepPunct{\mcitedefaultmidpunct}
{\mcitedefaultendpunct}{\mcitedefaultseppunct}\relax
\EndOfBibitem
\bibitem[Nilsing et~al.(2007)Nilsing, Persson, Lunell, and
  Ojam\"{a}e]{chptr3_phos_110}
Nilsing,~M.; Persson,~P.; Lunell,~S.; Ojam\"{a}e,~L. \emph{The Journal of
  Physical Chemistry C} \textbf{2007}, \emph{111}, 12116--12123\relax
\mciteBstWouldAddEndPuncttrue
\mciteSetBstMidEndSepPunct{\mcitedefaultmidpunct}
{\mcitedefaultendpunct}{\mcitedefaultseppunct}\relax
\EndOfBibitem
\bibitem[Thavasi et~al.(2009)Thavasi, Renugopalakrishnan, Jose, and
  Ramakrishna]{chptr3_NR_trans}
Thavasi,~V.; Renugopalakrishnan,~V.; Jose,~R.; Ramakrishna,~S. \emph{Materials
  Science and Engineering: R: Reports} \textbf{2009}, \emph{63}, 81 -- 99\relax
\mciteBstWouldAddEndPuncttrue
\mciteSetBstMidEndSepPunct{\mcitedefaultmidpunct}
{\mcitedefaultendpunct}{\mcitedefaultseppunct}\relax
\EndOfBibitem
\bibitem[Lv et~al.(2012)Lv, Zheng, Ye, Sun, Xiao, Guo, and
  Lin]{chptr3_NR_110maj1}
Lv,~M.; Zheng,~D.; Ye,~M.; Sun,~L.; Xiao,~J.; Guo,~W.; Lin,~C. \emph{Nanoscale}
  \textbf{2012}, \emph{4}, 5872--5879\relax
\mciteBstWouldAddEndPuncttrue
\mciteSetBstMidEndSepPunct{\mcitedefaultmidpunct}
{\mcitedefaultendpunct}{\mcitedefaultseppunct}\relax
\EndOfBibitem
\bibitem[Lv et~al.(2013)Lv, Zheng, Ye, Xiao, Guo, Lai, Sun, Lin, and
  Zuo]{chptr3_NR_1}
Lv,~M.; Zheng,~D.; Ye,~M.; Xiao,~J.; Guo,~W.; Lai,~Y.; Sun,~L.; Lin,~C.;
  Zuo,~J. \emph{Energy Environ. Sci.} \textbf{2013}, \emph{6}, 1615--1622\relax
\mciteBstWouldAddEndPuncttrue
\mciteSetBstMidEndSepPunct{\mcitedefaultmidpunct}
{\mcitedefaultendpunct}{\mcitedefaultseppunct}\relax
\EndOfBibitem
\bibitem[Laskova et~al.(2012)Laskova, Zukalova, Kavan, Chou, Liska, Wei, Bin,
  Kubat, Ghadiri, Moser, and Gr\"{a}tzel]{chptr3_001_DSSC1}
Laskova,~B. et~al. \emph{Journal of Solid State Electrochemistry}
  \textbf{2012}, \emph{16}, 2993--3001\relax
\mciteBstWouldAddEndPuncttrue
\mciteSetBstMidEndSepPunct{\mcitedefaultmidpunct}
{\mcitedefaultendpunct}{\mcitedefaultseppunct}\relax
\EndOfBibitem
\bibitem[Zhang et~al.(2010)Zhang, Han, Liu, Liu, Yu, Zhang, Yao, and
  Zhao]{chptr3_001_DSSC2}
Zhang,~H.; Han,~Y.; Liu,~X.; Liu,~P.; Yu,~H.; Zhang,~S.; Yao,~X.; Zhao,~H.
  \emph{Chem. Commun.} \textbf{2010}, \emph{46}, 8395--8397\relax
\mciteBstWouldAddEndPuncttrue
\mciteSetBstMidEndSepPunct{\mcitedefaultmidpunct}
{\mcitedefaultendpunct}{\mcitedefaultseppunct}\relax
\EndOfBibitem
\bibitem[Yu et~al.(2010)Yu, Fan, and Lv]{CHPTR3_001_DSSC3}
Yu,~J.; Fan,~J.; Lv,~K. \emph{Nanoscale} \textbf{2010}, \emph{2},
  2144--2149\relax
\mciteBstWouldAddEndPuncttrue
\mciteSetBstMidEndSepPunct{\mcitedefaultmidpunct}
{\mcitedefaultendpunct}{\mcitedefaultseppunct}\relax
\EndOfBibitem
\bibitem[Kresse and Furthm\"{u}ller(1996)]{chpt3_VASP}
Kresse,~G.; Furthm\"{u}ller,~J. \emph{Computational Materials Science}
  \textbf{1996}, \emph{6}, 15 -- 50\relax
\mciteBstWouldAddEndPuncttrue
\mciteSetBstMidEndSepPunct{\mcitedefaultmidpunct}
{\mcitedefaultendpunct}{\mcitedefaultseppunct}\relax
\EndOfBibitem
\bibitem[Perdew et~al.(1992)Perdew, Chevary, Vosko, Jackson, Pederson, Singh,
  and Fiolhais]{chpt3_PW91-2}
Perdew,~J.~P.; Chevary,~J.~A.; Vosko,~S.~H.; Jackson,~K.~A.; Pederson,~M.~R.;
  Singh,~D.~J.; Fiolhais,~C. \emph{Phys. Rev. B} \textbf{1992}, \emph{46},
  6671--6687\relax
\mciteBstWouldAddEndPuncttrue
\mciteSetBstMidEndSepPunct{\mcitedefaultmidpunct}
{\mcitedefaultendpunct}{\mcitedefaultseppunct}\relax
\EndOfBibitem
\bibitem[Vanderbilt(1990)]{chpt3_vanderbilt}
Vanderbilt,~D. \emph{Phys. Rev. B} \textbf{1990}, \emph{41}, 7892--7895\relax
\mciteBstWouldAddEndPuncttrue
\mciteSetBstMidEndSepPunct{\mcitedefaultmidpunct}
{\mcitedefaultendpunct}{\mcitedefaultseppunct}\relax
\EndOfBibitem
\bibitem[Pulay(1980)]{chpt3_RMM}
Pulay,~P. \emph{Chemical Physics Letters} \textbf{1980}, \emph{73}, 393 --
  398\relax
\mciteBstWouldAddEndPuncttrue
\mciteSetBstMidEndSepPunct{\mcitedefaultmidpunct}
{\mcitedefaultendpunct}{\mcitedefaultseppunct}\relax
\EndOfBibitem
\bibitem[Burdett et~al.(1987)Burdett, Hughbanks, Miller, Richardson, and
  Smith]{chpt3_anatase_exp}
Burdett,~J.~K.; Hughbanks,~T.; Miller,~G.~J.; Richardson,~J.~W.; Smith,~J.~V.
  \emph{Journal of the American Chemical Society} \textbf{1987}, \emph{109},
  3639--3646\relax
\mciteBstWouldAddEndPuncttrue
\mciteSetBstMidEndSepPunct{\mcitedefaultmidpunct}
{\mcitedefaultendpunct}{\mcitedefaultseppunct}\relax
\EndOfBibitem
\bibitem[Gong et~al.(2006)Gong, Selloni, and
  Vittadini]{chptr3_vitadini_form_001}
Gong,~X.-Q.; Selloni,~A.; Vittadini,~A. \emph{The Journal of Physical Chemistry
  B} \textbf{2006}, \emph{110}, 2804--2811, PMID: 16471889\relax
\mciteBstWouldAddEndPuncttrue
\mciteSetBstMidEndSepPunct{\mcitedefaultmidpunct}
{\mcitedefaultendpunct}{\mcitedefaultseppunct}\relax
\EndOfBibitem
\bibitem[Gong and Selloni(2005)]{chptr3_001_reactive}
Gong,~X.-Q.; Selloni,~A. \emph{The Journal of Physical Chemistry B}
  \textbf{2005}, \emph{109}, 19560--19562, PMID: 16853530\relax
\mciteBstWouldAddEndPuncttrue
\mciteSetBstMidEndSepPunct{\mcitedefaultmidpunct}
{\mcitedefaultendpunct}{\mcitedefaultseppunct}\relax
\EndOfBibitem
\bibitem[Ohno et~al.(2002)Ohno, Sarukawa, and Matsumura]{chptr3_001_reactive2}
Ohno,~T.; Sarukawa,~K.; Matsumura,~M. \emph{New J. Chem.} \textbf{2002},
  \emph{26}, 1167--1170\relax
\mciteBstWouldAddEndPuncttrue
\mciteSetBstMidEndSepPunct{\mcitedefaultmidpunct}
{\mcitedefaultendpunct}{\mcitedefaultseppunct}\relax
\EndOfBibitem
\bibitem[Yu et~al.(2011)Yu, Tian, and Zhang]{chptr3_001_photoc}
Yu,~H.; Tian,~B.; Zhang,~J. \emph{Chemistry - A European Journal}
  \textbf{2011}, \emph{17}, 5499--5502\relax
\mciteBstWouldAddEndPuncttrue
\mciteSetBstMidEndSepPunct{\mcitedefaultmidpunct}
{\mcitedefaultendpunct}{\mcitedefaultseppunct}\relax
\EndOfBibitem
\bibitem[Sun et~al.(2012)Sun, Zhao, Zhou, and Liu]{chptr3_001_photodegrad}
Sun,~L.; Zhao,~Z.; Zhou,~Y.; Liu,~L. \emph{Nanoscale} \textbf{2012}, \emph{4},
  613--620\relax
\mciteBstWouldAddEndPuncttrue
\mciteSetBstMidEndSepPunct{\mcitedefaultmidpunct}
{\mcitedefaultendpunct}{\mcitedefaultseppunct}\relax
\EndOfBibitem
\bibitem[Yu et~al.(2010)Yu, Fan, and Lv]{chptr3_Anatase001_3}
Yu,~J.; Fan,~J.; Lv,~K. \emph{Nanoscale} \textbf{2010}, \emph{2},
  2144--2149\relax
\mciteBstWouldAddEndPuncttrue
\mciteSetBstMidEndSepPunct{\mcitedefaultmidpunct}
{\mcitedefaultendpunct}{\mcitedefaultseppunct}\relax
\EndOfBibitem
\bibitem[Wu et~al.(2011)Wu, Chen, Lu, and Wang]{chptr3_001_dsscrecomb}
Wu,~X.; Chen,~Z.; Lu,~G. Q.~M.; Wang,~L. \emph{Advanced Functional Materials}
  \textbf{2011}, \emph{21}, 4167--4172\relax
\mciteBstWouldAddEndPuncttrue
\mciteSetBstMidEndSepPunct{\mcitedefaultmidpunct}
{\mcitedefaultendpunct}{\mcitedefaultseppunct}\relax
\EndOfBibitem
\bibitem[Zhang et~al.(2012)Zhang, Wang, Zhao, Yu, Feng, Yuan, Tang, Liu, Li,
  and Zou]{chptr3_dssc_001_ads}
Zhang,~J.; Wang,~J.; Zhao,~Z.; Yu,~T.; Feng,~J.; Yuan,~Y.; Tang,~Z.; Liu,~Y.;
  Li,~Z.; Zou,~Z. \emph{Phys. Chem. Chem. Phys.} \textbf{2012}, \emph{14},
  4763--4769\relax
\mciteBstWouldAddEndPuncttrue
\mciteSetBstMidEndSepPunct{\mcitedefaultmidpunct}
{\mcitedefaultendpunct}{\mcitedefaultseppunct}\relax
\EndOfBibitem
\bibitem[Kang et~al.(2008)Kang, Choi, Kang, Kim, Kim, Hyeon, and
  Sung]{chptr3_NR_trans2}
Kang,~S.; Choi,~S.-H.; Kang,~M.-S.; Kim,~J.-Y.; Kim,~H.-S.; Hyeon,~T.;
  Sung,~Y.-E. \emph{Advanced Materials} \textbf{2008}, \emph{20}, 54--58\relax
\mciteBstWouldAddEndPuncttrue
\mciteSetBstMidEndSepPunct{\mcitedefaultmidpunct}
{\mcitedefaultendpunct}{\mcitedefaultseppunct}\relax
\EndOfBibitem
\bibitem[Jung et~al.(2013)Jung, Park, Oh, Kim, and Hong]{chptr3_NR_eff}
Jung,~Y.~H.; Park,~K.-H.; Oh,~J.~S.; Kim,~D.-H.; Hong,~C.~K. \emph{Nanoscale
  Research Letters} \textbf{2013}, \emph{8}, 37\relax
\mciteBstWouldAddEndPuncttrue
\mciteSetBstMidEndSepPunct{\mcitedefaultmidpunct}
{\mcitedefaultendpunct}{\mcitedefaultseppunct}\relax
\EndOfBibitem
\bibitem[Guo et~al.(2012)Guo, Xu, Wang, Wang, Pan, Lin, and
  Wang]{chptr3_NR_110maj}
Guo,~W.; Xu,~C.; Wang,~X.; Wang,~S.; Pan,~C.; Lin,~C.; Wang,~Z.~L.
  \emph{Journal of the American Chemical Society} \textbf{2012}, \emph{134},
  4437--4441\relax
\mciteBstWouldAddEndPuncttrue
\mciteSetBstMidEndSepPunct{\mcitedefaultmidpunct}
{\mcitedefaultendpunct}{\mcitedefaultseppunct}\relax
\EndOfBibitem
\bibitem[Kiejna et~al.(2006)Kiejna, Pabisiak, and Gao]{rut_110_SE_struct}
Kiejna,~A.; Pabisiak,~T.; Gao,~S.~W. \emph{Journal of Physics: Condensed
  Matter} \textbf{2006}, \emph{18}, 4207\relax
\mciteBstWouldAddEndPuncttrue
\mciteSetBstMidEndSepPunct{\mcitedefaultmidpunct}
{\mcitedefaultendpunct}{\mcitedefaultseppunct}\relax
\EndOfBibitem
\bibitem[Lindsay et~al.(2005)Lindsay, Wander, Ernst, Montanari, Thornton, and
  Harrison]{chpt3_110_pos}
Lindsay,~R.; Wander,~A.; Ernst,~A.; Montanari,~B.; Thornton,~G.;
  Harrison,~N.~M. \emph{Phys. Rev. Lett.} \textbf{2005}, \emph{94},
  246102\relax
\mciteBstWouldAddEndPuncttrue
\mciteSetBstMidEndSepPunct{\mcitedefaultmidpunct}
{\mcitedefaultendpunct}{\mcitedefaultseppunct}\relax
\EndOfBibitem
\bibitem[Onishi and Iwasawa(1994)]{chptr3_110_(1x2)}
Onishi,~H.; Iwasawa,~Y. \emph{Surface Science} \textbf{1994}, \emph{313}, L783
  -- L789\relax
\mciteBstWouldAddEndPuncttrue
\mciteSetBstMidEndSepPunct{\mcitedefaultmidpunct}
{\mcitedefaultendpunct}{\mcitedefaultseppunct}\relax
\EndOfBibitem
\bibitem[Nunzi and De~Angelis(2011)]{chptr3_form(101)_bb}
Nunzi,~F.; De~Angelis,~F. \emph{The Journal of Physical Chemistry C}
  \textbf{2011}, \emph{115}, 2179--2186\relax
\mciteBstWouldAddEndPuncttrue
\mciteSetBstMidEndSepPunct{\mcitedefaultmidpunct}
{\mcitedefaultendpunct}{\mcitedefaultseppunct}\relax
\EndOfBibitem
\bibitem[Gong et~al.(2006)Gong, Selloni, and Vittadini]{chptr3_form_001}
Gong,~X.-Q.; Selloni,~A.; Vittadini,~A. \emph{The Journal of Physical Chemistry
  B} \textbf{2006}, \emph{110}, 2804--2811, PMID: 16471889\relax
\mciteBstWouldAddEndPuncttrue
\mciteSetBstMidEndSepPunct{\mcitedefaultmidpunct}
{\mcitedefaultendpunct}{\mcitedefaultseppunct}\relax
\EndOfBibitem
\bibitem[Bates et~al.(1998)Bates, Kresse, and Gillan]{chptr3_form_rut_110}
Bates,~S.; Kresse,~G.; Gillan,~M. \emph{Surface Science} \textbf{1998},
  \emph{409}, 336 -- 349\relax
\mciteBstWouldAddEndPuncttrue
\mciteSetBstMidEndSepPunct{\mcitedefaultmidpunct}
{\mcitedefaultendpunct}{\mcitedefaultseppunct}\relax
\EndOfBibitem
\bibitem[Luschtinetz et~al.(2009)Luschtinetz, Frenzel, Milek, and
  Seifert]{chptr3_phos_101_110_bbh}
Luschtinetz,~R.; Frenzel,~J.; Milek,~T.; Seifert,~G. \emph{The Journal of
  Physical Chemistry C} \textbf{2009}, \emph{113}, 5730--5740\relax
\mciteBstWouldAddEndPuncttrue
\mciteSetBstMidEndSepPunct{\mcitedefaultmidpunct}
{\mcitedefaultendpunct}{\mcitedefaultseppunct}\relax
\EndOfBibitem
\bibitem[Nilsing et~al.(2005)Nilsing, Lunell, Persson, and
  Ojam\"{a}e]{chptr3_phos(101)}
Nilsing,~M.; Lunell,~S.; Persson,~P.; Ojam\"{a}e,~L. \emph{Surface Science}
  \textbf{2005}, \emph{582}, 49 -- 60\relax
\mciteBstWouldAddEndPuncttrue
\mciteSetBstMidEndSepPunct{\mcitedefaultmidpunct}
{\mcitedefaultendpunct}{\mcitedefaultseppunct}\relax
\EndOfBibitem
\bibitem[Popova et~al.(2000)Popova, Andrushkevich, Chesalov, and
  Stoyanov]{chptr3_form_FTIR}
Popova,~G.; Andrushkevich,~T.; Chesalov,~Y.; Stoyanov,~E. \emph{Kinetics and
  Catalysis} \textbf{2000}, \emph{41}, 805--811\relax
\mciteBstWouldAddEndPuncttrue
\mciteSetBstMidEndSepPunct{\mcitedefaultmidpunct}
{\mcitedefaultendpunct}{\mcitedefaultseppunct}\relax
\EndOfBibitem
\bibitem[K\"{a}ckell and Terakura(2000)]{chptr3_form2_rut_110}
K\"{a}ckell,~P.; Terakura,~K. \emph{Surface Science} \textbf{2000}, \emph{461},
  191 -- 198\relax
\mciteBstWouldAddEndPuncttrue
\mciteSetBstMidEndSepPunct{\mcitedefaultmidpunct}
{\mcitedefaultendpunct}{\mcitedefaultseppunct}\relax
\EndOfBibitem
\bibitem[Rotzinger et~al.(2004)Rotzinger, Kesselman-Truttmann, Hug, Shklover,
  and Gr\"{a}tzel]{chptr3_110_form_FTIR}
Rotzinger,~F.~P.; Kesselman-Truttmann,~J.~M.; Hug,~S.~J.; Shklover,~V.;
  Gr\"{a}tzel,~M. \emph{The Journal of Physical Chemistry B} \textbf{2004},
  \emph{108}, 5004--5017\relax
\mciteBstWouldAddEndPuncttrue
\mciteSetBstMidEndSepPunct{\mcitedefaultmidpunct}
{\mcitedefaultendpunct}{\mcitedefaultseppunct}\relax
\EndOfBibitem
\bibitem[Herman et~al.(2000)Herman, Sievers, and Gao]{chptr3_001_recon}
Herman,~G.~S.; Sievers,~M.~R.; Gao,~Y. \emph{Phys. Rev. Lett.} \textbf{2000},
  \emph{84}, 3354--3357\relax
\mciteBstWouldAddEndPuncttrue
\mciteSetBstMidEndSepPunct{\mcitedefaultmidpunct}
{\mcitedefaultendpunct}{\mcitedefaultseppunct}\relax
\EndOfBibitem
\bibitem[Yang et~al.(2007)Yang, Sun, Qiao, Zou, Smith, Cheng, and
  Lu]{chptr3_nature_001}
Yang,~H.~G.; Sun,~C.~H.; Qiao,~S.~Z.; Zou,~G.,~Jinand~Liu; Smith,~S.~C.;
  Cheng,~H.~M.; Lu,~G.~Q. \emph{Nature} \textbf{2007}, \emph{453},
  638--641\relax
\mciteBstWouldAddEndPuncttrue
\mciteSetBstMidEndSepPunct{\mcitedefaultmidpunct}
{\mcitedefaultendpunct}{\mcitedefaultseppunct}\relax
\EndOfBibitem
\bibitem[Selcuk and Selloni(2013)]{chptr3_HF-001_reconstructs}
Selcuk,~S.; Selloni,~A. \emph{The Journal of Physical Chemistry C}
  \textbf{2013}, \emph{117}, 6358--6362\relax
\mciteBstWouldAddEndPuncttrue
\mciteSetBstMidEndSepPunct{\mcitedefaultmidpunct}
{\mcitedefaultendpunct}{\mcitedefaultseppunct}\relax
\EndOfBibitem
\bibitem[Hara et~al.(2003)Hara, Sato, Katoh, Furube, Ohga, Shinpo, Suga,
  Sayama, Sugihara, and Arakawa]{Hara_c343}
Hara,~K.; Sato,~T.; Katoh,~R.; Furube,~A.; Ohga,~Y.; Shinpo,~A.; Suga,~S.;
  Sayama,~K.; Sugihara,~H.; Arakawa,~H. \emph{The Journal of Physical Chemistry
  B} \textbf{2003}, \emph{107}, 597--606\relax
\mciteBstWouldAddEndPuncttrue
\mciteSetBstMidEndSepPunct{\mcitedefaultmidpunct}
{\mcitedefaultendpunct}{\mcitedefaultseppunct}\relax
\EndOfBibitem
\bibitem[Frontiera et~al.(2009)Frontiera, Dasgupta, and Mathies]{c343_femto}
Frontiera,~R.~R.; Dasgupta,~J.; Mathies,~R.~A. \emph{Journal of the American
  Chemical Society} \textbf{2009}, \emph{131}, 15630--15632, PMID:
  19860478\relax
\mciteBstWouldAddEndPuncttrue
\mciteSetBstMidEndSepPunct{\mcitedefaultmidpunct}
{\mcitedefaultendpunct}{\mcitedefaultseppunct}\relax
\EndOfBibitem
\bibitem[Sanchez-de Armas et~al.(2012)Sanchez-de Armas, San~Miguel, Oviedo, and
  Sanz]{c343_tddft}
Sanchez-de Armas,~R.; San~Miguel,~M.~A.; Oviedo,~J.; Sanz,~J.~F. \emph{Phys.
  Chem. Chem. Phys.} \textbf{2012}, \emph{14}, 225--233\relax
\mciteBstWouldAddEndPuncttrue
\mciteSetBstMidEndSepPunct{\mcitedefaultmidpunct}
{\mcitedefaultendpunct}{\mcitedefaultseppunct}\relax
\EndOfBibitem
\bibitem[Oviedo et~al.(2012)Oviedo, Zarate, Negre, Schott, Arratia-Pérez, and
  Sánchez]{c343_QMD}
Oviedo,~M.~B.; Zarate,~X.; Negre,~C. F.~A.; Schott,~E.; Arratia-Pérez,~R.;
  Sánchez,~C.~G. \emph{The Journal of Physical Chemistry Letters}
  \textbf{2012}, \emph{3}, 2548--2555\relax
\mciteBstWouldAddEndPuncttrue
\mciteSetBstMidEndSepPunct{\mcitedefaultmidpunct}
{\mcitedefaultendpunct}{\mcitedefaultseppunct}\relax
\EndOfBibitem
\bibitem[Perdew and Levy(1983)]{band-gap}
Perdew,~J.~P.; Levy,~M. \emph{Phys. Rev. Lett.} \textbf{1983}, \emph{51},
  1884--1887\relax
\mciteBstWouldAddEndPuncttrue
\mciteSetBstMidEndSepPunct{\mcitedefaultmidpunct}
{\mcitedefaultendpunct}{\mcitedefaultseppunct}\relax
\EndOfBibitem
\bibitem[Nilsing et~al.(2005)Nilsing, Persson, and Ojam\"{a}e]{FWHM}
Nilsing,~M.; Persson,~P.; Ojam\"{a}e,~L. \emph{Chemical Physics Letters}
  \textbf{2005}, \emph{415}, 375 -- 380\relax
\mciteBstWouldAddEndPuncttrue
\mciteSetBstMidEndSepPunct{\mcitedefaultmidpunct}
{\mcitedefaultendpunct}{\mcitedefaultseppunct}\relax
\EndOfBibitem
\bibitem[Ernstorfer et~al.(2006)Ernstorfer, Gundlach, Felber, Storck,
  Eichberger, and Willig]{cvp_elect}
Ernstorfer,~R.; Gundlach,~L.; Felber,~S.; Storck,~W.; Eichberger,~R.;
  Willig,~F. \emph{The Journal of Physical Chemistry B} \textbf{2006},
  \emph{110}, 25383--25391, PMID: 17165985\relax
\mciteBstWouldAddEndPuncttrue
\mciteSetBstMidEndSepPunct{\mcitedefaultmidpunct}
{\mcitedefaultendpunct}{\mcitedefaultseppunct}\relax
\EndOfBibitem
\bibitem[Brennan et~al.(2013)Brennan, Llansola~Portoles, Liddell, Moore, Moore,
  and Gust]{cvp2}
Brennan,~B.~J.; Llansola~Portoles,~M.~J.; Liddell,~P.~A.; Moore,~T.~A.;
  Moore,~A.~L.; Gust,~D. \emph{Phys. Chem. Chem. Phys.} \textbf{2013},
  \emph{15}, 16605--16614\relax
\mciteBstWouldAddEndPuncttrue
\mciteSetBstMidEndSepPunct{\mcitedefaultmidpunct}
{\mcitedefaultendpunct}{\mcitedefaultseppunct}\relax
\EndOfBibitem
\bibitem[Brewster et~al.(2013)Brewster, Konezny, Sheehan, Martini,
  Schmuttenmaer, Batista, and Crabtree]{Carb_vs_phos}
Brewster,~T.~P.; Konezny,~S.~J.; Sheehan,~S.~W.; Martini,~L.~A.;
  Schmuttenmaer,~C.~A.; Batista,~V.~S.; Crabtree,~R.~H. \emph{Inorganic
  Chemistry} \textbf{2013}, \emph{52}, 6752--6764\relax
\mciteBstWouldAddEndPuncttrue
\mciteSetBstMidEndSepPunct{\mcitedefaultmidpunct}
{\mcitedefaultendpunct}{\mcitedefaultseppunct}\relax
\EndOfBibitem
\bibitem[Sodeyama et~al.(2012)Sodeyama, Sumita, O'Rourke, Terranova, Islam,
  Han, Bowler, and Tateyama]{prot_stab}
Sodeyama,~K.; Sumita,~M.; O'Rourke,~C.; Terranova,~U.; Islam,~A.; Han,~L.;
  Bowler,~D.~R.; Tateyama,~Y. \emph{The Journal of Physical Chemistry Letters}
  \textbf{2012}, \emph{3}, 472--477\relax
\mciteBstWouldAddEndPuncttrue
\mciteSetBstMidEndSepPunct{\mcitedefaultmidpunct}
{\mcitedefaultendpunct}{\mcitedefaultseppunct}\relax
\EndOfBibitem
\bibitem[fig()]{figshare}
 \url{http://dx.doi.org/10.6084/m9.figshare.961771}\relax
\mciteBstWouldAddEndPuncttrue
\mciteSetBstMidEndSepPunct{\mcitedefaultmidpunct}
{\mcitedefaultendpunct}{\mcitedefaultseppunct}\relax
\EndOfBibitem
\end{mcitethebibliography}
\end{document}